\documentclass{aa}

\usepackage[T1]{fontenc}

\usepackage{graphicx}   

\usepackage{amsmath}
\usepackage{amssymb}
\usepackage{mathptmx}
\usepackage{txfonts}

\usepackage[colorlinks=true,linkcolor=blue, citecolor=blue, filecolor=blue, urlcolor=blue]{hyperref} 

\usepackage{float}
\usepackage{color}
\usepackage{enumitem}

\usepackage[none]{hyphenat} %Para no romper palabras

\begin{document}

\title{Arm morphology in off-centre barred galaxies}

% The list of authors, and the short list which is used in the headers.
% If you need two or more lines of authors, add an extra line using \newauthor
\author{
P. S\'anchez-Mart\'{i}n\inst{1}\thanks{E-mail: patricia.sanchez.martin@upc.edu} \and
M. L\'opez-Vilamaj\'o\inst{2} \and M. Romero-G\'omez \inst{2,3,5}
\and J. J. Masdemont\inst{1,4,5}
}

% List of institutions
\institute{Dept. de Matem\`{a}tiques, Universitat Polit\`{e}cnica de Catalunya, Barcelona, Spain
%\email{patricia.sanchez.martin@upc.edu}
\and
Departament de F\'isica Qu\`antica i Astrof\'isica (FQA), Universitat de Barcelona (UB), Barcelona, Spain
\and
Institut de Ci\`encies del Cosmos (ICCUB), Universitat de Barcelona, Barcelona, Spain
\and
IMTEch, Universitat Polit\`{e}cnica de Catalunya, Barcelona, Spain %\email{josep.masdemont@upc.edu}
\and
Institut d’Estudis Espacials de Catalunya (IEEC), Barcelona, Spain
}

\date{Accepted XXX. Received YYY; in original form ZZZ}

\abstract
  % context heading (optional)
   {Many barred galaxies, including the Large Magellanic Cloud (LMC), display strong lopsidedness and off-centre bars. The dynamical connection between bar–disc misalignments, internal mass asymmetries, and arm morphology is not yet fully characterised.}
  % aims heading (mandatory)
   {We investigate how internal mass imbalances within the bar and global offsets between the bar and the centre of mass of the system modify the equilibrium-point structure and the invariant manifolds that organise arms.}
  % methods heading (mandatory)
   {We construct a barred galaxy model which includes an off-centred and asymmetric in shape bar. Using numerical continuation, we track the position, stability, and bifurcations of the Lagrangian equilibrium points as functions of the displacement of the asymmetric mass component along the bar and of the offset between the bar and the system’s centre of mass. For representative configurations we compute the invariant manifolds of planar Lyapunov orbits around unstable points and analyse the resulting arm structures.}
   % results heading (mandatory)
   {Internal bar lopsidedness and modest bar–disc offsets that keep the centre of mass inside the bar preserve the classical configuration with five equilibrium points, but strongly distort the associated invariant manifolds, producing two arms with different densities and shapes. The bar–disc offset reaches a threshold at the point at which the galactic centre of mass exits the bar ellipsoid, in which a pitchfork bifurcation removes the collinear unstable points and the system transitions to a three–equilibrium–point configuration in which a single unstable point and its associated manifold supports one arm.}
   % conclusions heading (optional)
   {Internal inhomogeneities in the bar's density and modest bar–disc offsets dynamically support two asymmetric arms, while larger offsets drive a transition to one-armed structures.This framework is compatible with the observed correlation between off-centre bars and photometric lopsidedness, and it provides a dynamical explanation for the strongly asymmetric arm morphology of galaxies such as the LMC.}
     

% \keywords{
% galaxies: kinematics and dynamics -- galaxies: structure -- galaxies: spiral -- methods: numerical
% }

% \titlerunning{}

\maketitle
\nolinenumbers

%%%%%%%%%%%%%%%%%%%%%%%%%%%%%%%%%%%%%%%%%%%%%%%%%%

%%%%%%%%%%%%%%%%% BODY OF PAPER %%%%%%%%%%%%%%%%%%

%====================================
% \mainmatter
%\input{1-intro}
\section{Introduction}
\label{sec:intro}

While theoretical models often idealize barred galaxies as bisymmetric systems centred on the centre of the galaxy, observational evidence indicates that deviations from this symmetry are common. Several barred systems display off-centre or intrinsically asymmetric bars. Classic examples such as NGC 4027 and NGC 55 show a bar centre clearly displaced from their outer disc \citep{deVaucouleurs1970}. Deep infrared imaging has revealed a similarly offset bar in NGC 3906, where the stellar bar centre matches with the dynamical centre while the photometric disc centre lags by approximately $0.9$ kpc \citep{deSwardt2015}.

Systematic studies corroborate that these are not isolated cases. \citet{Kruk2017} conducted a search using the Sloan Digital Sky Survey \citep{York2000SDSS} and the Galaxy Zoo project \citep{Willett2013Zoo}, finding that approximately 270 local barred galaxies possess measurable bar–disc offsets of $0.2$--$2.5$ kpc. Notably, the authors found that 90\% of galaxies with off-centre bars are lopsided.

Intrinsic asymmetries within the bar mass distribution itself are also significant. \citet{Lokas2021} studied lopsided bars using the IllustrisTNG simulations \citep{Springel2018}, showing that mass asymmetries are not restricted to a specific position but can occur anywhere from the centre to the semi-major axis edge of the bar. Similarly, \citet{Patra2019} analyzed DDO 168, a system presenting a clearly lopsided bar.

Perhaps the most prominent local example of such asymmetries is the Large Magellanic Cloud (LMC). The LMC possesses a bar that is both off-centre with respect to the disc and intrinsically lopsided \citep[see e.g.][]{Jacys2017, VanDerMarel2001}. Recent studies of the LMC have focused on the analysis of its velocity maps using Gaia DR3 data \citep{JimenezArranz2023, JimenezArranz2025, Scholch2025}, further highlighting the complexity of its central dynamics. Furthermore, in a recent set of simulations modelling the interaction between the LMC and the SMC, \citet{JimenezArranz2024Kratos} showed that the bar presents a clear offset with respect to the centre of mass defined by the baryonic matter.

The origin of these asymmetries remains a subject of debate. While interactions with companions can produce offset bars \citep{AthanPuerari1997, Pardy2016, Scholch2025}, isolated galaxies also exhibit relevant lopsidedness \citep{deSwardt2015, JogCombes2009, Kruk2017, Zaritsky2013}, suggesting internal mechanisms or long-lived modes may be at play. For instance, \citet{JogCombes2009} point out that the continuous accretion of gas is another viable mechanism capable of exciting $m=1$ lopsided instabilities. \citet{Kruk2017} posit that a possible explanation could be a misalignment between the dark matter halo and the disc centres, as modelled by \citet{Levine1998} using N-body simulations. \citet{Zaritsky2013} reached similar conclusions, suggesting that lopsidedness is fundamentally related to asymmetries in the mass component distributions. Regardless of how the asymmetry is produced, the end result is an asymmetric mass distribution that establishes a specific gravitational potential. The use of an idealized asymmetric potential serves as a necessary mathematical tool to compute and represent the invariant manifolds that outline the arm structures. Therefore, this work focuses on isolating the underlying dynamical skeleton of the system once this asymmetric potential is in place.

The goal of this work is to analyse the effect produced by an offset bar and/or a bar with an asymmetric mass distribution on the number and density of the galactic arms. In a previous study \citep{Asymmetry}, we examined how a modest bar asymmetry, amounting to 15\% of the bar's mass and confined near its centre, induces a clear density imbalance in the galactic arms. Building on that work, the present study extends the analysis in two directions: first, by distributing the asymmetry along the entire bar length; and second, by allowing the centre of the bar to shift away from the centre of mass (CM) of the system. These configurations mimic the features observed in lopsided galaxies and produce bifurcations in the Lagrangian equilibrium points of the system, which in turn generate asymmetries in the resulting arm morphology. In fact, morphological imbalances can become quite extreme; the emergence of a single spiral arm is essentially a manifestation of this lopsidedness, dynamically understood as an $m=1$ Fourier mode with a radially dependent phase \citep{Jog1997, JogCombes2009}.

We perform a detailed parameter study of these bifurcations as functions of the displacement of the asymmetric mass component within the bar and of the bar-CM offset. The models studied in this paper can be considered, to some extent, as an extension of the models proposed by \citet{Colin1989} and \citet{deVaucouleurs1970}, which contemplated the possibility of an offset bar. In our work, we consider this case and, moreover, that of an intrinsically asymmetric bar, analysing in depth the dynamical features of the system in both scenarios.

To conduct this analysis, we rely on the theory of Invariant Manifolds \citep{Efthym2010, Harsoula2016, Romero1, Romero2, Rom09, Warps, Asymmetry, Tsigaridi2013, Tsoutsis2008}. This theory establishes that, when modelling a barred galaxy as a dynamical system \citep{BinneyTremaine}, the unstable equilibrium points (which appear in the frame of reference corotating with the bar) have associated Lyapunov periodic orbits. From these orbits emanate invariant manifolds (both stable and unstable) that delineate the galactic arms. Because these invariant manifolds are dynamical structures intrinsically tied to the bar, they provide the causal link between the displacement or intrinsic asymmetry of the bar and the resulting arm morphology. Under this dynamical framework, in the case of an offset bar, regardless of the mechanism that displaces it, the associated invariant manifolds would naturally be dragged along by this displacement. The material guided by these shifted manifolds would then form the observed two-armed or one-armed structure. \citet{Athan2012, Efthym2019, BarDetection} showed the strong relation between this theory and N-body simulations. Similar conclusions were reached in \citet{Martinez2011} by comparing manifold models with a subsample of barred galaxies provided by the Ohio State University Bright Galaxy Survey \citep{Eskridge2002}.

The paper is structured as follows: Section~\ref{sec:model} introduces our dynamical model, including potential components and parameter choices. In Section~\ref{sec:bifurcations}, we track how equilibrium points evolve and bifurcate under varying asymmetry parameters using numerical continuation methods. Section~\ref{sec:manifolds} analyses the invariant manifolds of planar Lyapunov orbits and shows how their geometry governs global arm structure across our three model configurations. Finally, Section~\ref{sec:discussion} places our results in an observational context, while Section~\ref{sec:conclusions} summarises our main findings.

%%%%%%%%%%%%%%%%%%%%%%%%%%%%%%%%%%%%%%%%%%%%%%%%%%%%%%%%%%%%%%%%%%%%%%%%%%%%%%%%%%%%%%
%\input{2-model}
\section{Dynamical system models}
\label{sec:model}

We describe the dynamics of the galaxy in a rotating frame of reference whose origin is located at the centre of mass (CM) of the system. In this non-inertial frame, the equations of motion for a test particle with position vector $\mathbf{r}=(x, y, z)$ subject to a total gravitational potential $\phi(x,y,z)$ are given by:
\begin{equation}
\left\{\begin{aligned}
  \ddot{x} &=  2\,\Omega\, \dot{y} + \Omega^2\, x - \phi_{x}, \\
  \ddot{y} &= -2\,\Omega\, \dot{x} + \Omega^2\, y - \phi_{y}, \\
  \ddot{z} &= -\phi_{z}.
\end{aligned}
\right.
\label{eqn:systmodelLMC}
\end{equation}
where the subscripts in $\phi$ denote partial derivatives with respect to the coordinates, and $\Omega$ represents the constant angular velocity (pattern speed) of the bar about the $z$-axis. It is convenient to introduce the effective potential, $\phi_{\text{\scriptsize eff}}$, defined as the sum of the gravitational and centrifugal potentials:
\[
\phi_{\text{\scriptsize eff}}(x,y,z) = \phi(x,y,z) - \frac{1}{2}\,\Omega^2\, (x^2 + y^2).
\]

This dynamical system admits a Hamiltonian formulation and possesses a first integral of motion known as the Jacobi integral, $C_J$, which is related to the total energy in the rotating frame. We define the Jacobi constant as:
\begin{equation}\label{eqn:PreCJAC}
   C_J(x,y,z,\dot{x},\dot{y},\dot{z}) = -\,(\dot{x}^2+\dot{y}^2+\dot{z}^2)-2\,\phi_{\text{\scriptsize eff}}(x,y,z).
\end{equation}
Regions of forbidden motion for a particle with a given $C_J$ are those where $\phi_{\text{\scriptsize eff}} > C_J$.

The total potential $\phi$ is modelled as the sum of contributions from the components of the system. Specifically, we consider $\phi = \phi_d + \phi_b + \phi_h$, where each term represents the potential due to the disc, bar (which includes an asymmetric mass component), and halo, respectively.

The disc potential, $\phi_d$, follows the Miyamoto-Nagai model \citep{Miyamoto}:
\begin{equation}\label{eqn:Miyamoto}
 \phi_d = - \frac{GM_d}{\sqrt{R^2+(A+\sqrt{B^2+z^2})^2}},
\end{equation}
where $R^2 = x^2+y^2$ is the cylindrical radial coordinate, $z$ is the vertical height, $M_d$ is the total mass of the disc, and $A$ and $B$ are scale lengths determining the radial and vertical profiles, respectively.

The halo ($\phi_h$) is modelled using a spherical Plummer potential \citep{Plummer1911}, which can be derived from the Miyamoto-Nagai potential by setting the radial scale length $A = 0$.

In this study we model the bar potential, $\phi_b$, as a superposition of a Ferrers ellipsoid, with homogeneity index of $n_h = 2$ \citep{Ferrers}, and a spherical Plummer potential. This superposition allows us to model an asymmetric bar density distribution, by displacing the spherical distribution along the bar major axis. The Ferrers ellipsoid density is expressed by:
\begin{equation}\label{eqn:Ferrers}
 \rho = 
 \left\lbrace
 \begin{array}{ll}
  \rho_0(1-m^2)^{n_h}, & m\leq 1 \\
  0,   & m > 1 \\
 \end{array}
 \right.
\end{equation}
where $m^2=x^2/a^2 + y^2/b^2 + z^2/c^2$, with $a$, $b$, and $c$ representing the semi-axes of the bar. We adopt a homogeneity index of $n_h = 2$. The central density is then given by $\rho_0 = \frac{105}{32\pi}\frac{GM_b}{abc}$, where $M_b$ is the bar mass. The spherical Plummer bulge has radial scale-length of $1$\,kpc and it represents $18\%$ of the bar total mass.

\begin{figure}
\centering
\includegraphics[width=0.3\textwidth]{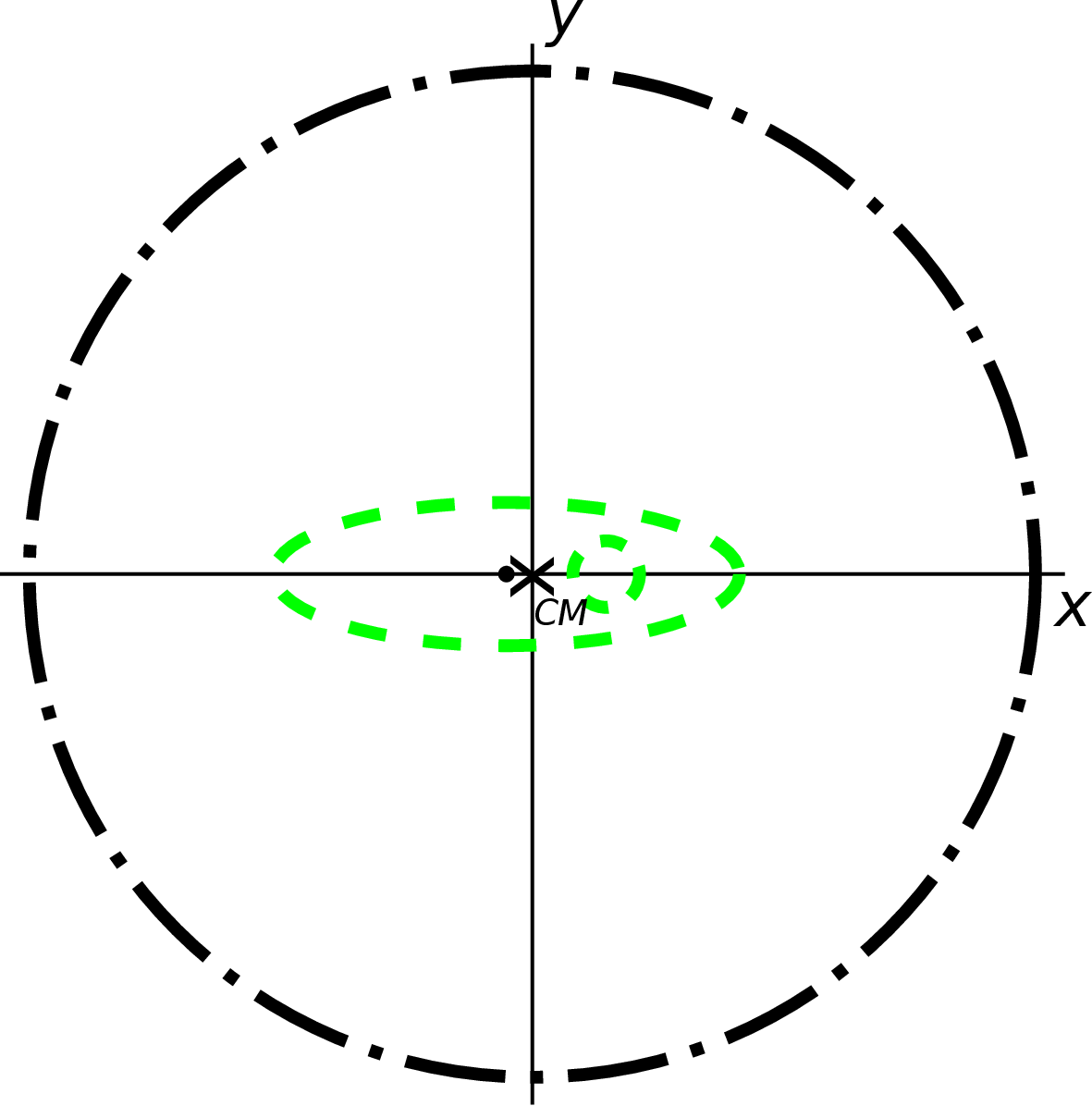}
\includegraphics[width=0.3\textwidth]{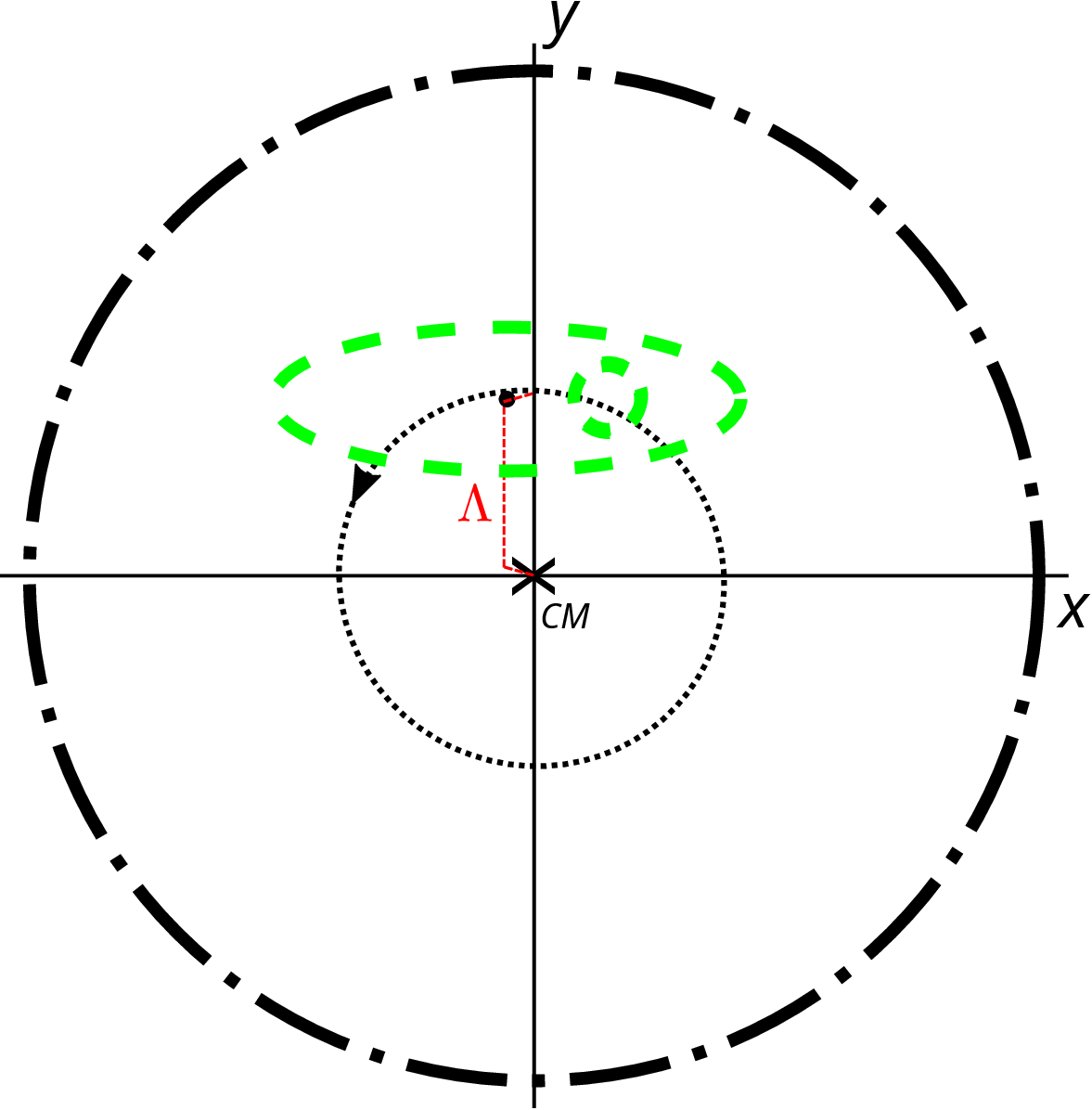}
\caption{Schematic of the lopsided galaxy configurations. Top: Model A, where the bar major axis is aligned with the system’s centre of mass (CM). Bottom: Models B and C, where the bar is offset from the CM, with $\Lambda$ representing the distance from the CM to the bar major axis. The bar and the asymmetric mass component are outlined by green dashed lines; the disc contour is shown as a black dash-dotted line; the black cross marks the system CM; and the black filled circle indicates the geometric centre of the bar.
} \label{fig:models}
\end{figure}

The adopted numerical parameters for the galactic components are summarized below:

\begin{center}
\resizebox{\columnwidth}{!}{%
\begin{tabular}{@{}l l@{}}
\hline
\textbf{Component} & \textbf{Parameters} \\
\hline
Disc ($d$) & $A=3$, $B=1$~kpc, $GM_d=0.34$~kpc$^3$/$u_t^2$ \\
Bar ($b$) & $a=6$, $b=1.5$, $c=0.4$~kpc, $GM_b=0.4$~kpc$^3$/$u_t^2$ \\
Asymmetric mass & $B_{am} = 1$~kpc, $GM_{am}=0.18\, GM_b$\\
component ($am$) &  \\
Halo ($h$) & $B_h=100$~kpc, $GM_h=0.4$~kpc$^3$/$u_t^2$ \\
\hline
\end{tabular}%
}
\end{center}

We adopt a system of units where $G(M_d+M_b+M_{am}+M_h) = 1$. The bar pattern speed is fixed to $\Omega=0.05$ [u$_t$]$^{-1}$ (approx. $24.46$~km/s/kpc), with a time unit of $u_t = 2 \times 10^6$ years.

To systematically study the impact of asymmetry, we define three distinct model configurations (see Fig.~\ref{fig:models}):
\begin{description}
\item[Model A:] The bar's major axis passes through the system's centre of mass (CM). The asymmetry is purely internal to the bar (via the asymmetric mass component displacement).
\item[Model B:] The bar's major axis is offset from the CM, but the CM remains strictly \textit{inside} the bar ellipsoid. The offset parameter $\Lambda$ (distance from CM to the major axis) satisfies $0 < \Lambda \leq 1.5$ kpc (where $1.5$ kpc is the bar's semi-minor axis $b$).
\item[Model C:] The offset is sufficiently large that the CM lies \textit{outside} the bar, i.e., $\Lambda > 1.5$ kpc.
\end{description}

These configurations are chosen to cover the full range of morphological distortions observed in lopsided systems, where the dynamical centre (kinematic centre) and the photometric centre of the bar do not coincide.

Note that in all models the geometric centre of the disc coincides with the centre of mass of the system. This setup follows theoretical frameworks for asymmetric galaxies \citep[e.g.,][]{deVaucouleurs1972, Colin1989}. Observational support for this choice is found in systems like the LMC, where the stellar dynamical centre (which defines the rotating frame) aligns precisely with the H~{\sc i} gas dynamical centre rather than the bar \citep{vanderMarel2014}. Furthermore, observations of galaxies with off-centre bars, such as NGC 3906 \citep{deSwardt2015}, reveal physical offsets between the photometric centre of the bar and the dynamical centre of the system, an asymmetry that falls within our parameter space (Model B). It is important to emphasize that, unlike in purely symmetric galaxies, the global centre of mass in systems with asymmetric mass distributions does not mathematically have to coincide with the global minimum of the gravitational potential, as the latter is locally dominated by the dense bar structure. Numerically, we enforce this condition by displacing the halo component. Due to the low density of the halo in comparison to the other components of the galaxy, this displacement has a negligible effect on the dynamics of the studied systems, which are mainly driven by the offset of the bar.

In all models, the internal mass asymmetry of the bar is modelled by displacing the asymmetric mass component along the bar's major axis. This displacement is quantified by the parameter $\delta$, representing the distance from the bar's geometric centre to the centre of this asymmetric component. Model A with small $\delta$ was analyzed in \citet{Asymmetry}; here we extend $\delta$ to the full length of the bar and introduce the $\Lambda$-offset to capture the more complex dynamics of strongly interacting or perturbed galaxies. The dynamics of the symmetric reference case ($\delta=0, \Lambda=0$) are well documented in \citet[e.g.][]{Romero1, Romero2, Warps}.

%%%%%%%%%%%%%%%%%%%%%%%%%%%%%%%%%%%%%%%%%%%%%%%%%%%%%%%%%%%%%%%%%%%%%%%%%%%%%%%%%
%\input{3-bifurc}
\section{Equilibrium points and bifurcations}
\label{sec:bifurcations}

\begin{figure}
\centering
\includegraphics[width=0.4\textwidth]{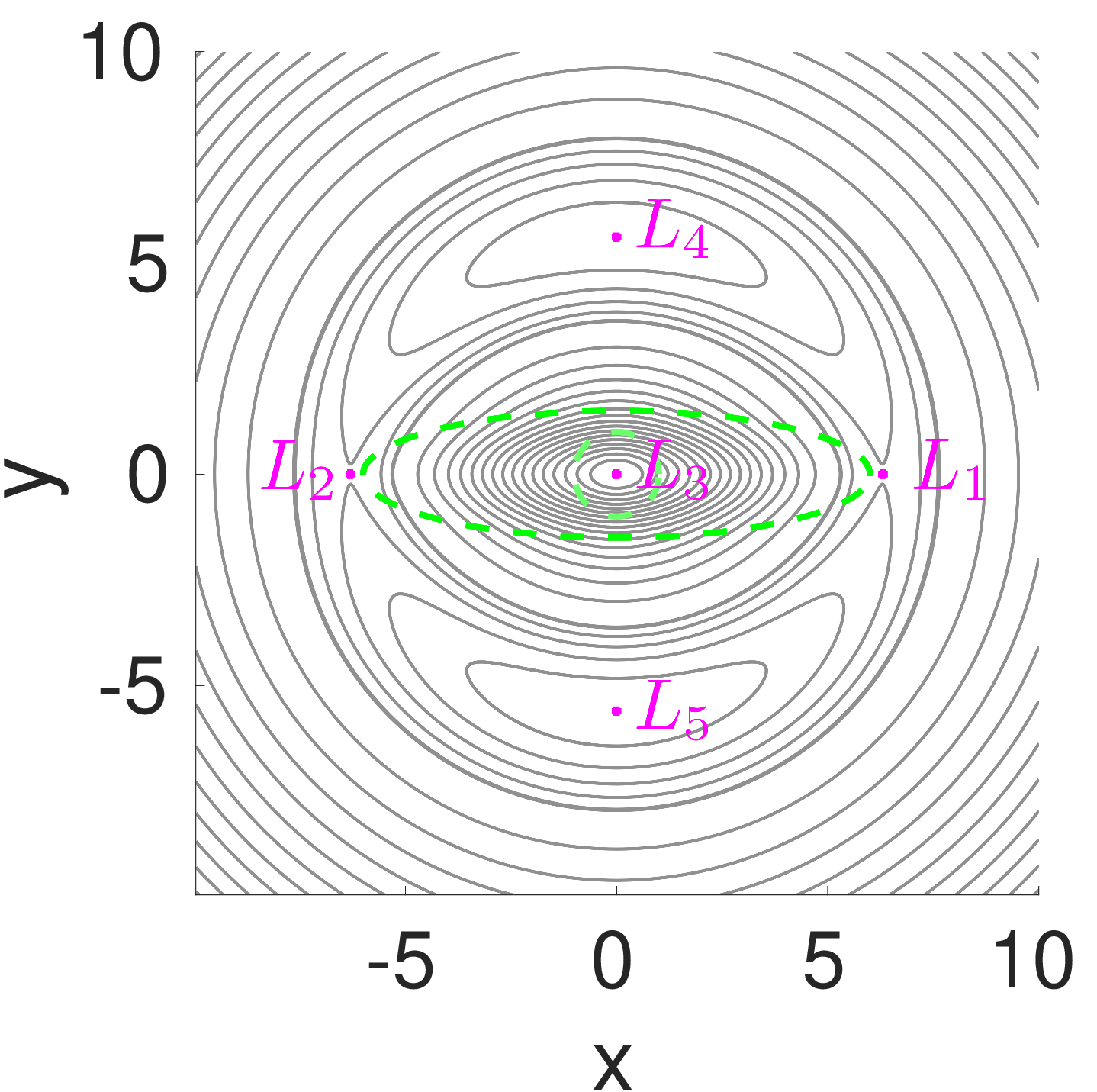}
\caption{Equilibrium points in the symmetric reference model. Bar and the asymmetric mass component outlined by green dashed lines. Five Lagrangian points shown as magenta dots. Contour levels of the effective potential highlight the dynamics near the unstable points L$_1$ and L$_2$.}
\label{fig:eqpoints}
\end{figure}

The Lagrangian equilibrium points of the system in the rotating frame are defined as critical points of the effective potential, satisfying $\nabla \phi_{\text{eff}} = 0$. These points are fundamental to understanding galactic dynamics: they organize the phase space structure and their stability determines whether material can be efficiently transported across the disc. In the classical symmetric case (Model A with $\delta=0$), five equilibrium points exist: the collinear points L$_1$ and L$_2$ located at the bar ends, the central point L$_3$, and the triangular points L$_4$ and L$_5$. Under standard parameter values, L$_1$ and L$_2$ are unstable points, while L$_3$, L$_4$, and L$_5$ are linearly stable. The configuration of these five points is shown in Fig.~\ref{fig:eqpoints}, which displays the effective potential that supports this equilibrium structure. Detailed analysis of the local dynamics around these points is available in prior work \citep[see e.g.][]{Athan1983, Romero1}.

When the bar becomes asymmetric or offset from the system's centre of mass, the five-point configuration may not persist. Bifurcations, qualitative changes in the number and stability of equilibrium points, can substantially alter the dynamical structure available for particle transport. To systematically understand how asymmetries reshape the equilibrium configuration and generate bifurcations, we employ numerical continuation methods using the \textsc{COCO} package \citep{COCO}, which tracks position and stability of equilibrium points as functions of our two key parameters: the internal mass imbalance $\delta$ (the asymmetric mass component displacement along the bar) and the bar offset $\Lambda$ (perpendicular distance of the bar's major axis from the system centre of mass). This bifurcation analysis provides the foundation for understanding how the arm structure, delineated by the invariant manifolds that emanate from Lyapunov periodic orbits around the unstable points, evolves in asymmetric systems (as will be discussed in Section~\ref{sec:manifolds}).

\subsection{Continuation with respect to the $\delta$ parameter: The role of internal mass imbalance}

To assess the individual contribution of the bar's internal mass imbalance, we evaluate the continuation of the equilibrium points across all three models exclusively as a function of the displacement of the asymmetric mass component ($\delta$ parameter).

\subsubsection{Model A: Bar with internal mass imbalance aligned with the centre of mass}

We begin by isolating the effect of internal asymmetry while keeping the bar axis along the system's CM ($\Lambda = 0$, Model A). This allows us to observe how the displacement of the asymmetric mass component along the major axis of the bar directly perturbs the potential structure.

\begin{figure}
\centering
\includegraphics[width=0.49\linewidth]{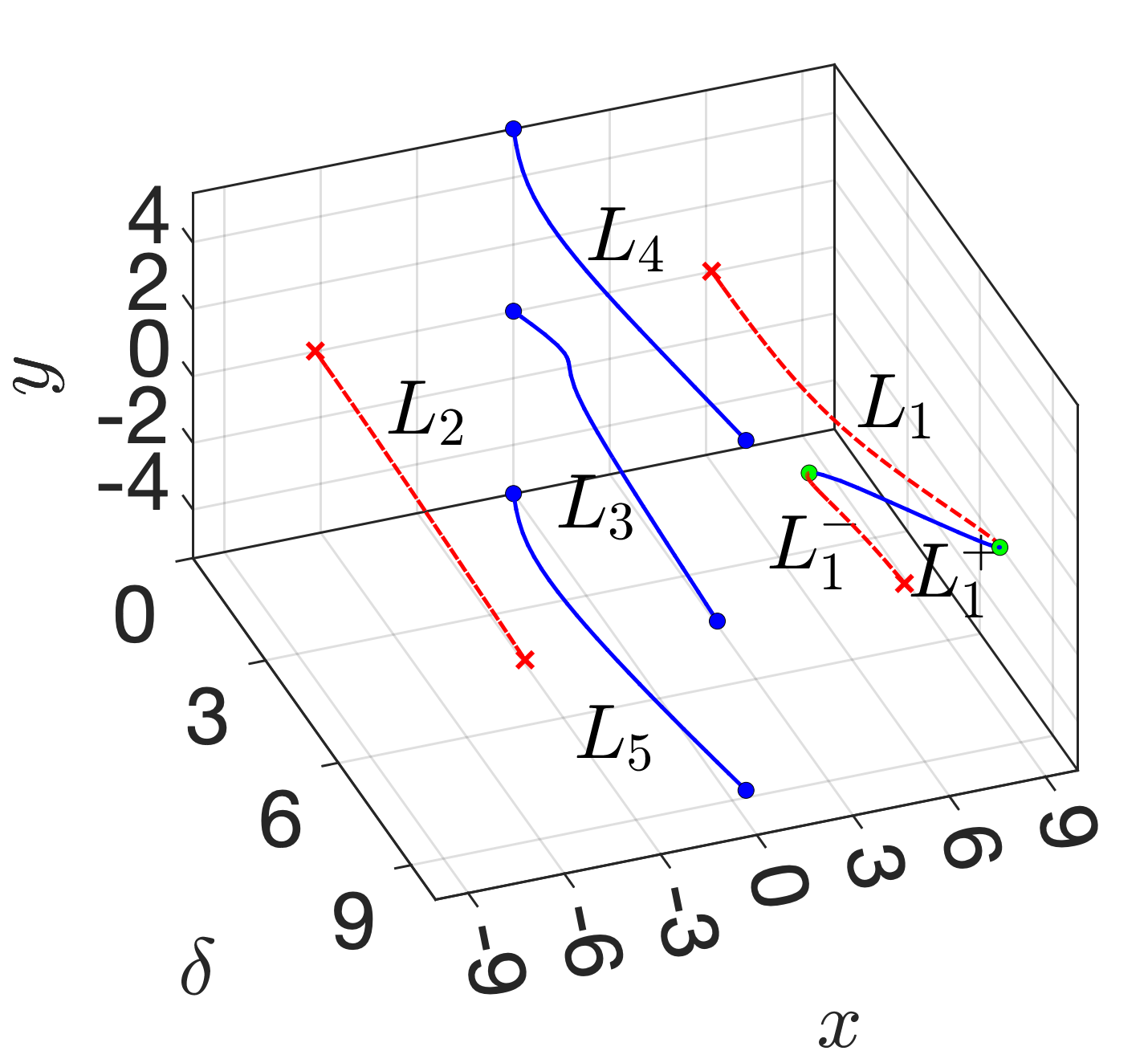}
\includegraphics[width=0.49\linewidth]{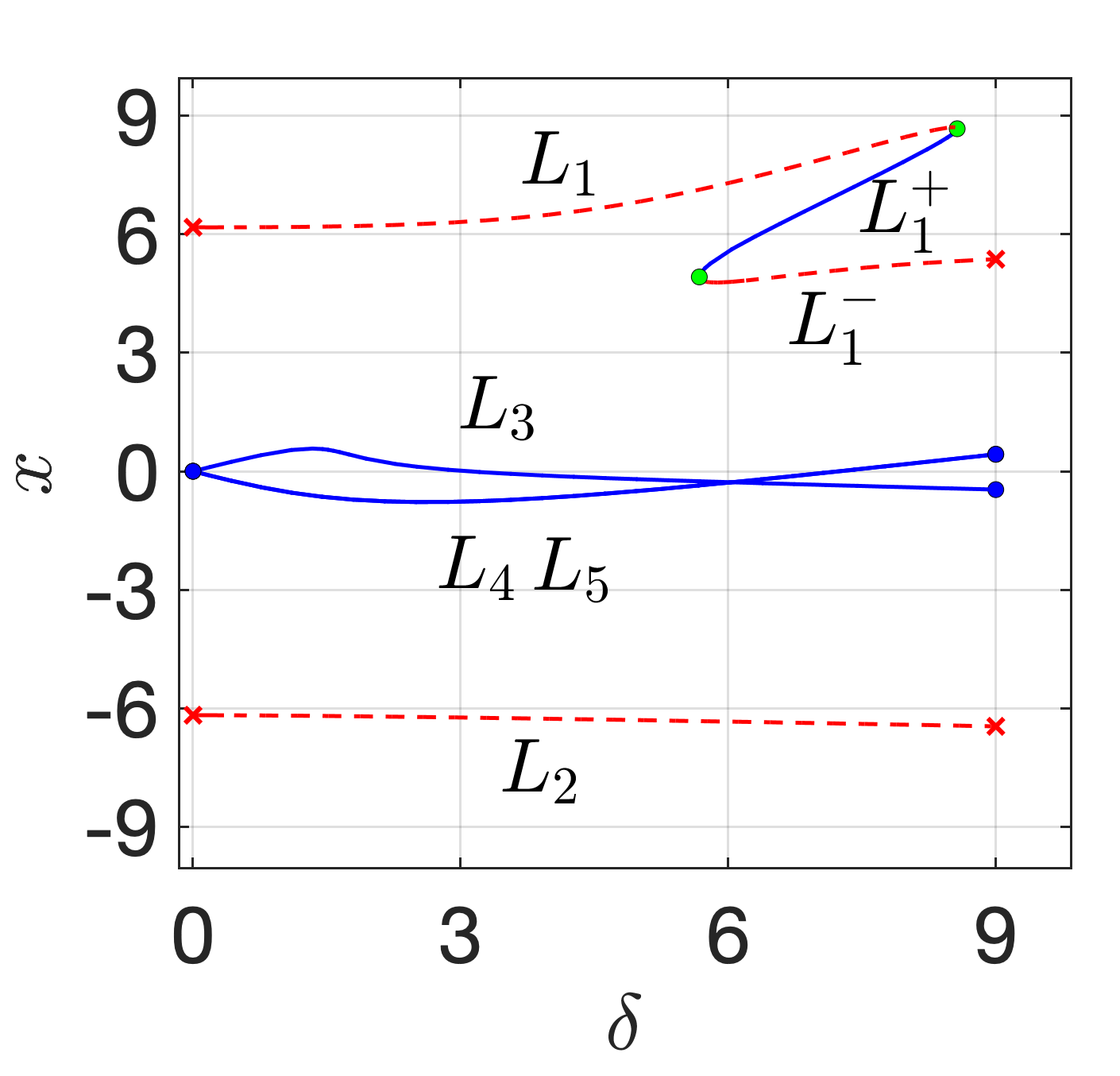} \\
\includegraphics[width=0.49\linewidth]{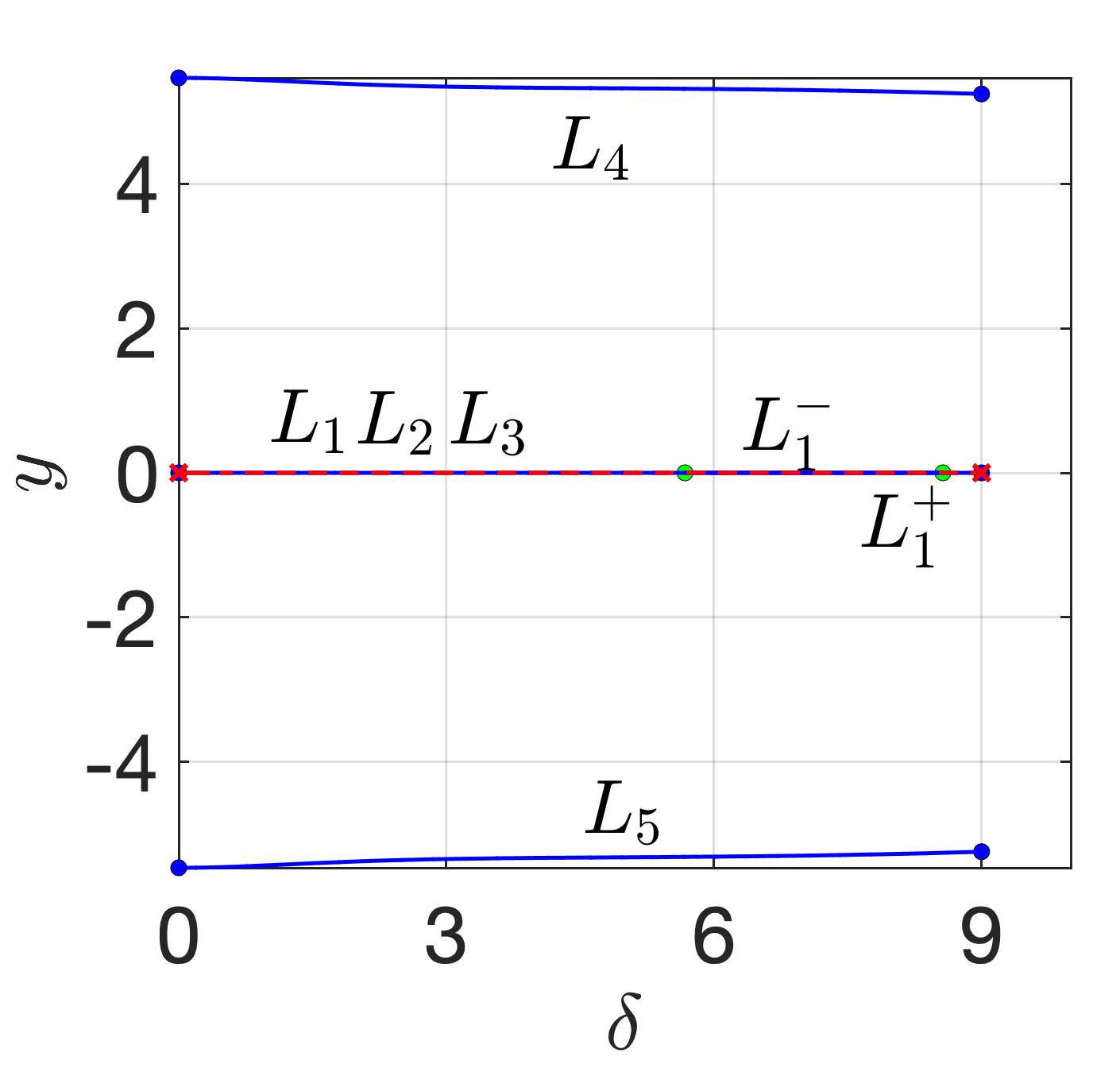}
\includegraphics[width=0.49\linewidth]{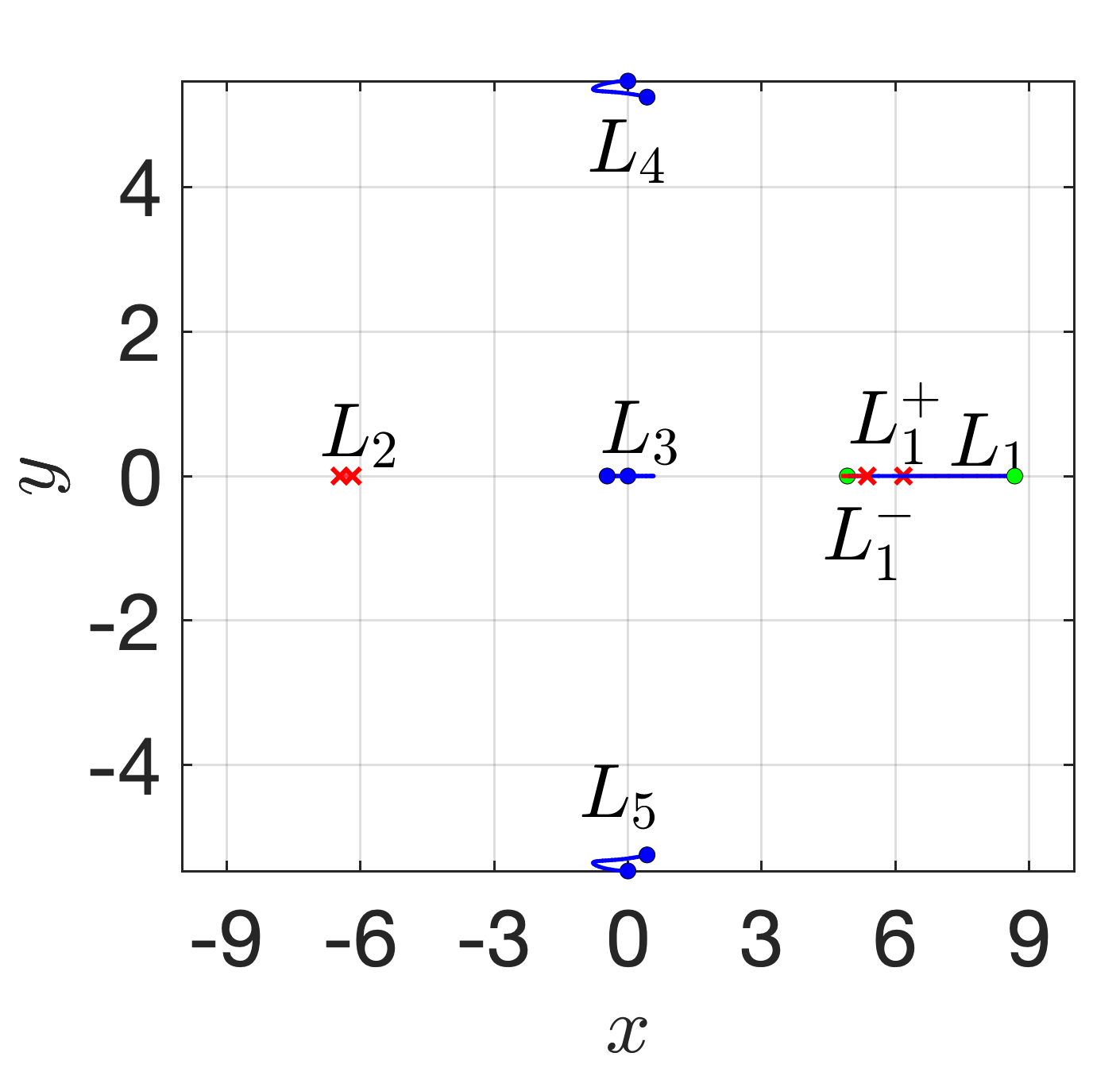}
\caption{Equilibrium point continuation for Model A ($\Lambda = 0$) as a function of $\delta$. Top panels: 3D representation of the continuation curves in $(x,y,\delta)$ space and the $\delta$--$x$ projection, showing a pair of turning points at $\delta \approx 5.7$ kpc and at $\delta \approx 8.6$ kpc displaying two saddle-node or fold bifurcations with branches  $L_1^+$ (stable) and $L_1^-$ (unstable) at the corresponding $x$-values. Bottom panels: $\delta$--$y$ projection showing collinear points remain on $y = 0$ while triangular points drift toward it, and spatial $(x$--$y)$ projection. Solid blue: stable branches; dashed red: unstable branches. Green dot marks the bifurcation.}
\label{fig:Bifur_xdb_A}
\end{figure}

Figure~\ref{fig:Bifur_xdb_A} shows the evolution of the equilibrium points as $\delta$ varies from 0 to 9 kpc ($\Lambda = 0$, towards $L_1$). The four panels, displaying the continuation in the full $(x, y, \delta)$ space together with three orthogonal projections, reveal that most equilibrium points undergo only modest displacements as the asymmetric component moves along the bar. In contrast, L$_1$ exhibits a pronounced response: at the critical values $\delta \approx 5.7$ kpc and $\delta \approx 8.6$ kpc the solution branch experiences two saddle–node (fold) bifurcations, corresponding to turning points in parameter space, where L$_1$ splits into two distinct equilibrium points, $L_1^-$ and $L_1^+$, with $x(L_1^-) < x(L_1^+)$. This sequence of bifurcations is significant because it temporarily produces a three–point subsystem that coexists with the original four equilibrium points.

For $\delta \in [5.7, 8.6]$~kpc, seven equilibrium points coexist in the system. This configuration is noteworthy because the displaced mass has moved beyond the edge of the bar, effectively creating a decoupled subsystem. At $\delta \approx 8.6$~kpc, $L_1^+$ merges with the original L$_1$ and both annihilate, returning the system to five equilibrium points: $L_1^-$, L$_2$, L$_3$, L$_4$, and L$_5$. The collinear points L$_1$, L$_2$, and L$_3$ remain confined to the $y = 0$ symmetry axis throughout the continuation, while the triangular points gradually approach this axis as the mass displacement increases.

\subsubsection{Model B: Small bar offset}

\begin{figure}
\centering
\includegraphics[width=0.49\linewidth]{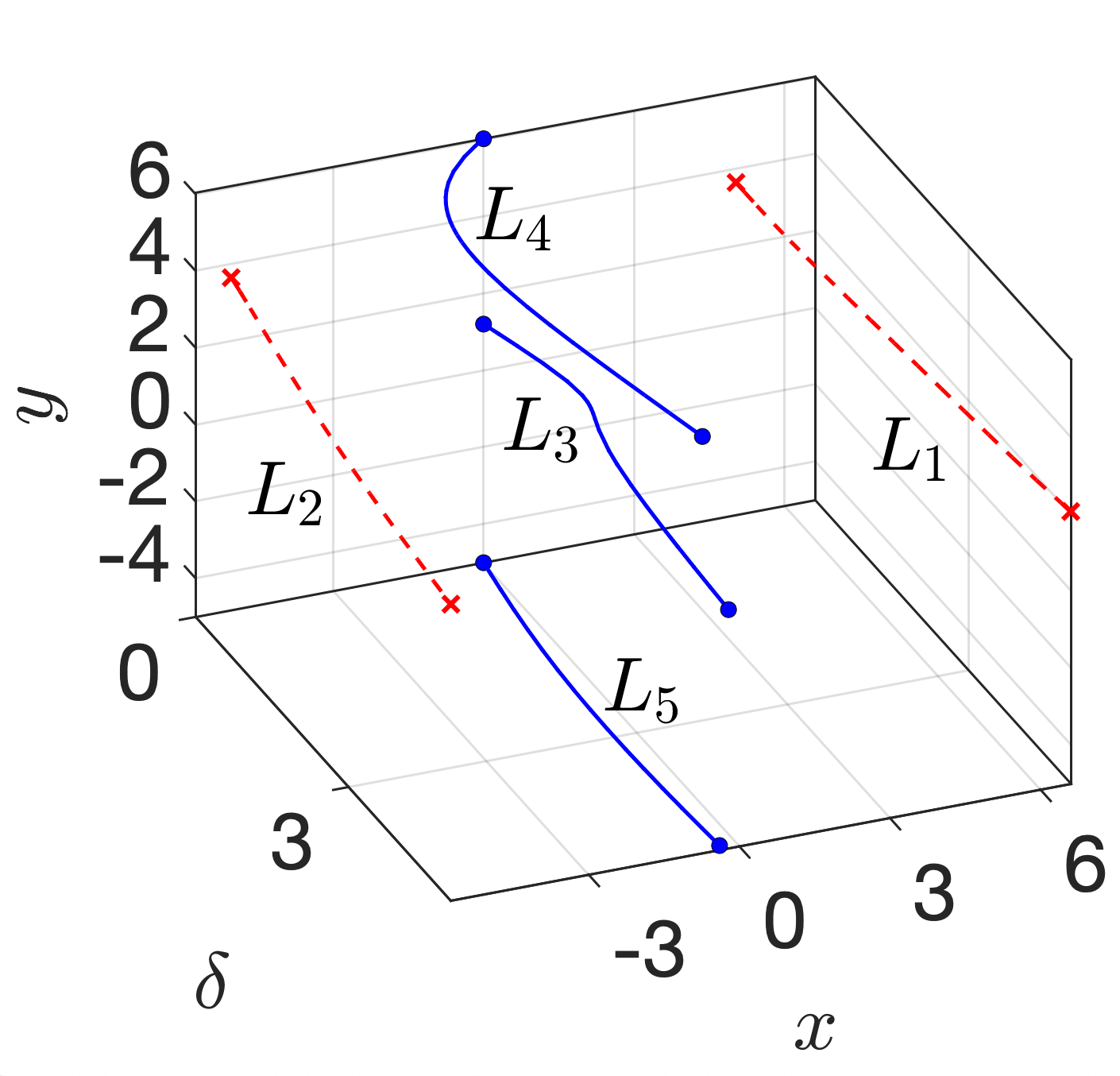}
\includegraphics[width=0.49\linewidth]{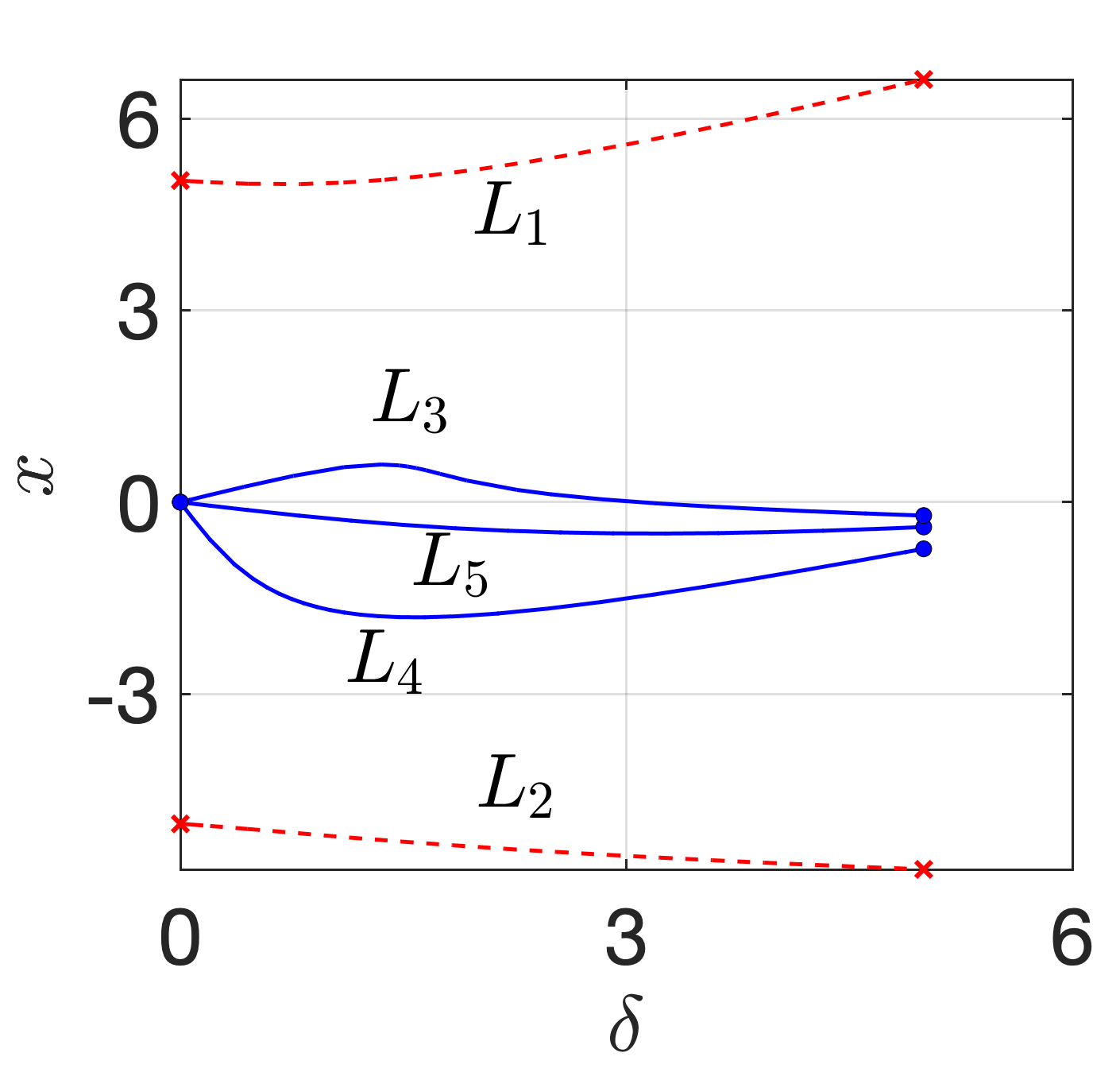} \\
\includegraphics[width=0.49\linewidth]{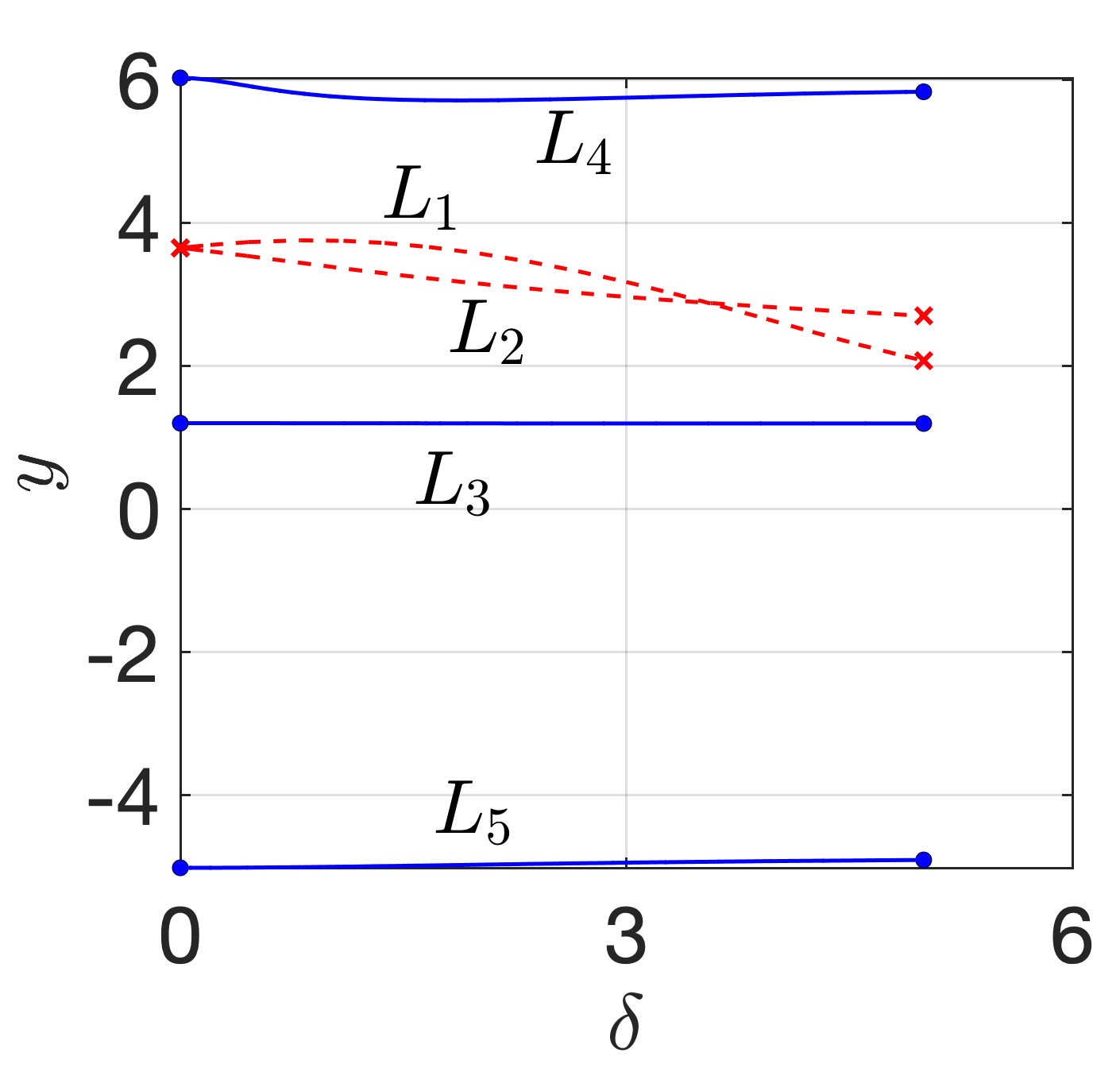}
\includegraphics[width=0.49\linewidth]{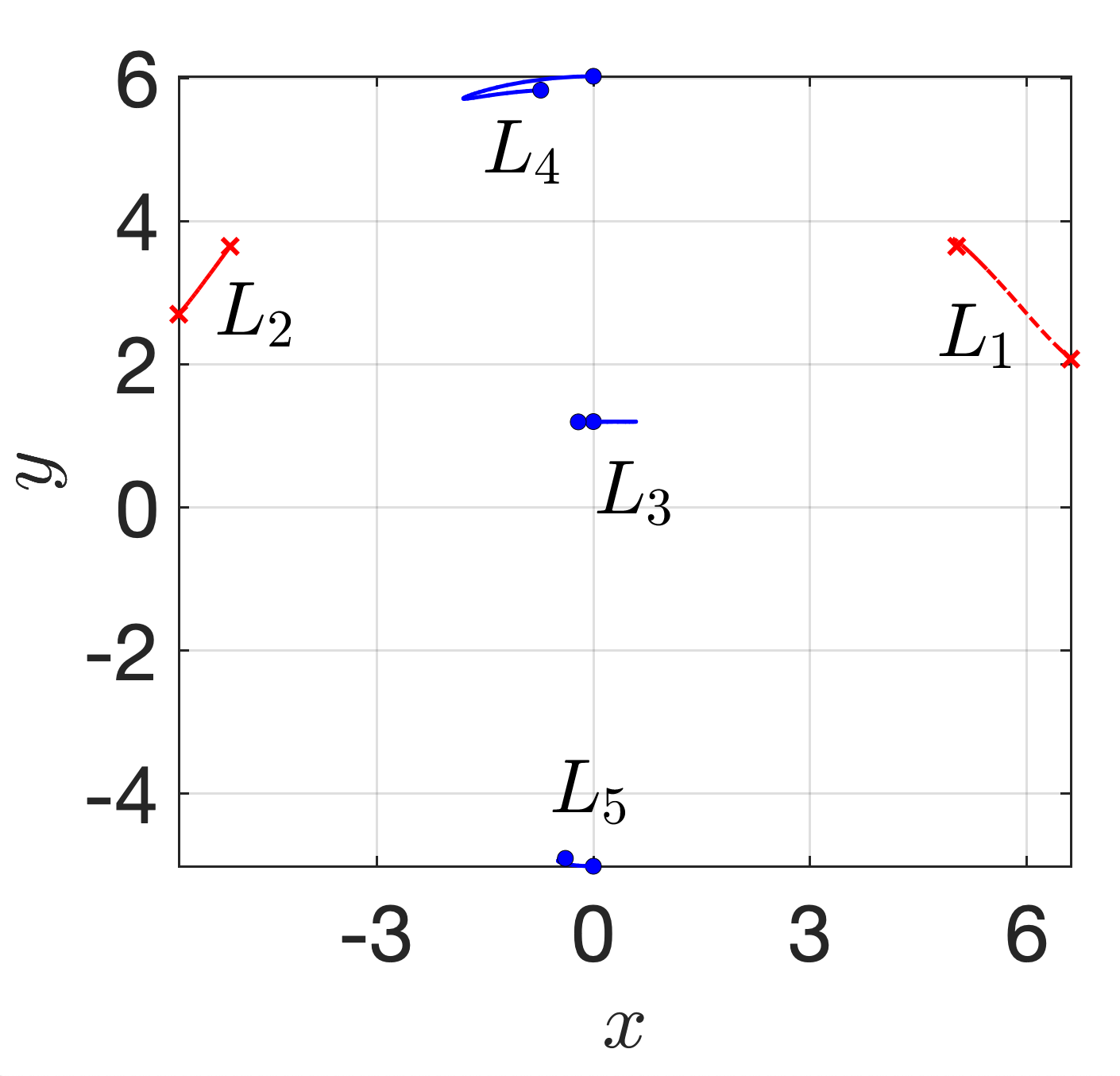}
\caption{Equilibrium point continuation for Model B ($\Lambda = 1.21$ kpc) as a function of $\delta$. Top panels: smooth 3D continuation curves with no bifurcations, and the $\delta$--$x$ projection showing monotonic evolution of L$_1$ and L$_2$ away from the centre of mass. Bottom panels: $\delta$--$y$ projection where linearly stable points remain approximately constant while unstable points shift toward the bar region, and spatial $(x$--$y)$ view. All equilibrium points are offset to higher $y$-values compared to Model A. Solid blue: stable branches; dashed red: unstable branches.}
\label{fig:Bifur_xdb_B}
\end{figure}

When a small bar offset ($\Lambda = 1.21$ kpc for this case) is introduced while maintaining the centre of mass interior to the bar, the bifurcation landscape changes qualitatively. Figure~\ref{fig:Bifur_xdb_B} shows that all equilibrium points follow smooth, continuous paths across the $\delta$ range from 0 to 5 kpc; no bifurcation occurs. This robustness is physically significant: despite the internal asymmetry, the offset acts to stabilize the five-point configuration, preserving the two unstable points essential for generating two arm structure. The offset simply translates all equilibrium points to higher $y$-values, effectively shifting the configuration perpendicularly to the bar without qualitative restructuring. As shown in Section~\ref{sec:manifolds}, this offset configuration modifies the geometry of the invariant manifolds, resulting in arms of unequal shape and density.

The collinear points L$_1$ and L$_2$ move progressively further from the centre of mass with increasing $\delta$, a trend consistent with Model A. The linearly stable equilibrium points L$_3$, L$_4$, and L$_5$ initially move away from the $x$-axis but reverse direction near $\delta \approx 1.5$ kpc, indicating a balance between competing effects of the offset and internal asymmetry. Importantly, the five-point configuration persists without bifurcation across all explored values of $\delta$, which has direct consequences for the manifold structure: the preservation of L$_1$ and L$_2$ as unstable points maintains the architecture required for two arms.

\subsubsection{Model C: Large bar offset}

\begin{figure}
\centering
\includegraphics[width=0.49\linewidth]{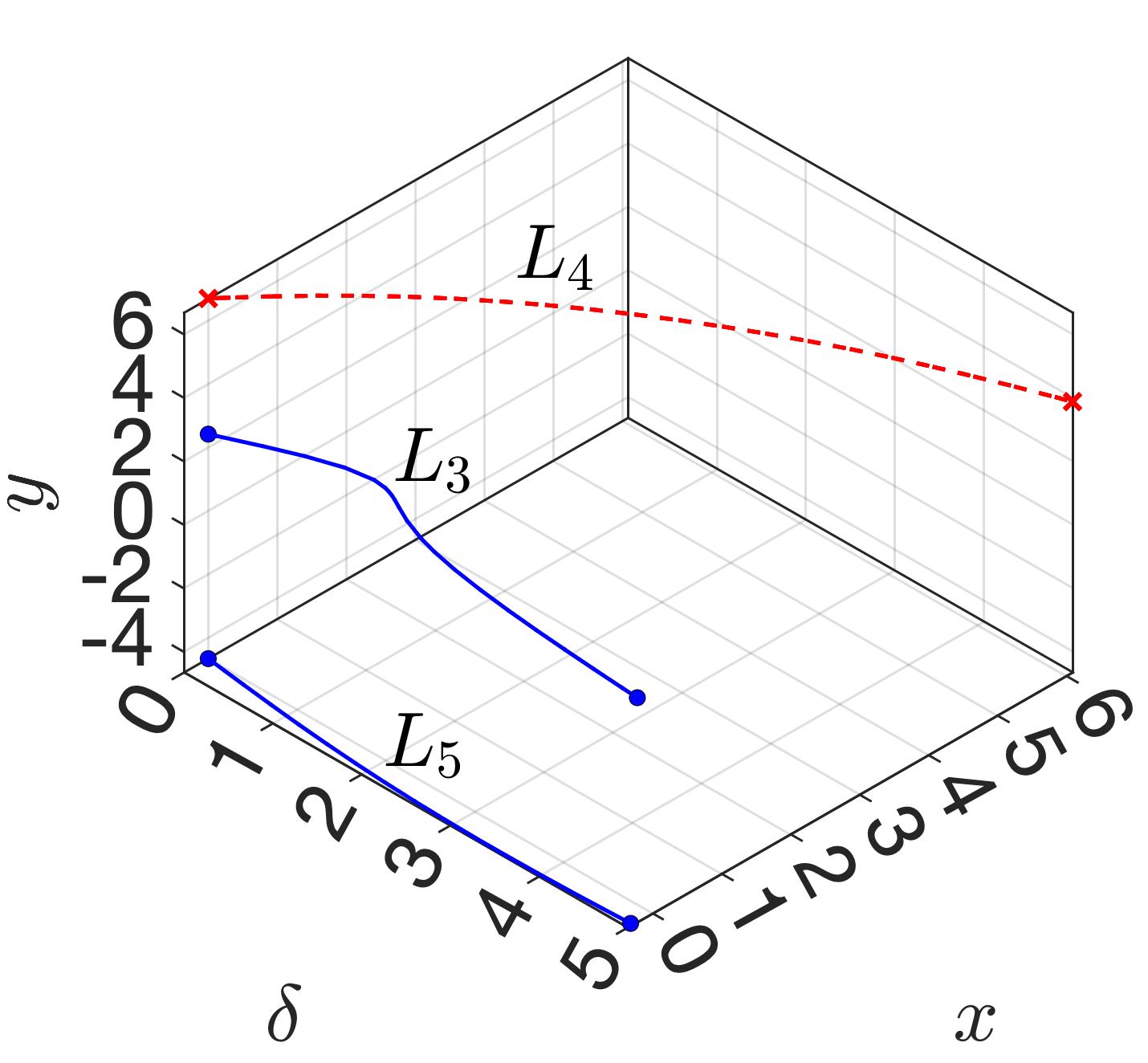}
\includegraphics[width=0.49\linewidth]{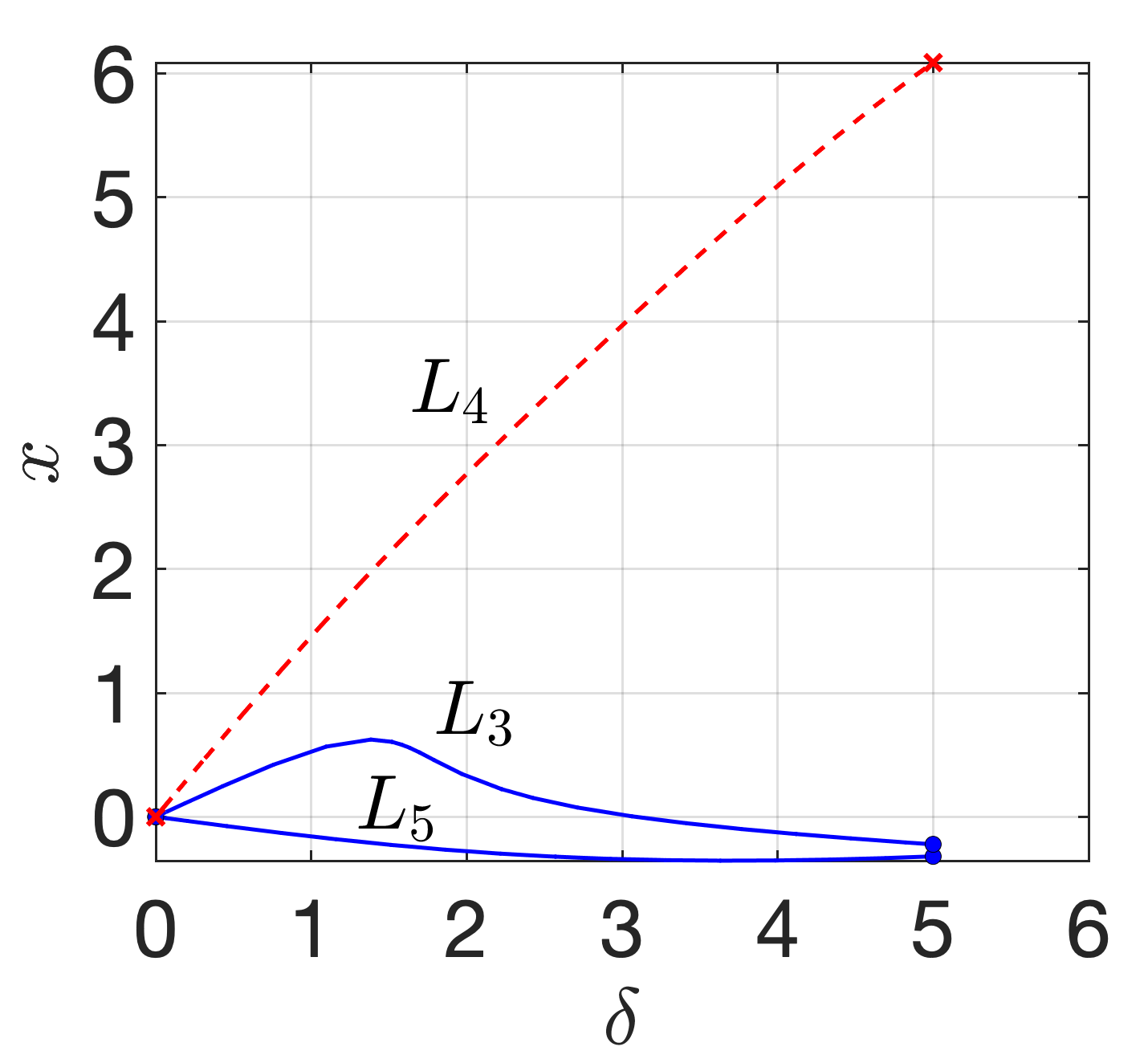} \\
\includegraphics[width=0.49\linewidth]{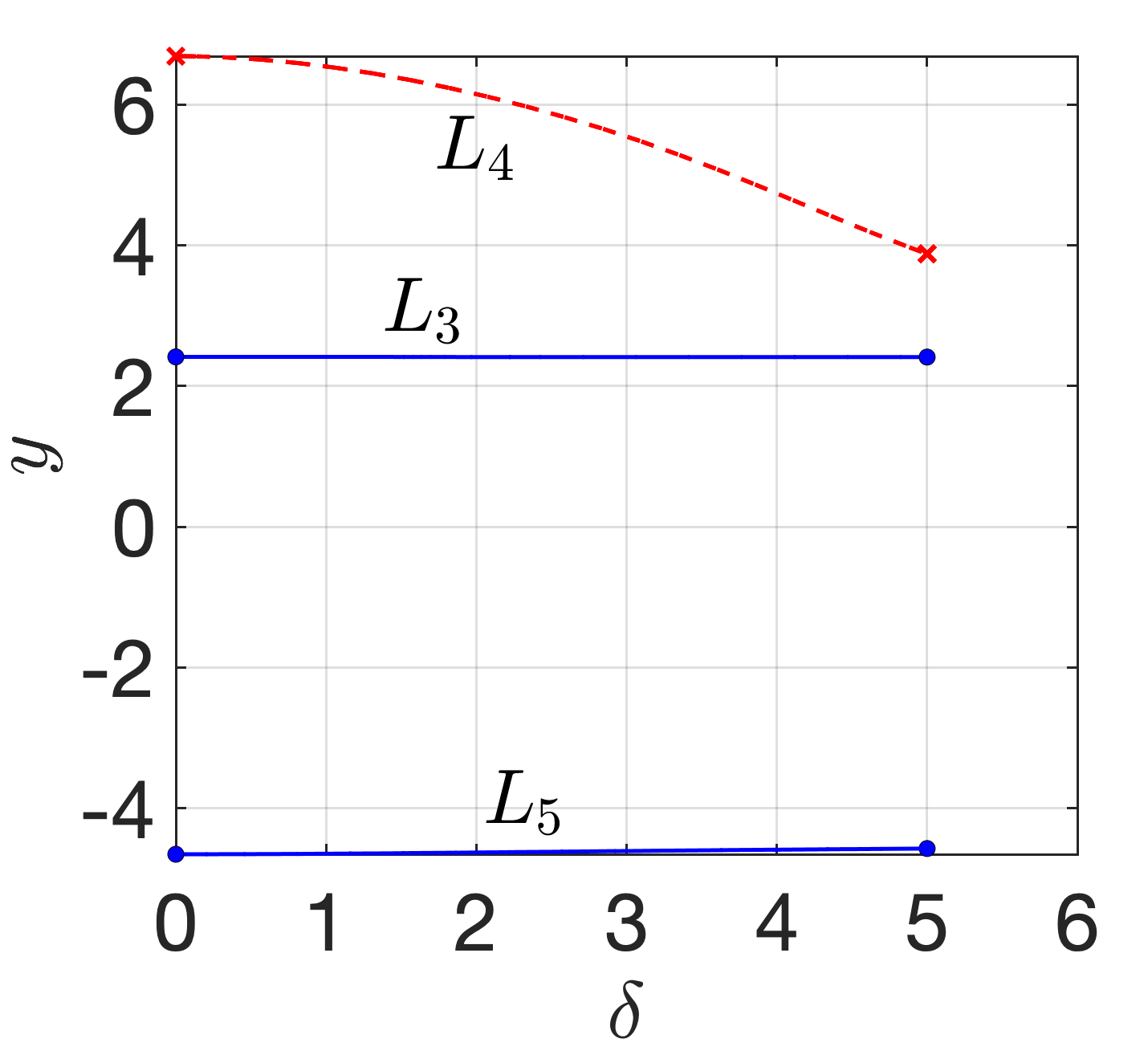}
\includegraphics[width=0.49\linewidth]{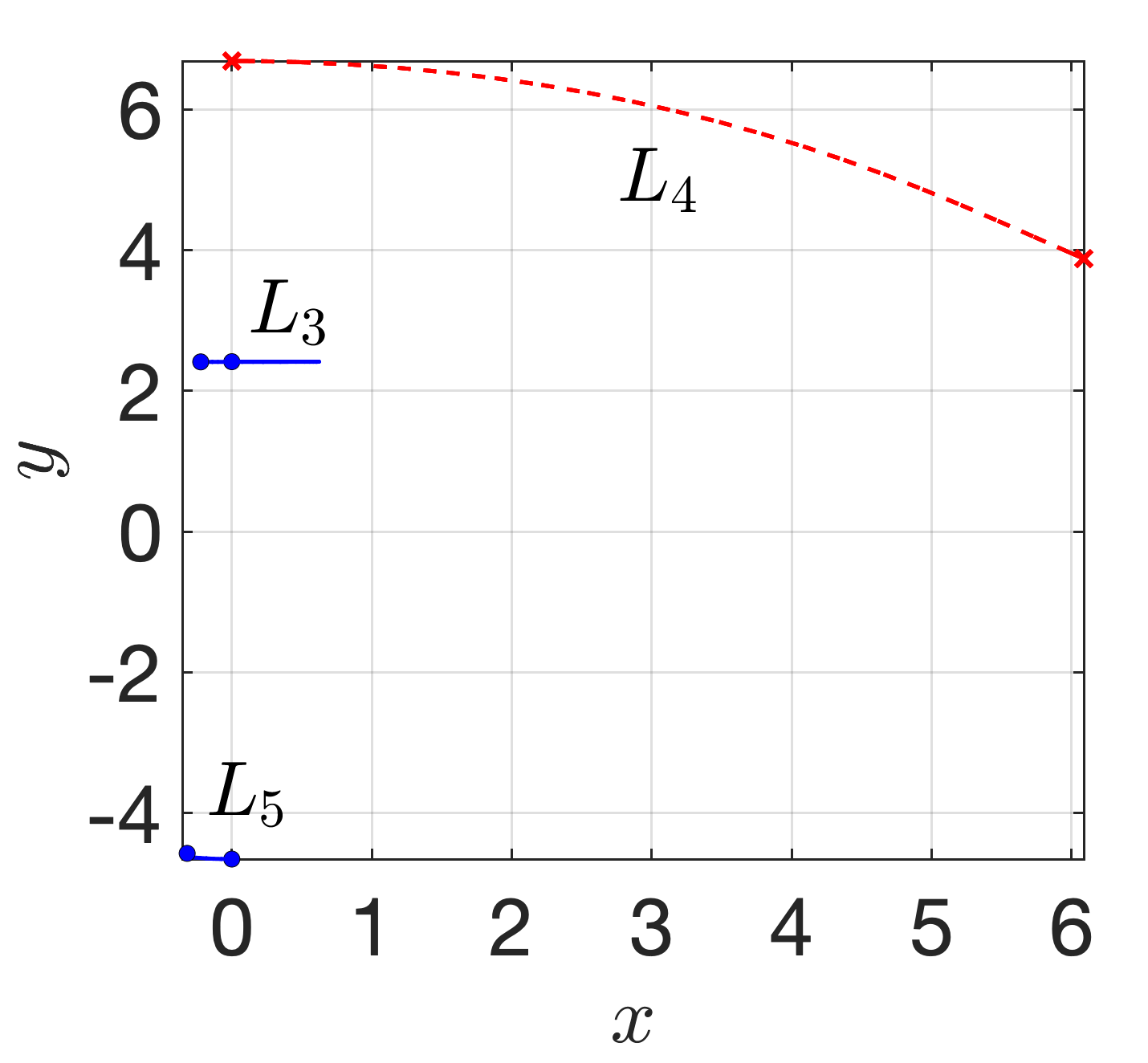}
\caption{Equilibrium point continuation for Model C ($\Lambda = 2.42$ kpc) as a function of $\delta$. Only three equilibrium points exist: L$_4$, L$_3$, and L$_5$ (shown from top to bottom in spatial projection). Top panels: 3D curves and $\delta$--$x$ projection showing the high sensitivity of L$_4$ to displacement and the change in stability of L$_4$ relative to Models A and B. Bottom panels: $\delta$--$y$ projection where L$_4$ approaches the $y$-coordinate of L$_3$, and spatial $(x$–$y)$ view where L$_5$ progressively shifts toward the original position of L$_3$. Solid blue: stable branches; dashed red: unstable branches.}
\label{fig:Bifur_xdb_C}
\end{figure}

For offsets sufficiently large to place the centre of mass exterior to the bar ellipsoid ($\Lambda = 2.42$ kpc), the system undergoes a fundamental topological transformation. Figure~\ref{fig:Bifur_xdb_C} reveals that only three equilibrium points exist across the entire $\delta$ range ($\delta\in[0,5]$ kpc): L$_4$ (upper), L$_3$ (middle), and L$_5$ (lower). The collinear unstable points L$_1$ and L$_2$ are completely absent. L$_4$ is now unstable, in contrast to Models A and B. This is not merely a quantitative difference from Models A and B; it represents a qualitative change in the available phase space structure. The point L$_4$ exhibits substantially higher sensitivity to displacement than in Model B, with the $x$-coordinate increasing approximately linearly as $x \approx \delta + 1$ kpc. The absence of two unstable points fundamentally constrains the system's ability to sustain two arm morphology.

The spatial projection $(x$–$y)$ and the $\delta$--$y$ projection reveal that L$_4$ moves significantly along the $y$-direction, approaching the $y$-coordinate of L$_3$, while the linearly stable points maintain approximate constancy. These dynamic changes indicate how the geometry of the equilibrium configuration responds to increasing internal asymmetry in an already offset system.

\subsection{Continuation with respect to the $\Lambda$ parameter}

\begin{figure}
\centering
\includegraphics[width=0.49\linewidth]{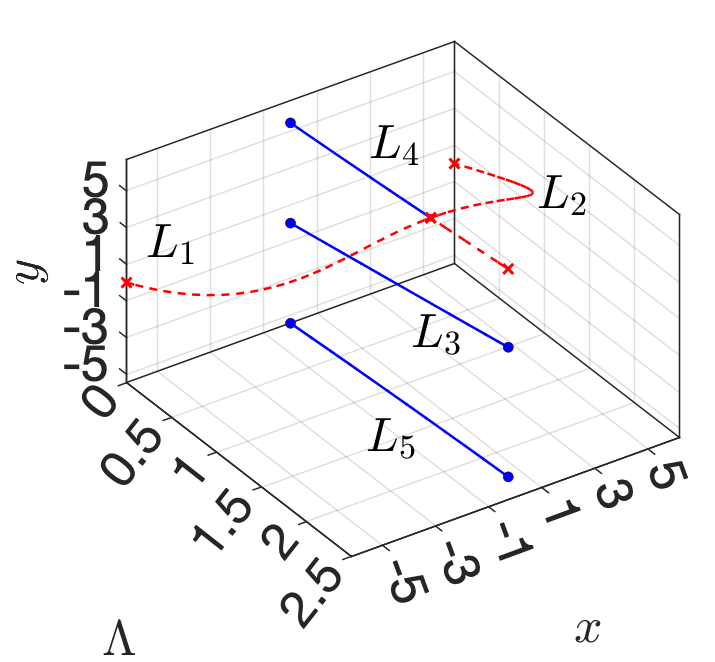}
\includegraphics[width=0.49\linewidth]{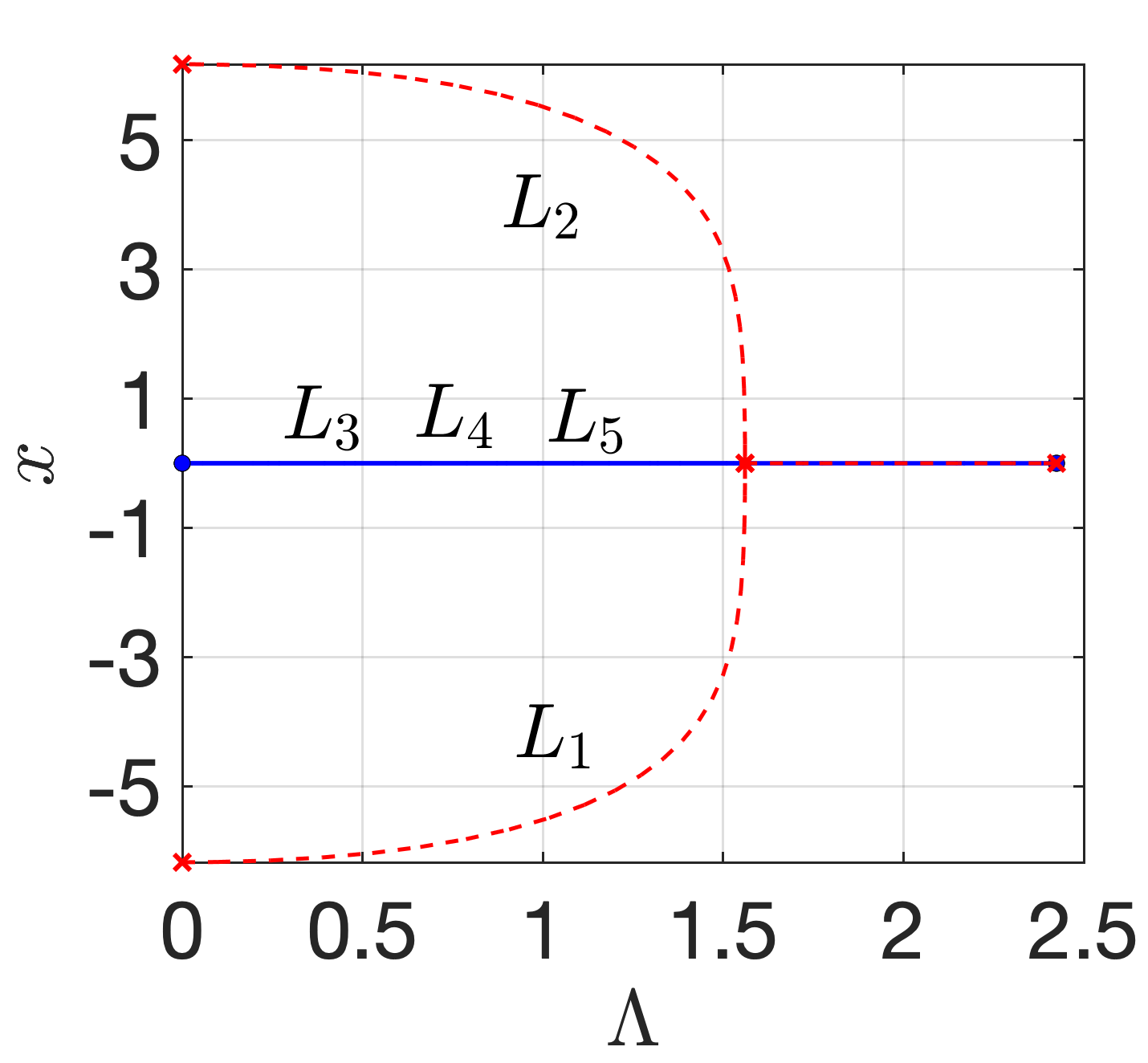} \\
\includegraphics[width=0.49\linewidth]{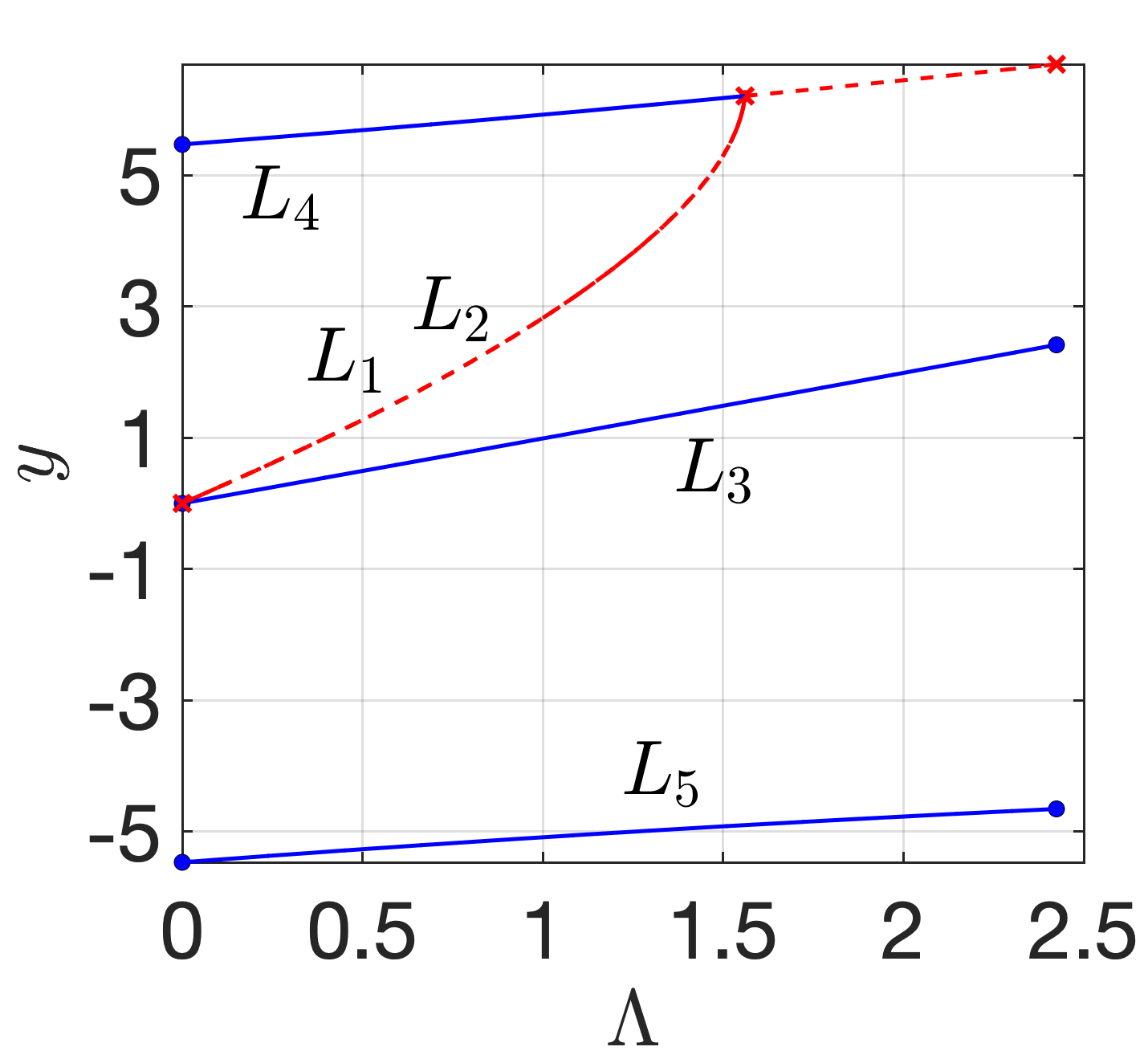}
\includegraphics[width=0.49\linewidth]{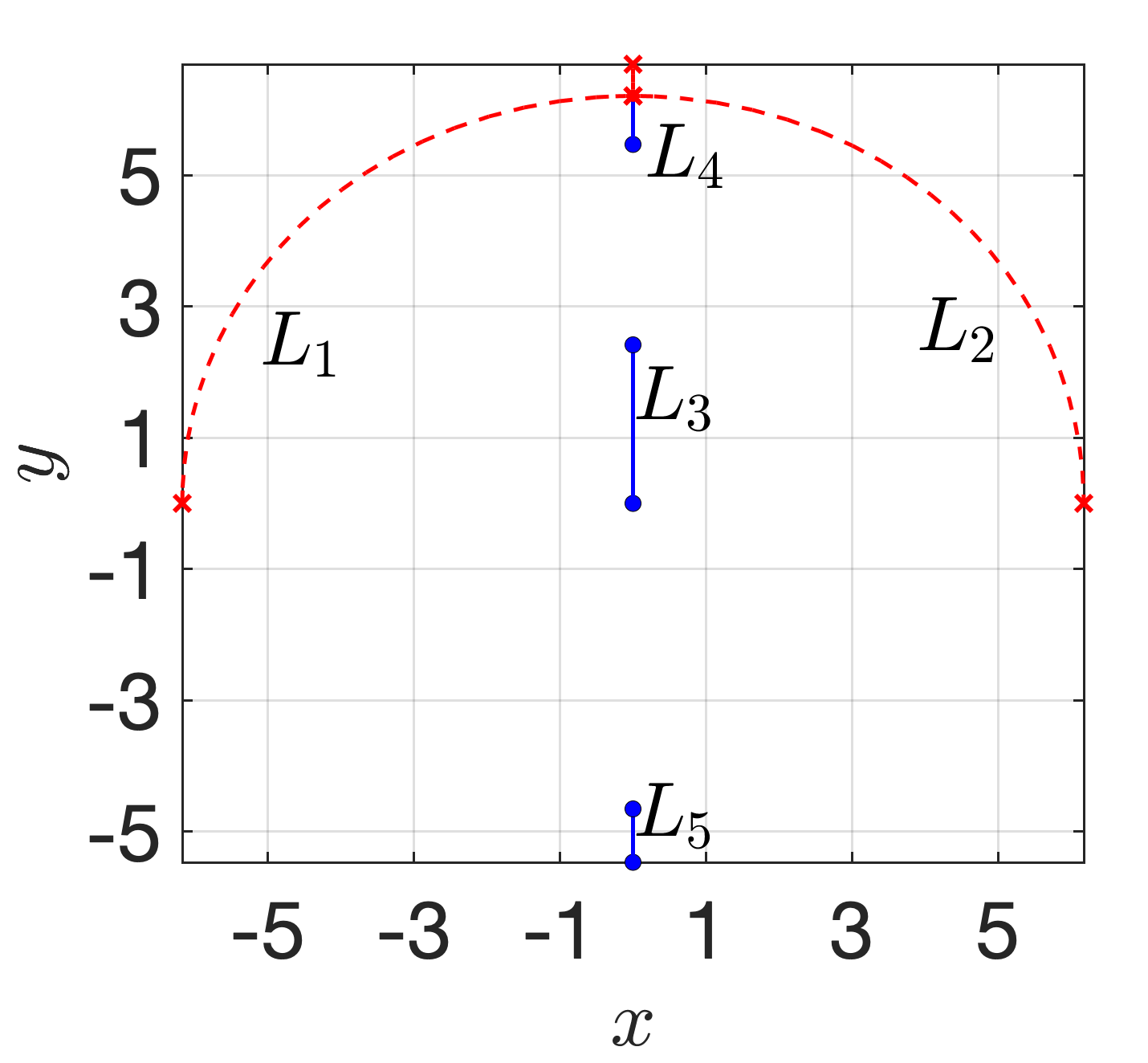}
\caption{Equilibrium point continuation as $\Lambda$ varies with $\delta = 0$. The pitchfork bifurcation at $\Lambda \approx 1.5$ kpc (bar semi-minor axis) marks the structural transition from Models B to C. Top panels: 3D paths in $(x,y,\Lambda)$ space showing the clear coalescence, and $\Lambda$--$x$ projection displaying how L$_1$ and L$_2$ approach and merge with L$_4$ at the critical value. Bottom panels: $\Lambda$--$y$ projection showing remaining equilibrium points primarily shift along $y$ following the bar displacement, and spatial $(x$–$y)$ view emphasizing the merging. Stable and unstable branches indicated by solid blue and dashed red lines, resp.}
\label{fig:Bifur_ydh}
\end{figure}

The transition between Models B and C occurs at a critical bar offset. This transition is physically important because it determines the regime in which the system can support a two arm structure. Figure~\ref{fig:Bifur_ydh} presents the evolution as $\Lambda$ varies with $\delta = 0$, isolating the pure effect of the bar offset on equilibrium configuration.

Beginning from $\Lambda = 0$ (Model A), the equilibrium points follow smooth paths until reaching $\Lambda \approx 1.5$ kpc, at which a pitchfork bifurcation occurs. The critical value $\Lambda = 1.5$ kpc coincides precisely with the bar semi-minor axis $b$, suggesting that the bifurcation is geometrically triggered when the centre of mass reaches the boundary of the bar ellipsoid. At this point, the unstable points L$_1$ and L$_2$ merge with the linearly stable point L$_4$; all three coalesce and exchange stability properties. Beyond this critical value, L$_4$ becomes unstable while the collinear points cease to exist.

The structural consequences are significant: Models A and B maintain five equilibrium points (three linearly stable, two unstable) for $\Lambda < 1.5$ kpc, whereas Model C contains only three equilibrium points (two linearly stable, one unstable) for $\Lambda > 1.5$ kpc. The remaining linearly stable points (L$_3$ and L$_5$) respond to increasing offset by shifting predominantly along the $y$-direction, tracking the displacement of the bar. This preservation of L$_3$ and L$_5$ but loss of the collinear unstable points dramatically alters the manifold configuration and arm structure supported by the system, as the invariant manifolds emanating from the missing unstable points can no longer support two arm transport.

\subsection{Pattern speed dependence}

\begin{figure}
\centering
\includegraphics[width=0.49\linewidth]{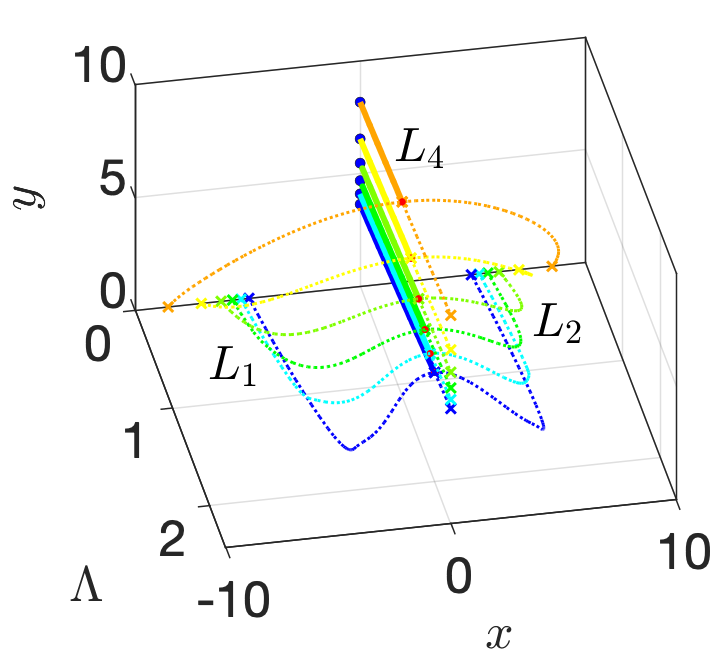}
\includegraphics[width=0.49\linewidth]{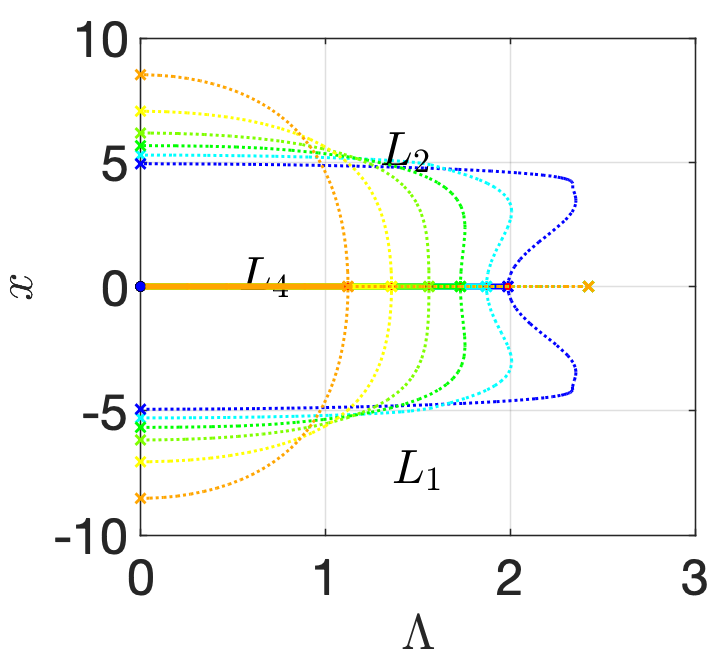}\\
\includegraphics[width=0.49\linewidth]{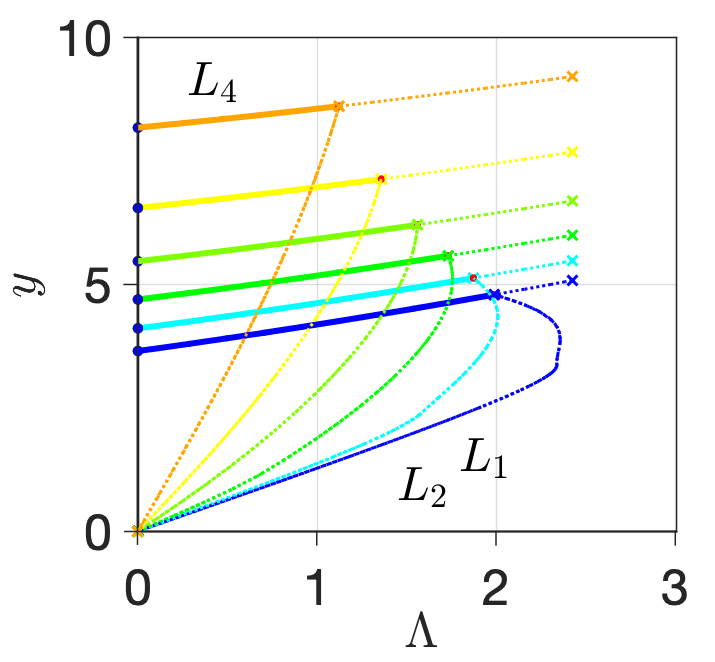}
\includegraphics[width=0.49\linewidth]{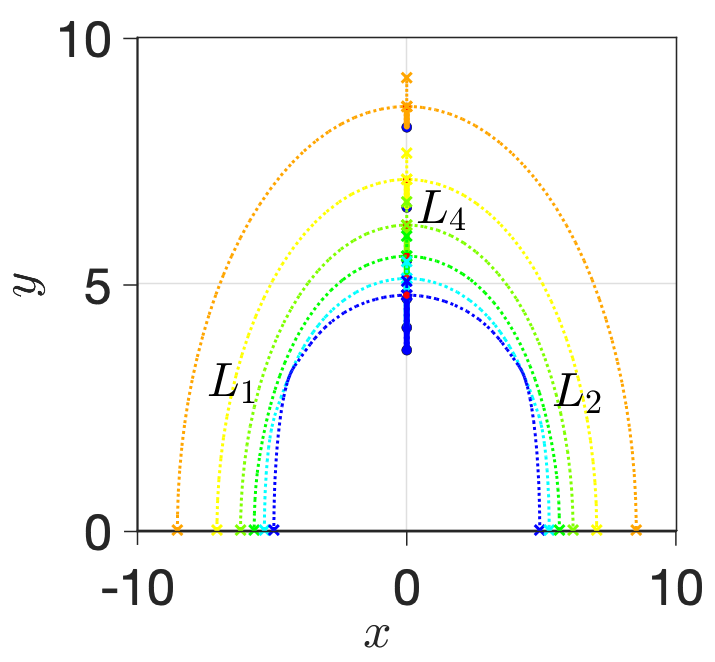}
\caption{Bifurcation location as a function of bar pattern speed $\Omega$. Colours span from orange ($\Omega = 0.03$ [u$_t$]$^{-1}$) to dark blue ($\Omega = 0.08$ [u$_t$]$^{-1}$) in 0.01 increments. Top panels: 3D view of all bifurcation curves and the $\Lambda$--$x$ projection showing how bifurcation location shifts to higher $\Lambda$ with increasing $\Omega$. Bottom panels: $\Lambda$--$y$ and spatial $(x$–$y)$ projections. Blue dots: extrema of stable branches; crosses: extrema of unstable branches; coloured dots: bifurcation points. Stable and unstable branches indicated by solid and dashed lines, resp.}
\label{fig:Bifur_vbar}
\end{figure}

The bifurcation location at $\Lambda \approx 1.5$ kpc was obtained with standard pattern speed $\Omega = 0.05$ [u$_t$]$^{-1}$. To assess whether this critical transition is robust or depends on dynamical parameters, Fig.~\ref{fig:Bifur_vbar} explores the bifurcation across a range of pattern speeds, $\Omega \in [0.03, 0.08]$ [u$_t$]$^{-1}$. Each colour represents a different $\Omega$ value, progressing from orange ($\Omega = 0.03$) to dark blue ($\Omega = 0.08$) in 0.01 increments; only L$_1$, L$_2$, and L$_4$ are shown for clarity.

For low-to-intermediate speeds ($\Omega \in [0.03, 0.06]$), the bifurcation behaviour is consistent with the standard case shown in Fig.~\ref{fig:Bifur_ydh}, confirming the five-to-three equilibrium transition is a robust feature across this velocity range. However, the bifurcation location exhibits systematic dependence on pattern speed: as $\Omega$ increases, the bifurcation occurs at progressively higher values of $\Lambda$. This shift reflects the increasing stabilization provided by centrifugal effects at higher pattern speeds, which resist the coalescence of the collinear points with L$_4$.

A qualitatively different bifurcation structure emerges in the high-speed regime ($\Omega \geq 0.07$). At $\Omega = 0.08$, instead of annihilating the collinear points, a subcritical pitchfork bifurcation occurs near $\Lambda \approx 2$ kpc, creating two new equilibrium points $L_1'$ and $L_2'$ rather than destroying existing ones. These transient points coexist with L$_1$ and L$_2$ until approximately $\Lambda \approx 2.4$ kpc, where they merge with their progenitors. During this intermediate interval, the system temporarily contains four unstable equilibrium points. Their existence may modulate the overall phase-space geometry and material circulation patterns during this regime.

\vspace{0.2cm}

The continuation analysis presented in this section shows how internal mass imbalance and bar offsets qualitatively reshape the equilibrium-point configuration of the system. In particular, it reveals when the classical five-point configuration is preserved and when it collapses to a three-point configuration, thereby setting the stage for the invariant manifold structures analysed in Section~\ref{sec:manifolds}.

%%%%%%%%%%%%%%%%%%%%%%%%%%%%%%%%%%%%%%%%%%%%%%%%%%%%%%%%%%%%%%%%%%%%%%%%%%%%%%%%%%%%%%%
%\input{4-manifolds}
\section{Arm morphology from invariant manifolds}
\label{sec:manifolds}

The bifurcation structure analyzed in Section~\ref{sec:bifurcations} determines the fundamental dynamical organization of the galactic system. Around the unstable equilibrium points L$_1$ and L$_2$ in the classical symmetric model (Fig.~\ref{fig:eqpoints}), there exist planar and vertical families of periodic orbits called Lyapunov orbits. Of particular interest is the planar family: from each planar Lyapunov orbit emanate hyperbolic invariant manifolds (both stable and unstable) at the same energy level, which create transport tubes of matter between inner and outer galactic regions. These manifolds delineate the arm structure and provide the fundamental mechanism for organizing galactic transport. Details of these families of periodic orbits, their associated invariant manifolds, and dynamical implications can be found in \citet{Efthym2019, Romero2, Rom09, Warps, Tsoutsis2008}.

The loss or preservation of the collinear unstable points L$_1$ and L$_2$, determined by the bifurcations identified in Section~\ref{sec:bifurcations}, directly controls whether the system can sustain a two-armed or one-armed morphology. When both of these unstable points exist (Models A and B with $\Lambda < 1.5$ kpc), the emanating manifolds organize a two-armed structure. When the bifurcation at $\Lambda \approx 1.5$ kpc eliminates these unstable points and only one unstable point exists (Model C with $\Lambda > 1.5$ kpc), only a single one-armed configuration remains dynamically possible.

In this section, we analyze isodensity contours, isopotential curves, and invariant manifolds for representative configurations of each model. Isodensity curves clarify the density distribution of the full system, allowing us to specify which configuration we are discussing. Isopotential curves emphasize the dynamical features and highlight the approximate location and behavior of equilibrium points. Invariant manifolds delineate the arms and reveal the transport pathways that organize particle dynamics.

For each model, we analyze the $(y,\dot{y})$ projection of the closed curve formed by the Poincaré map intersection of stable manifolds with the hyperplane $S$ defined as $x=0$. Initial conditions located within these curves correspond to transit orbits at a fixed energy level.

The transfer of matter from inner to outer galactic regions (delimited by the zero velocity curves) is carried out by transit orbits trapped inside the invariant manifold tubes. These orbits are responsible for transporting matter and constitute an indicator of arm density \citep{GideaMasdem2007, Romero1, LCS2018}. The forward in time integration of initial conditions $(y,\dot{y})$ within the closed curve establishes the trajectories of particles confined to the invariant manifold tubes. This procedure enables quantification of matter transferred through each manifold at a given energy level and, consequently, the density distribution of each arm.

The dynamics of the system~\eqref{eqn:systmodelLMC} occur in a four-dimensional phase space when we consider orbits with $z=\dot{z}=0$ (the $z=0$ plane is invariant). A pair $(y,\dot{y})$ inside the closed curve at energy level $C_{J,{L_i}}+\alpha$ (slightly above the equilibrium point's energy) defines a point on $S$ with coordinates $(0,y,0,\dot{x},\dot{y},0)$, where $\dot{x}$ is determined by the energy condition from the Jacobi integral~\eqref{eqn:PreCJAC}:
\begin{equation*}
\dot{x} = \sqrt{-\dot{y}^2 -2\phi_{\text{\scriptsize eff}}(0,y,0)-(C_{J,{L_i}}+\delta)} \, ,
\end{equation*}
\noindent with the sign of the square root determined by the sense of crossing $S$.

\subsection{Model A: Bar with internal mass imbalance aligned with the centre of mass. Two-armed structure}

\begin{figure}
\includegraphics[width=0.24\textwidth]{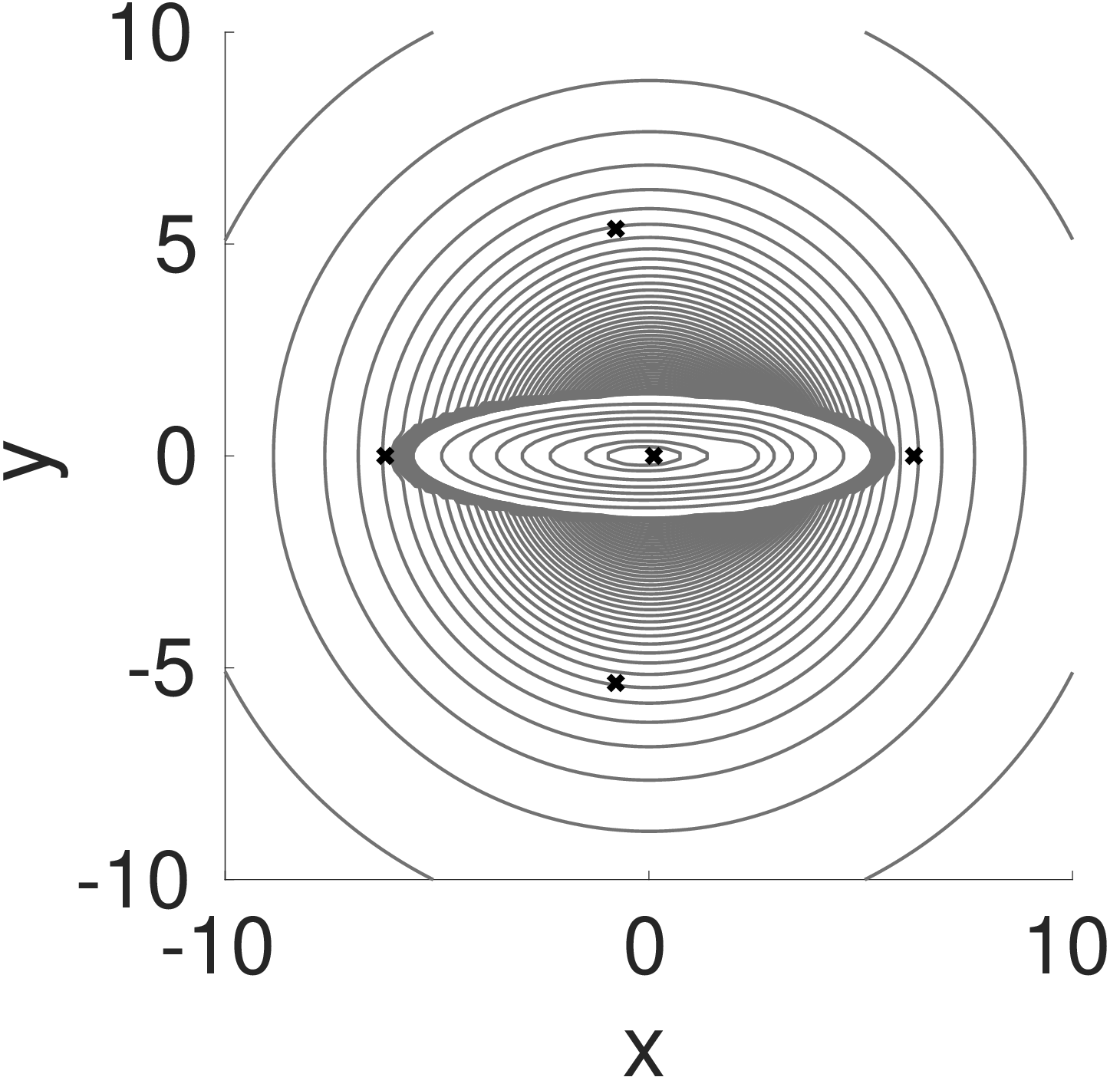}
\includegraphics[width=0.24\textwidth]{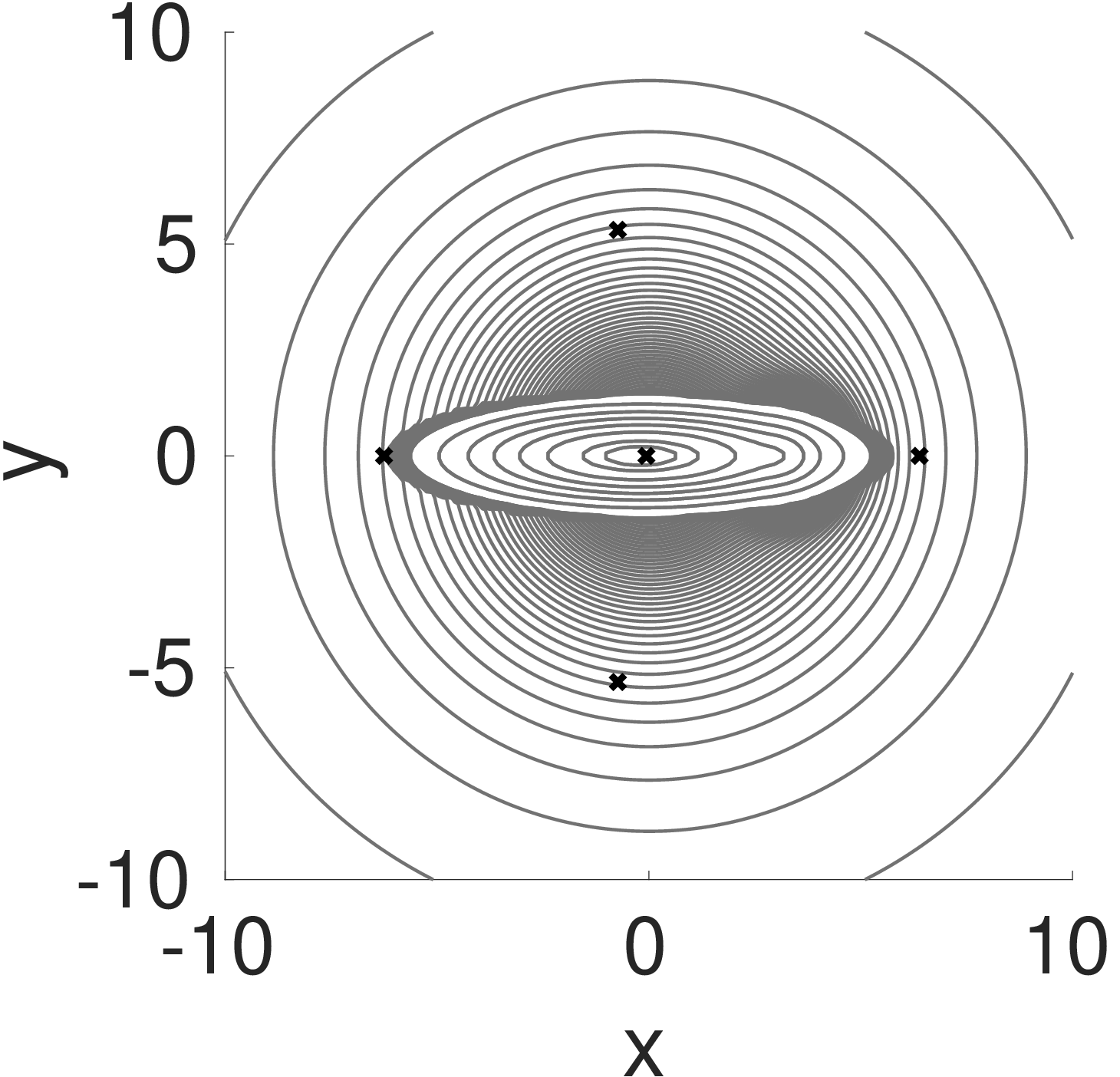} \\
\includegraphics[width=0.24\textwidth]{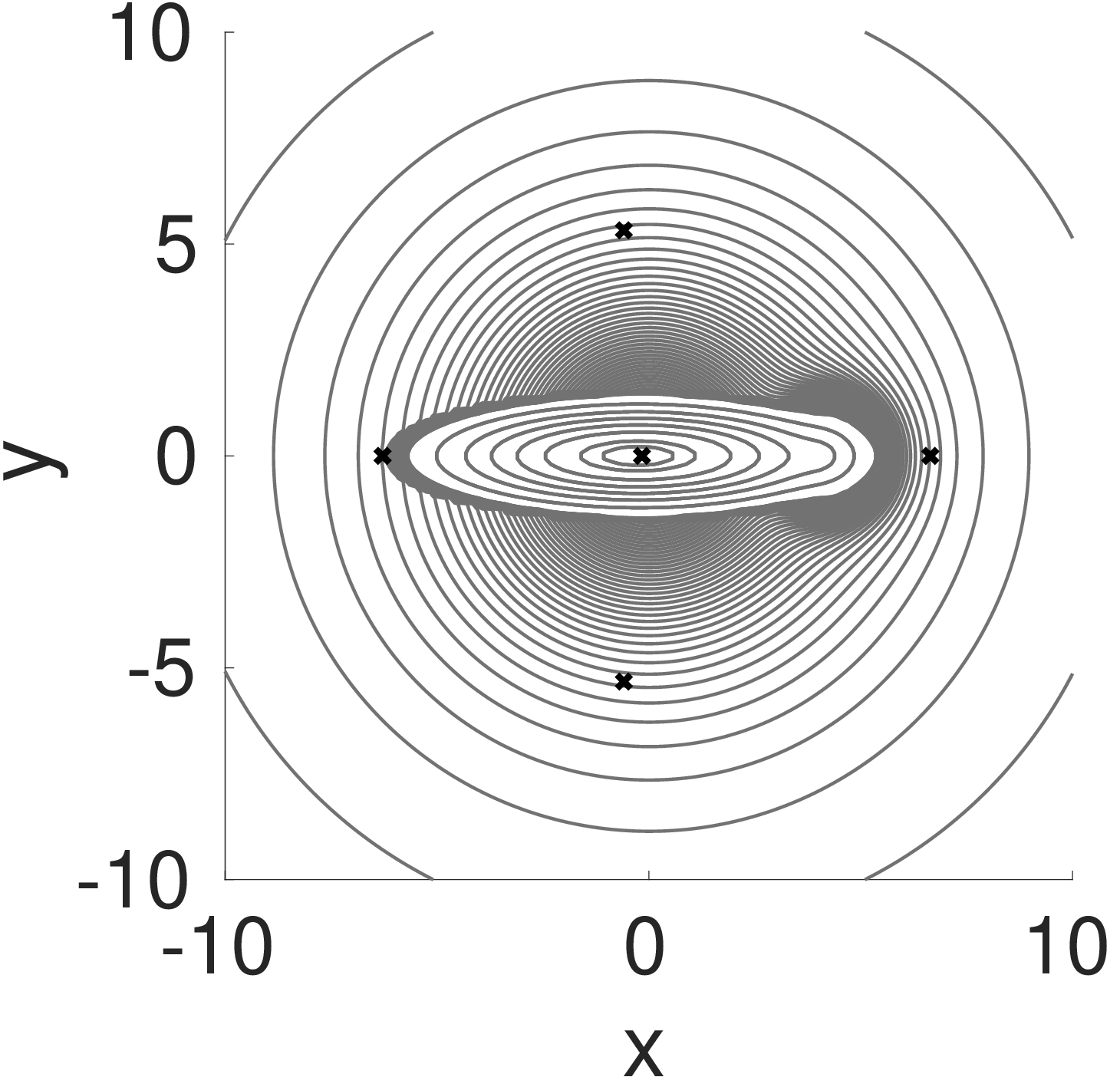}
\includegraphics[width=0.24\textwidth]{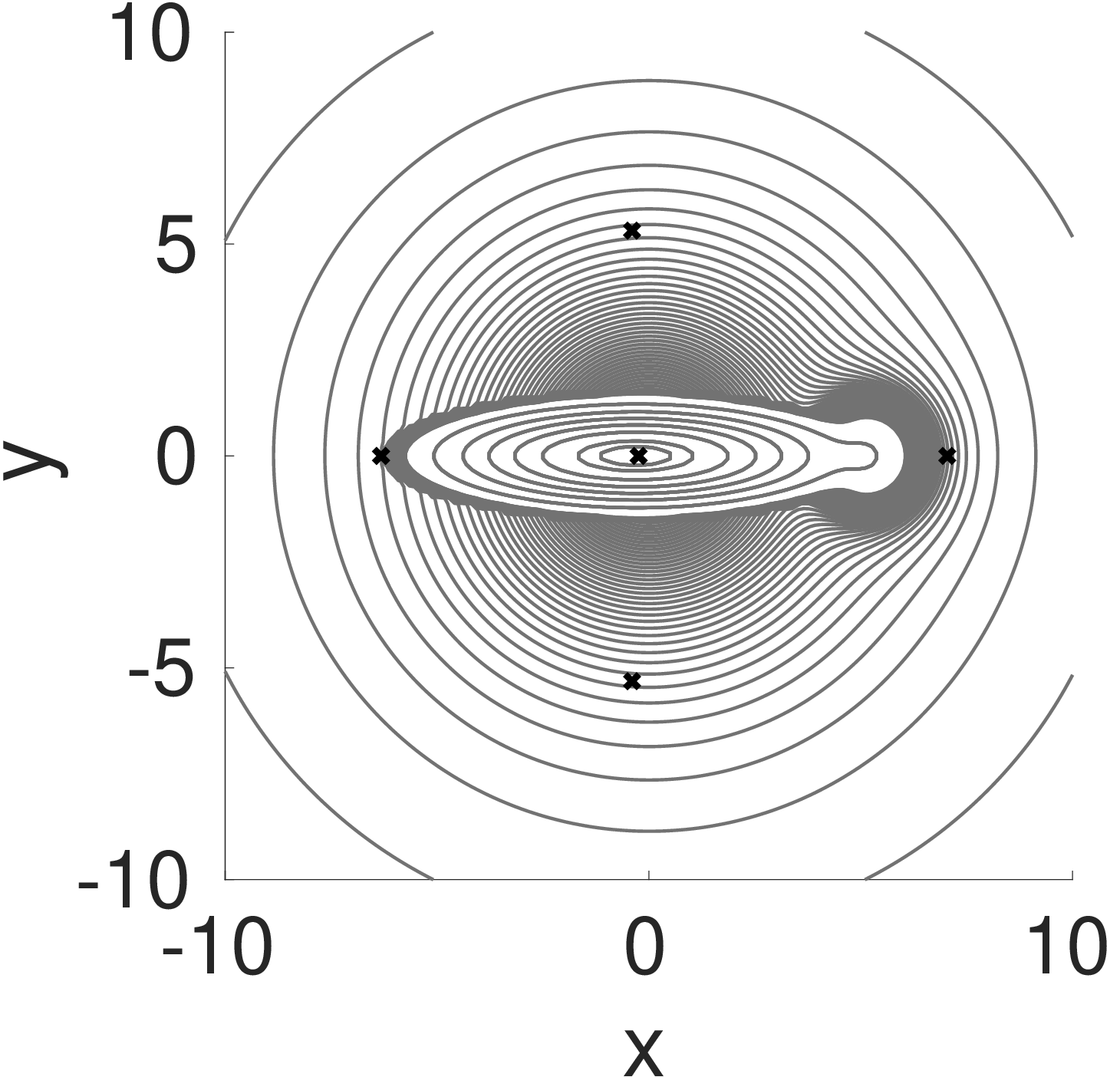}
\caption{Model A isodensity contours ($\Lambda=0$, main axis of the bar on the CM of the system). Equilibrium points marked by black crosses. From top left to bottom right: $\delta = 2.5, 3.5, 4.5, 5.5$ kpc. Progressive asymmetric mass component displacement creates increasing asymmetry in the density distribution, particularly evident around the unstable equilibrium points.}
\label{fig:ModelA_isodens}
\end{figure}

The isodensity contours for Model A ($\Lambda=0$) are shown in Fig.~\ref{fig:ModelA_isodens}. As the asymmetric mass component displaces along the major axis of the bar (increasing $\delta$), the density distribution develops systematic asymmetry, particularly in the right region and around the unstable points. This asymmetric potential structure directly influences the manifold geometry, as discussed below.

\begin{figure}
\includegraphics[width=0.24\textwidth]{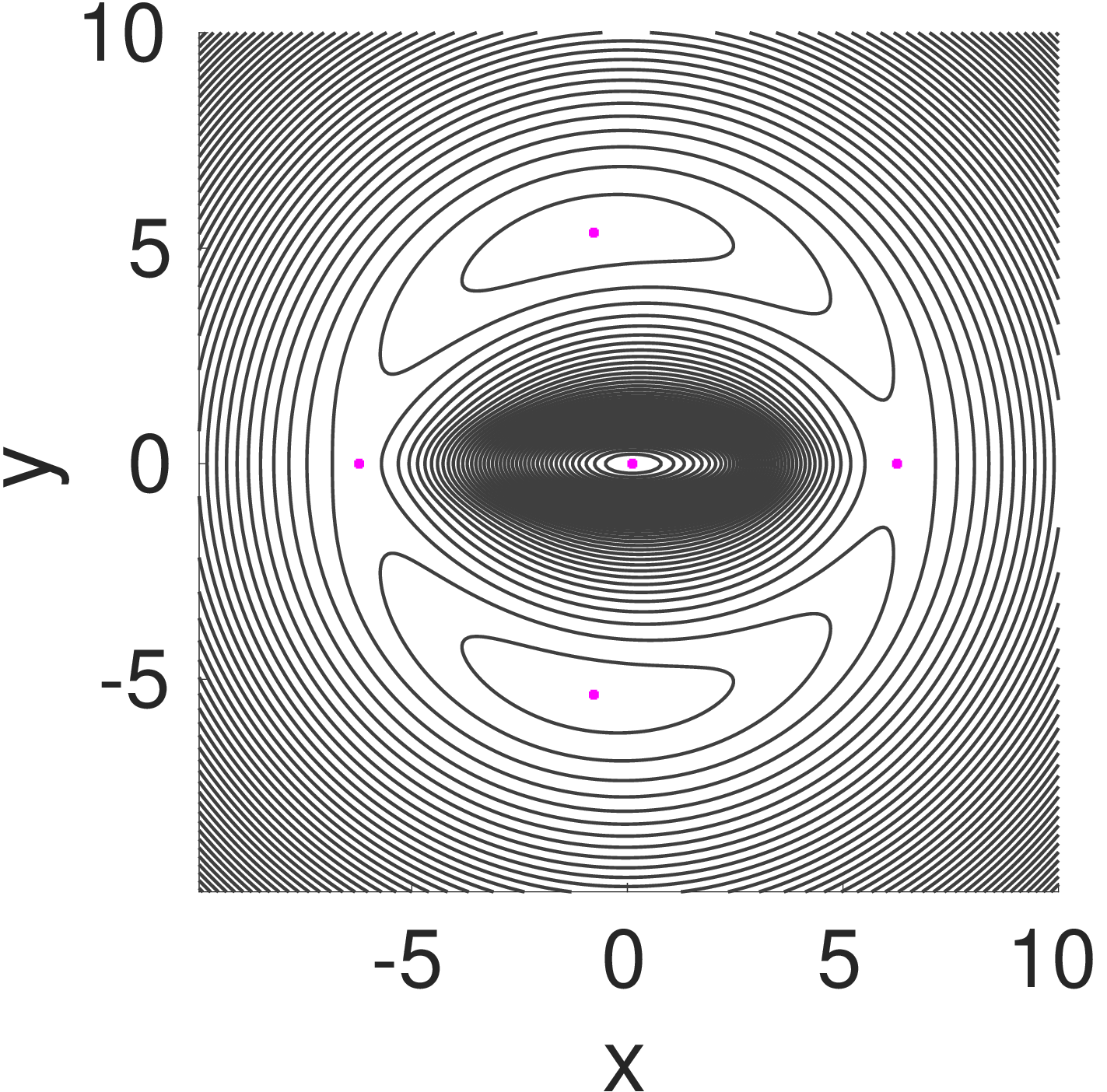}
\includegraphics[width=0.24\textwidth]{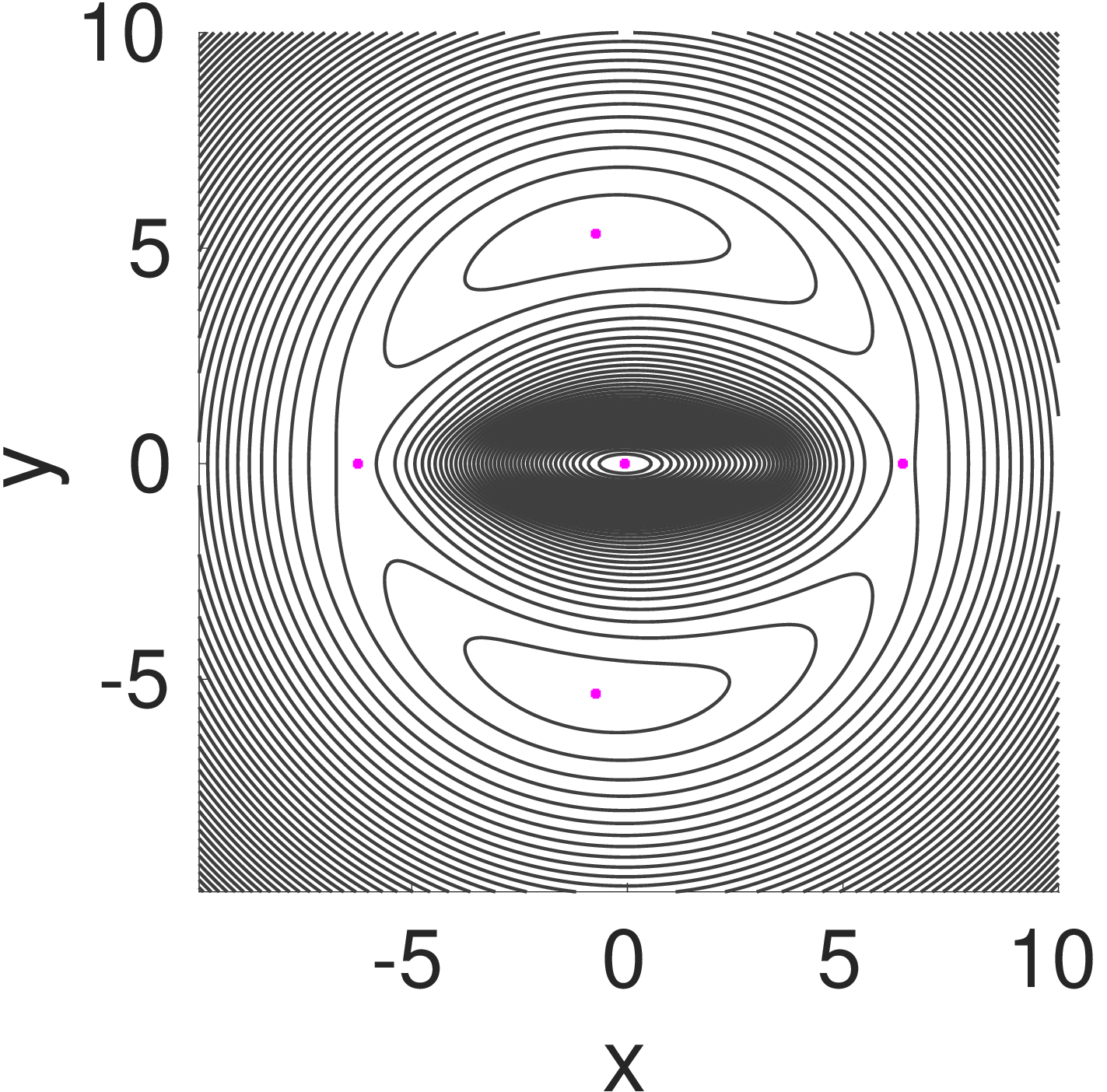} \\
\includegraphics[width=0.24\textwidth]{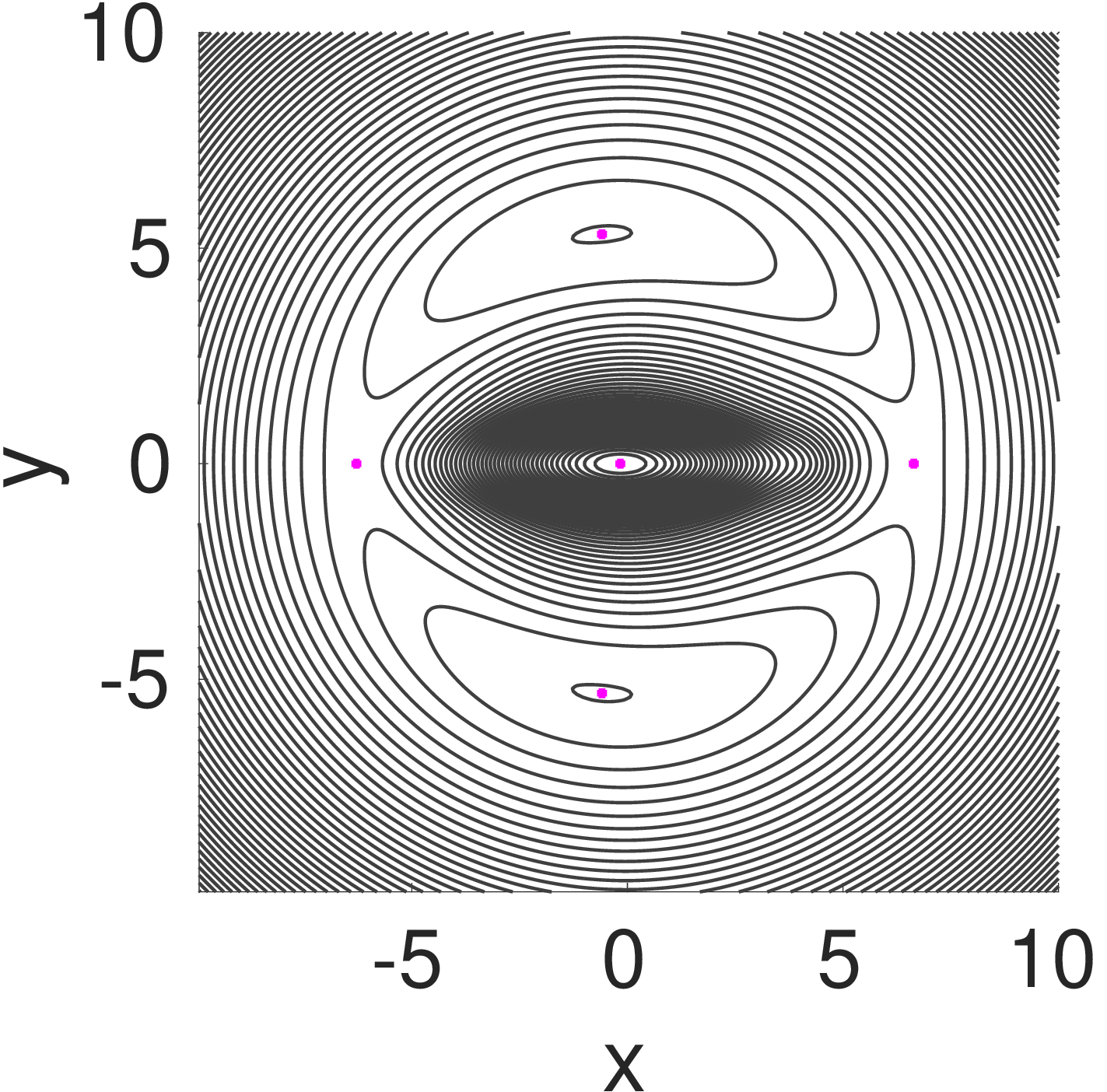}
\includegraphics[width=0.24\textwidth]{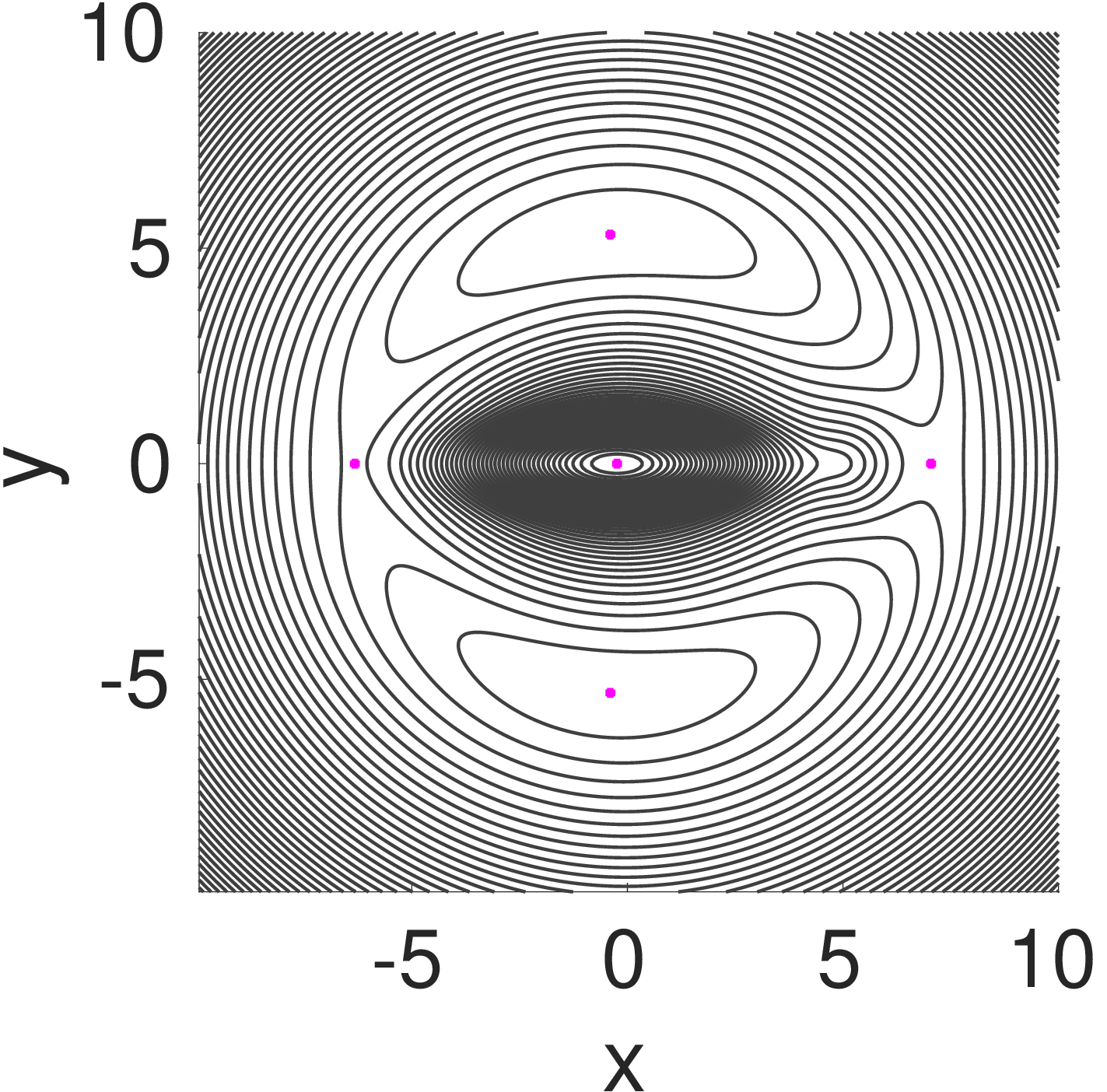}
\caption{Model A isopotential contours ($\Lambda=0$). Equilibrium points marked by magenta dots. From top left to bottom right: $\delta = 2.5, 3.5, 4.5, 5.5$ kpc. The level structure of the potential deforms with increasing $\delta$, reflecting the asymmetric distribution of the displaced mass component.}
\label{fig:ModelA_isopot}
\end{figure}

The isopotential contours (Fig.~\ref{fig:ModelA_isopot}) reveal the potential landscape structure. Although the stability of the equilibrium points L$_1$ and L$_2$ does not change relative to the symmetric model, the potential contours deform progressively, indicating how the manifold geometry responds to internal asymmetry.

\begin{figure}
\includegraphics[width=0.24\textwidth]{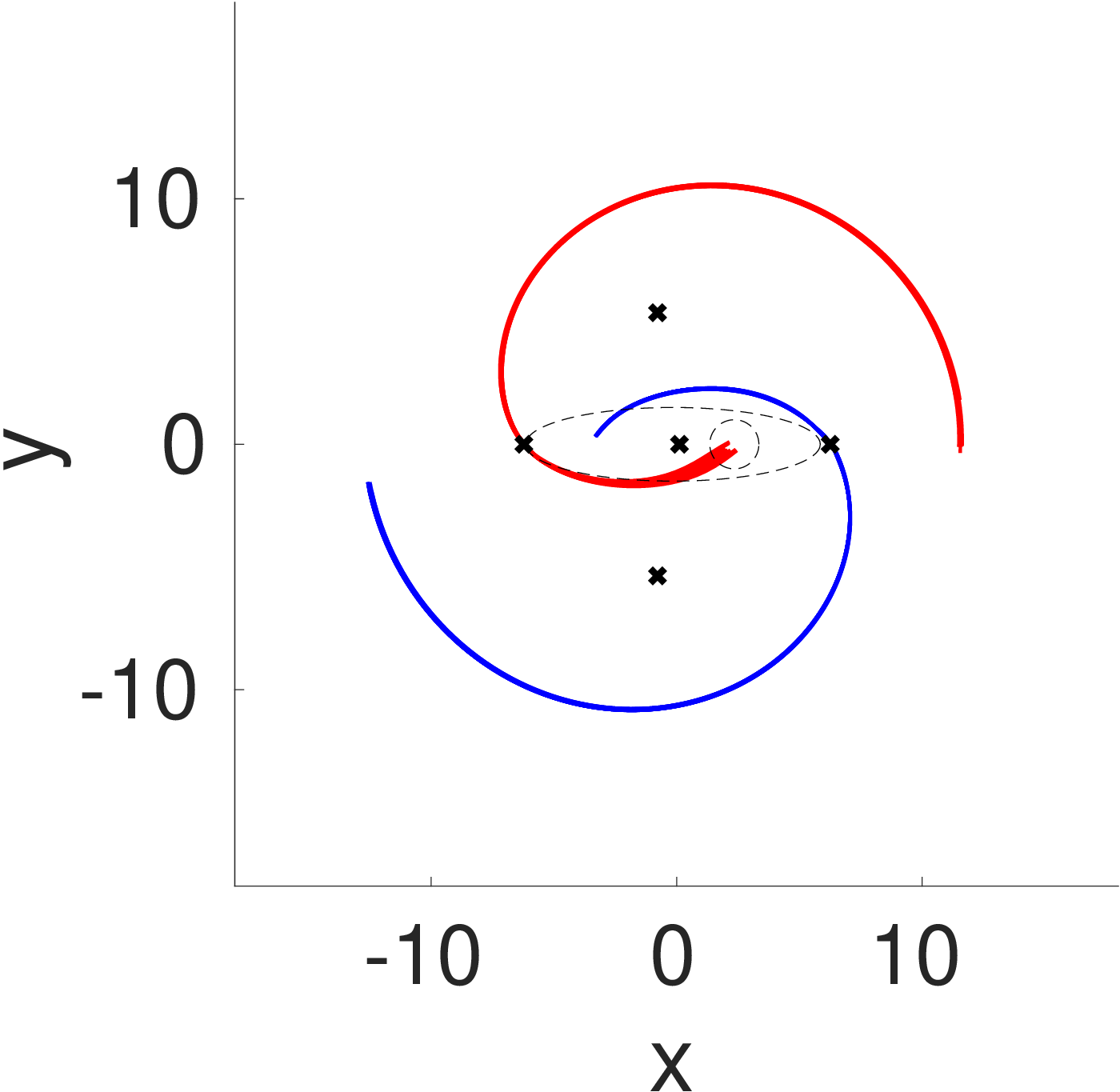}
\includegraphics[width=0.24\textwidth]{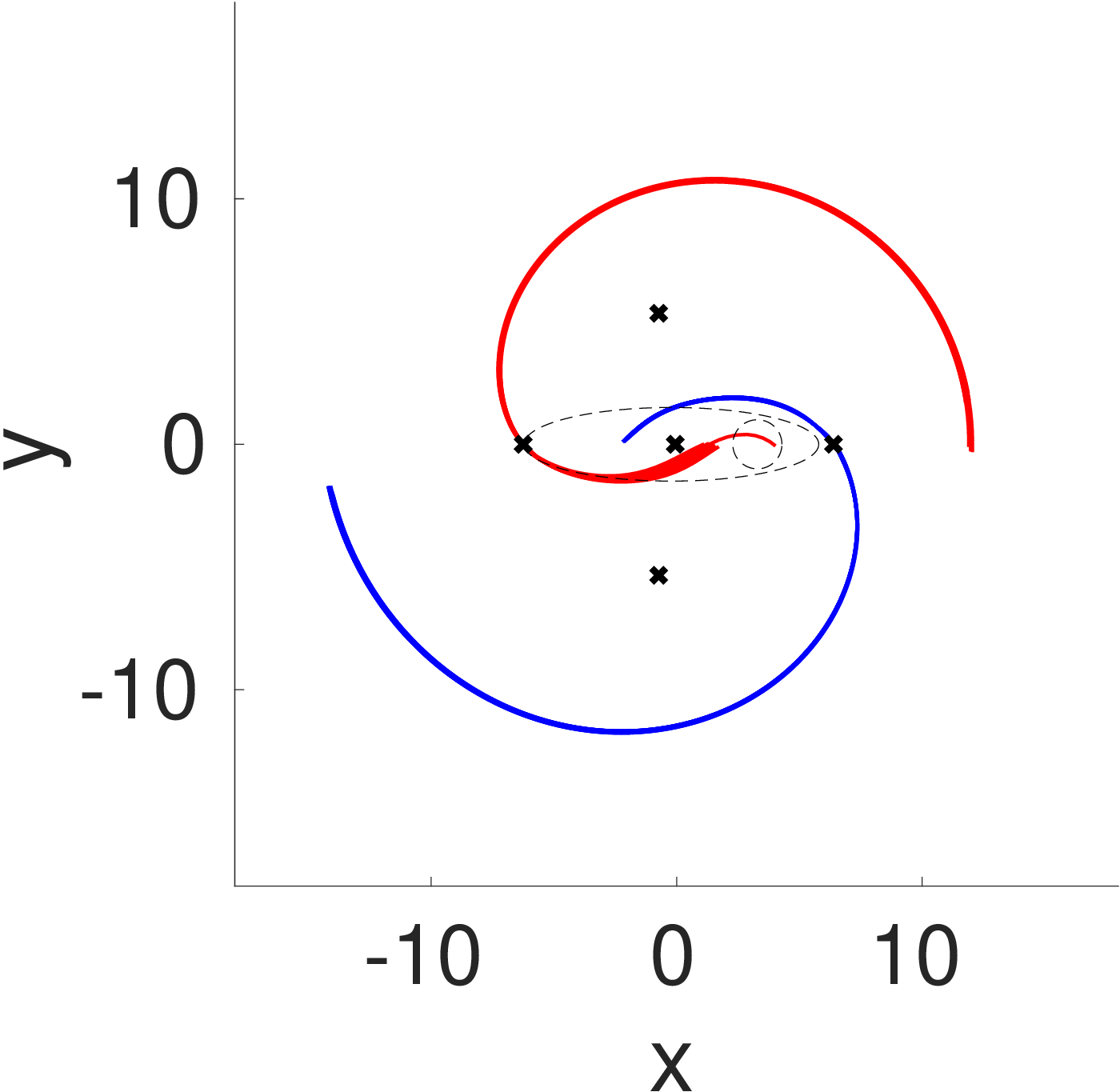}\\
\includegraphics[width=0.24\textwidth]{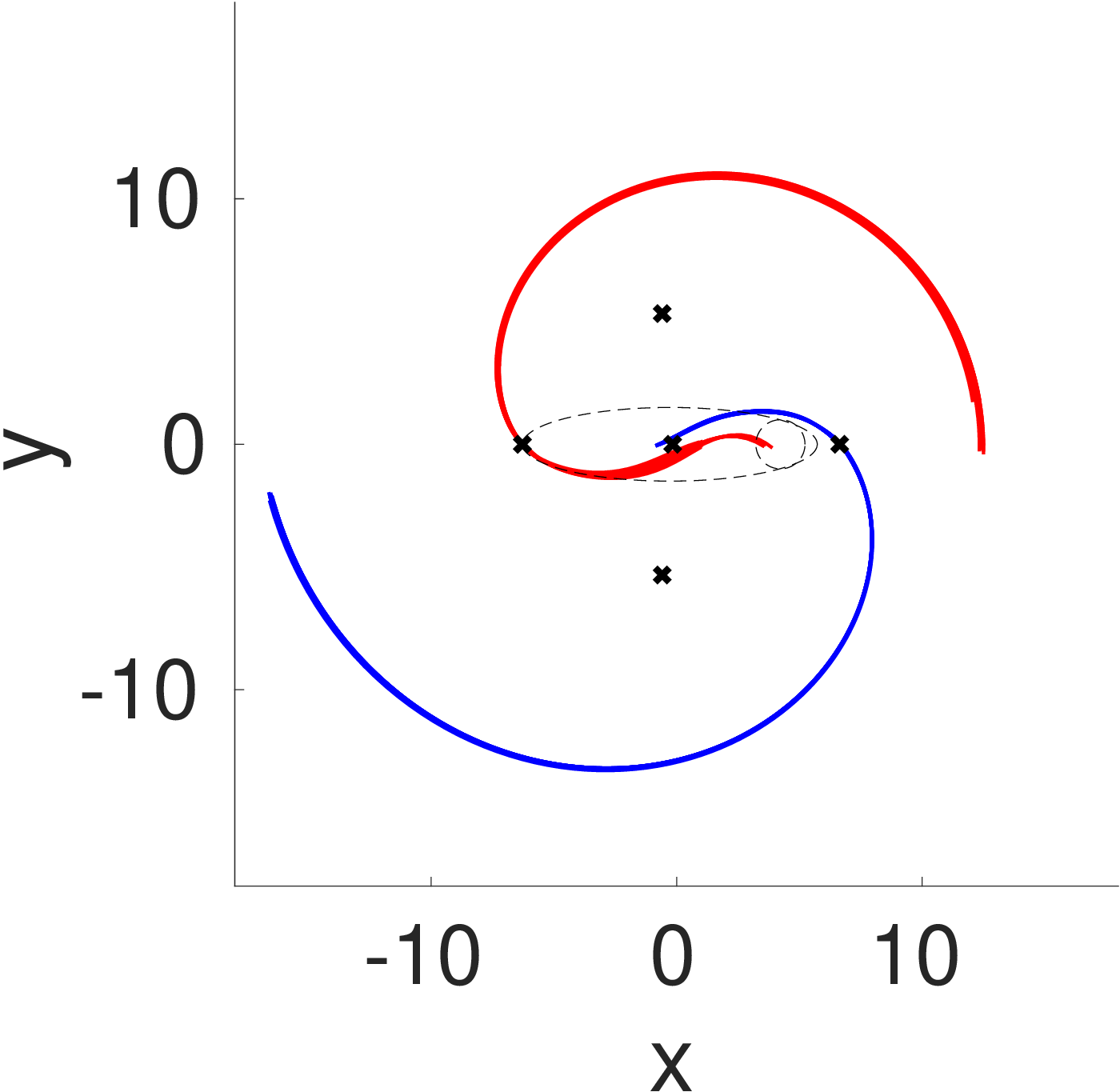}
\includegraphics[width=0.24\textwidth]{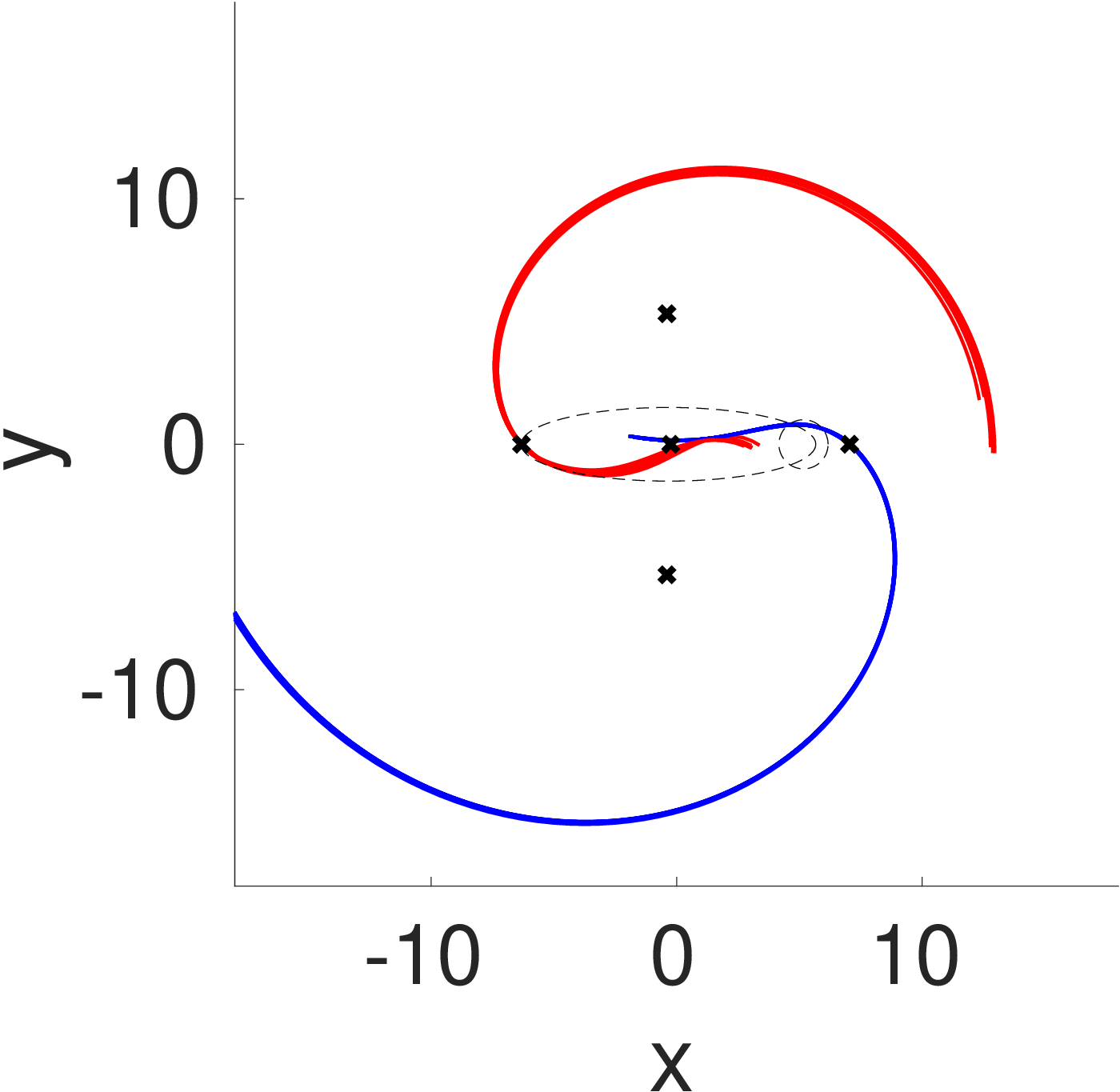}
\caption{Model A unstable invariant manifolds associated with Lyapunov periodic orbits of L$_1$ and L$_2$ ($\Lambda=0$). Equilibrium points marked by crosses. Bar and the asymmetric mass component outlined by dotted black curves. From top left to bottom right: $\delta = 2.5, 3.5, 4.5, 5.5$ kpc. The arm structure retains a two-armed pattern with asymmetric arm opening, becoming more pronounced as $\delta$ increases.}
\label{fig:ModelA_manifolds}
\end{figure}

The unstable invariant manifolds emanating from L$_1$ and L$_2$ retain a two-armed configuration (Fig.~\ref{fig:ModelA_manifolds}), consistent with the preservation of both unstable points across the entire $\delta$ range (Section~\ref{sec:bifurcations}). However, the manifold geometry deforms progressively: one arm has a larger pitch angle than the other, and this asymmetry increases with $\delta$. This behavior is the direct geometric consequence of the asymmetric potential landscape created by the internal mass displacement within the bar.

\begin{figure*}
\centering
\includegraphics[width=0.245\textwidth]{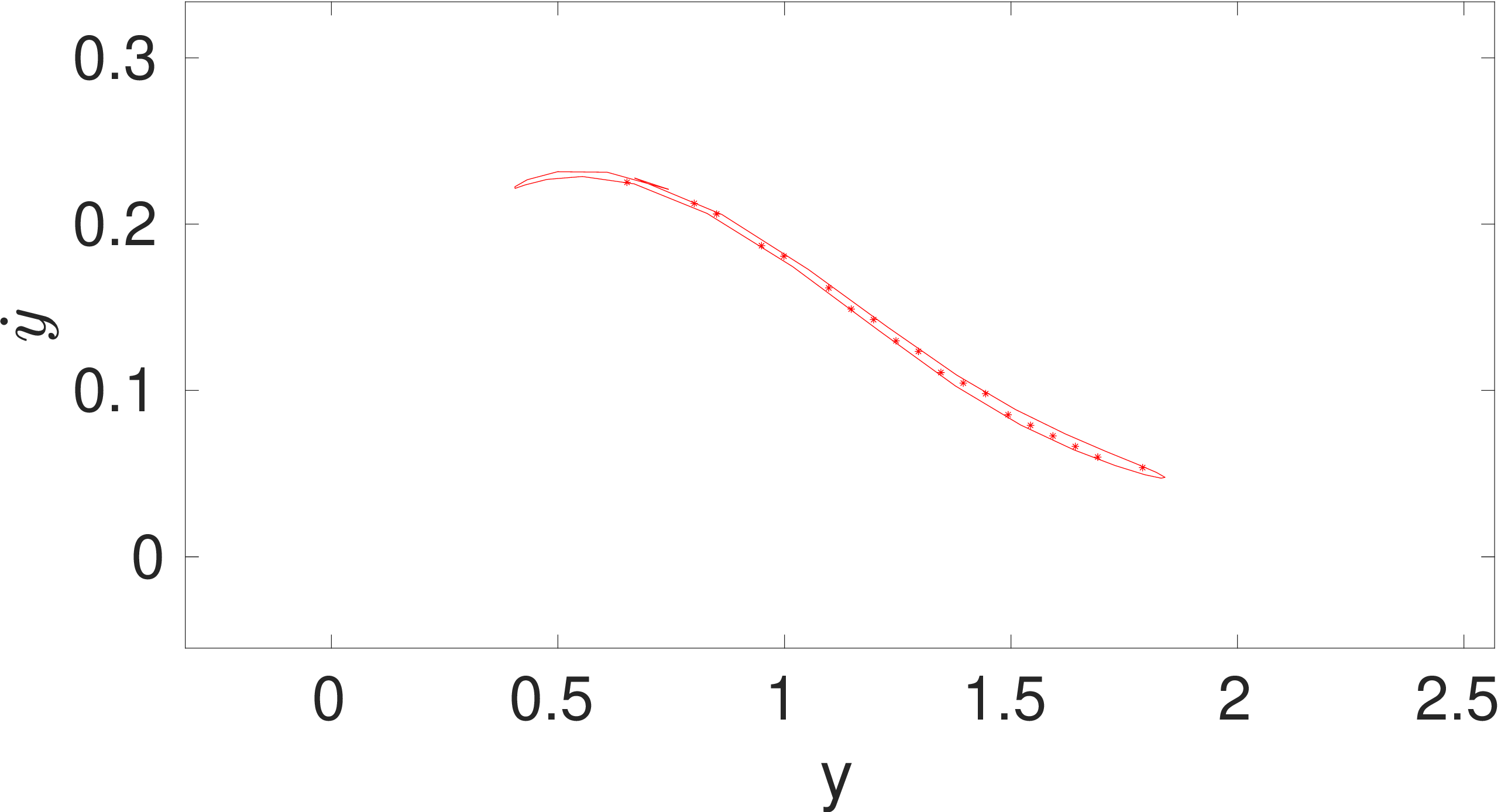}
\includegraphics[width=0.245\textwidth]{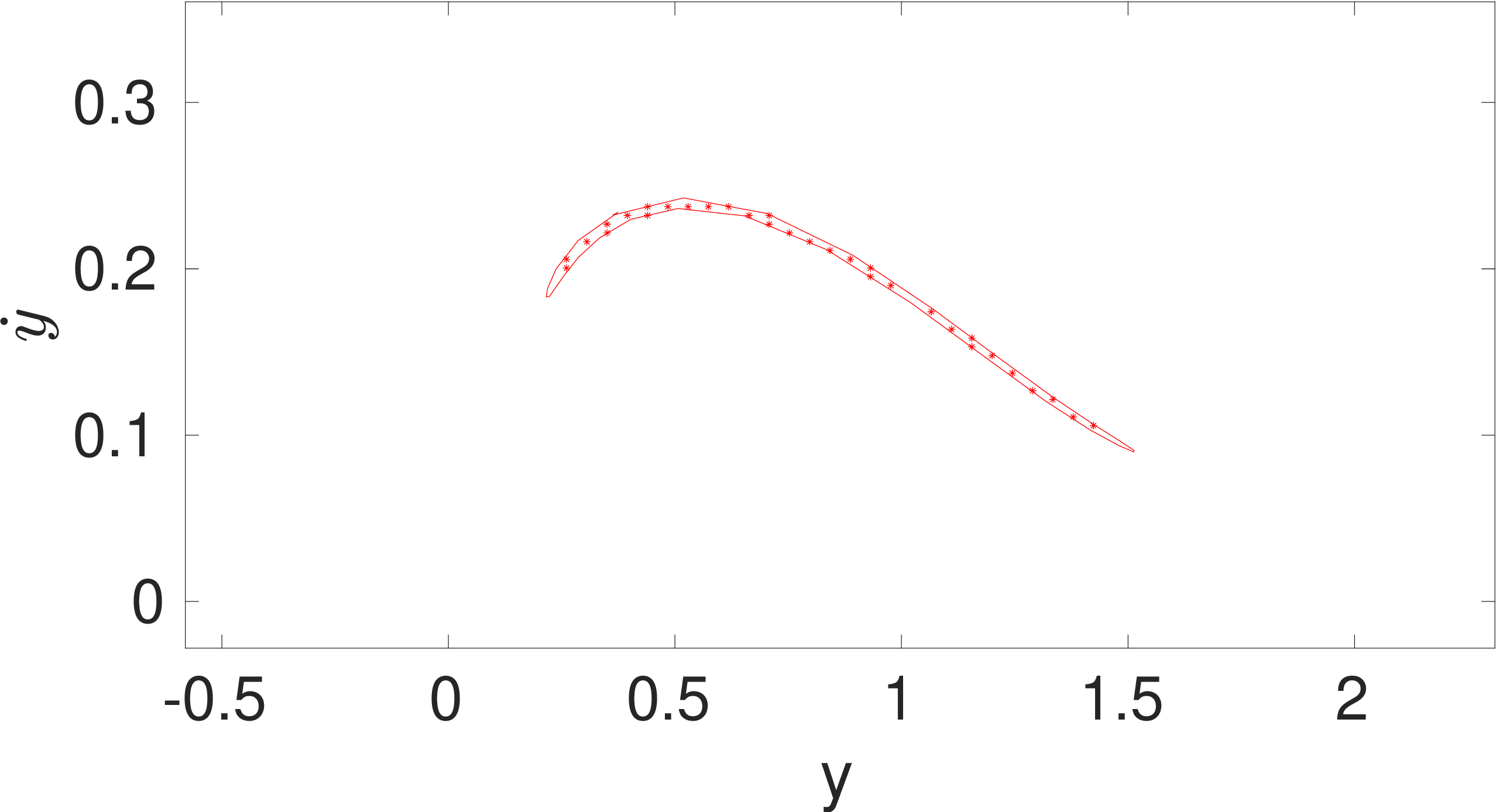}
\includegraphics[width=0.245\textwidth]{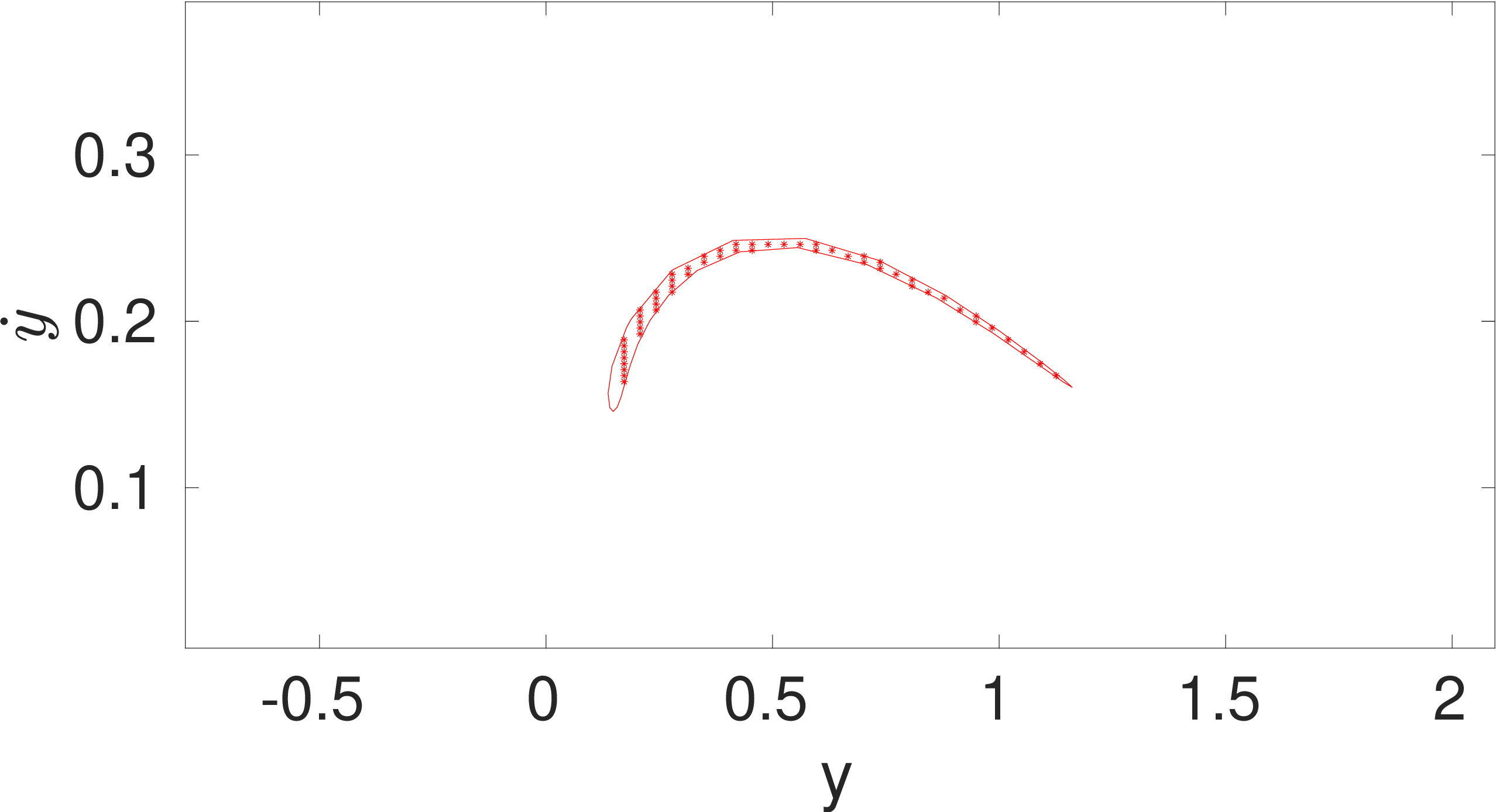}
\includegraphics[width=0.245\textwidth]{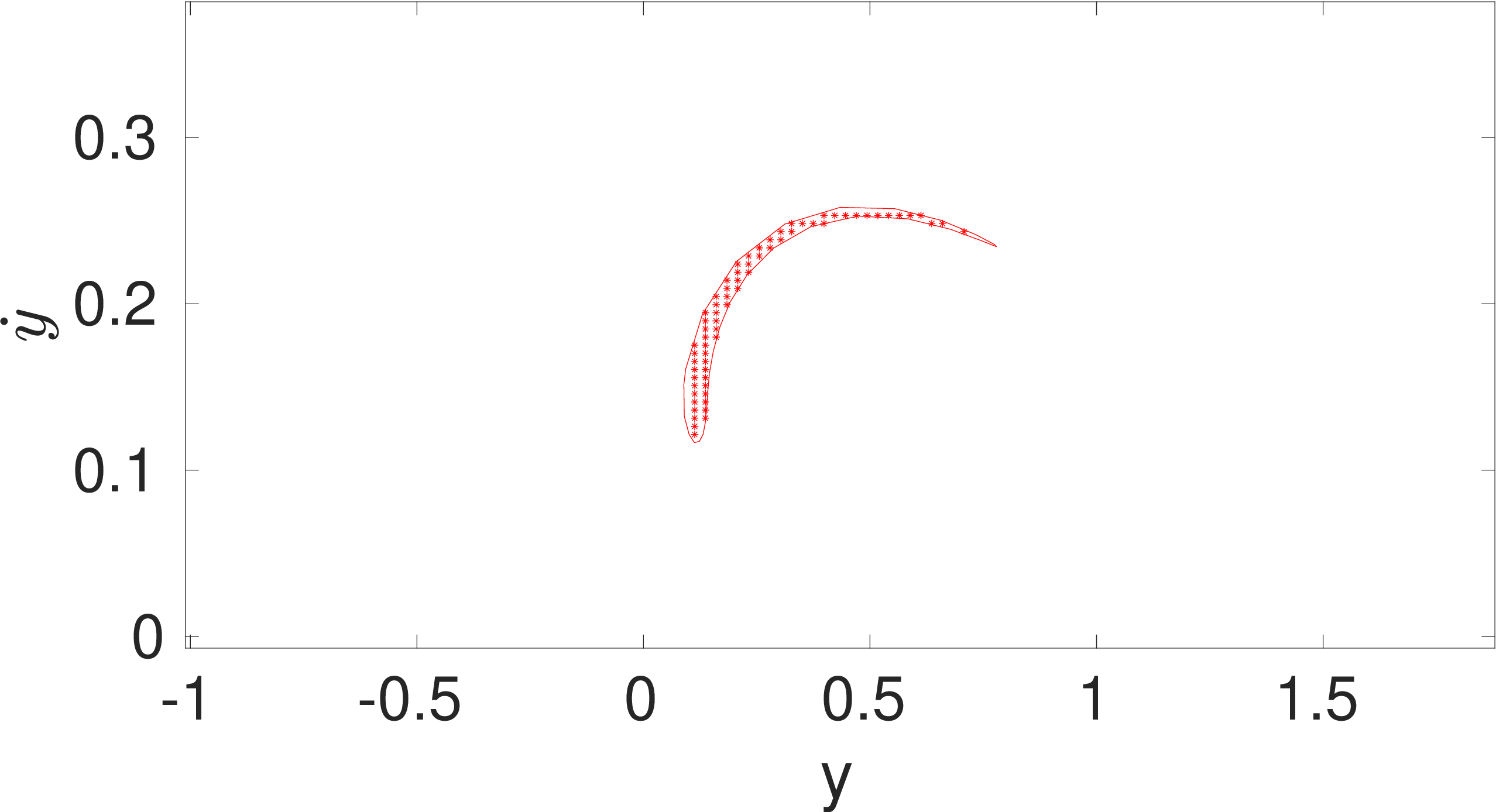}

\includegraphics[width=0.245\textwidth]{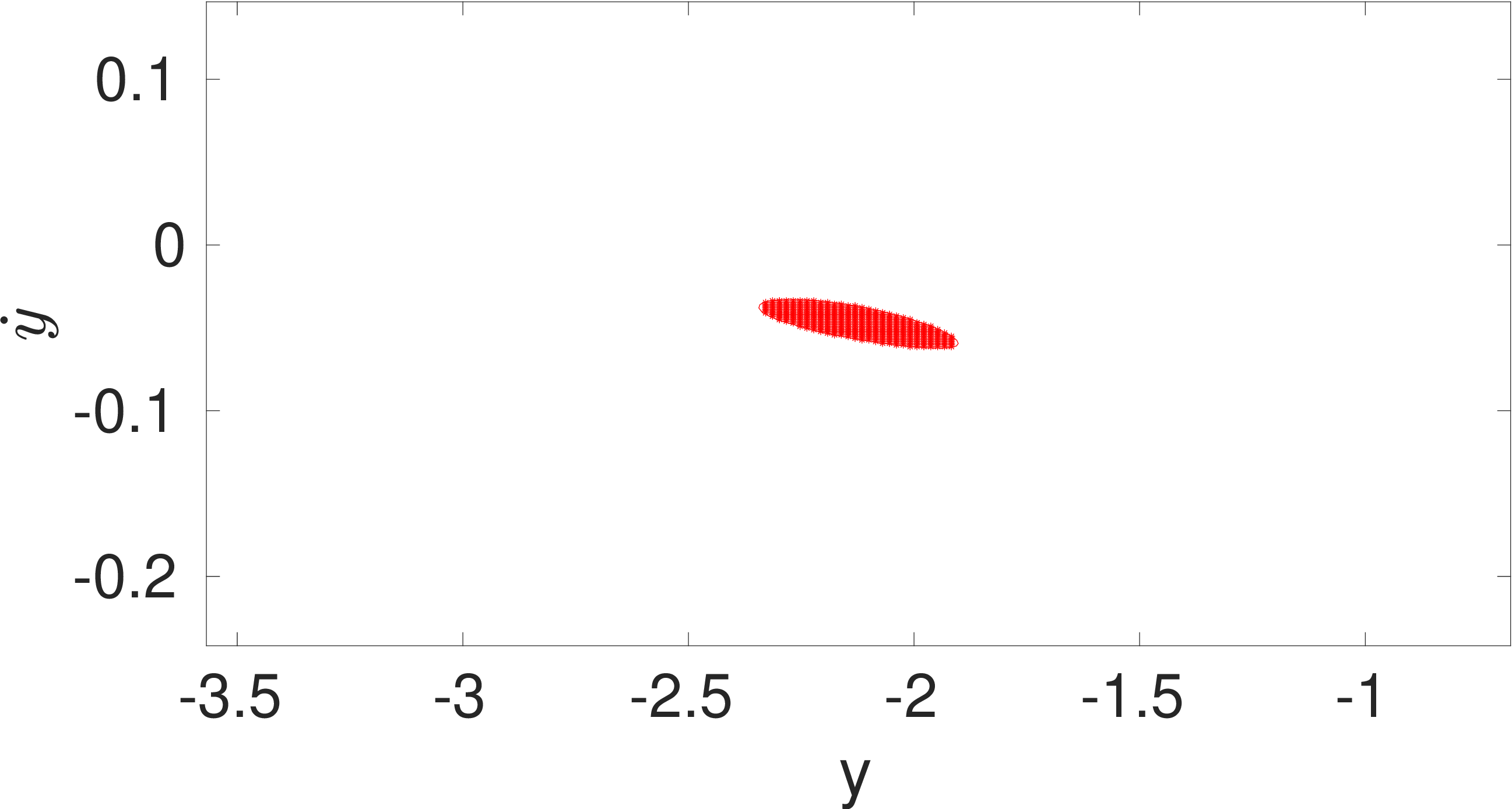}
\includegraphics[width=0.245\textwidth]{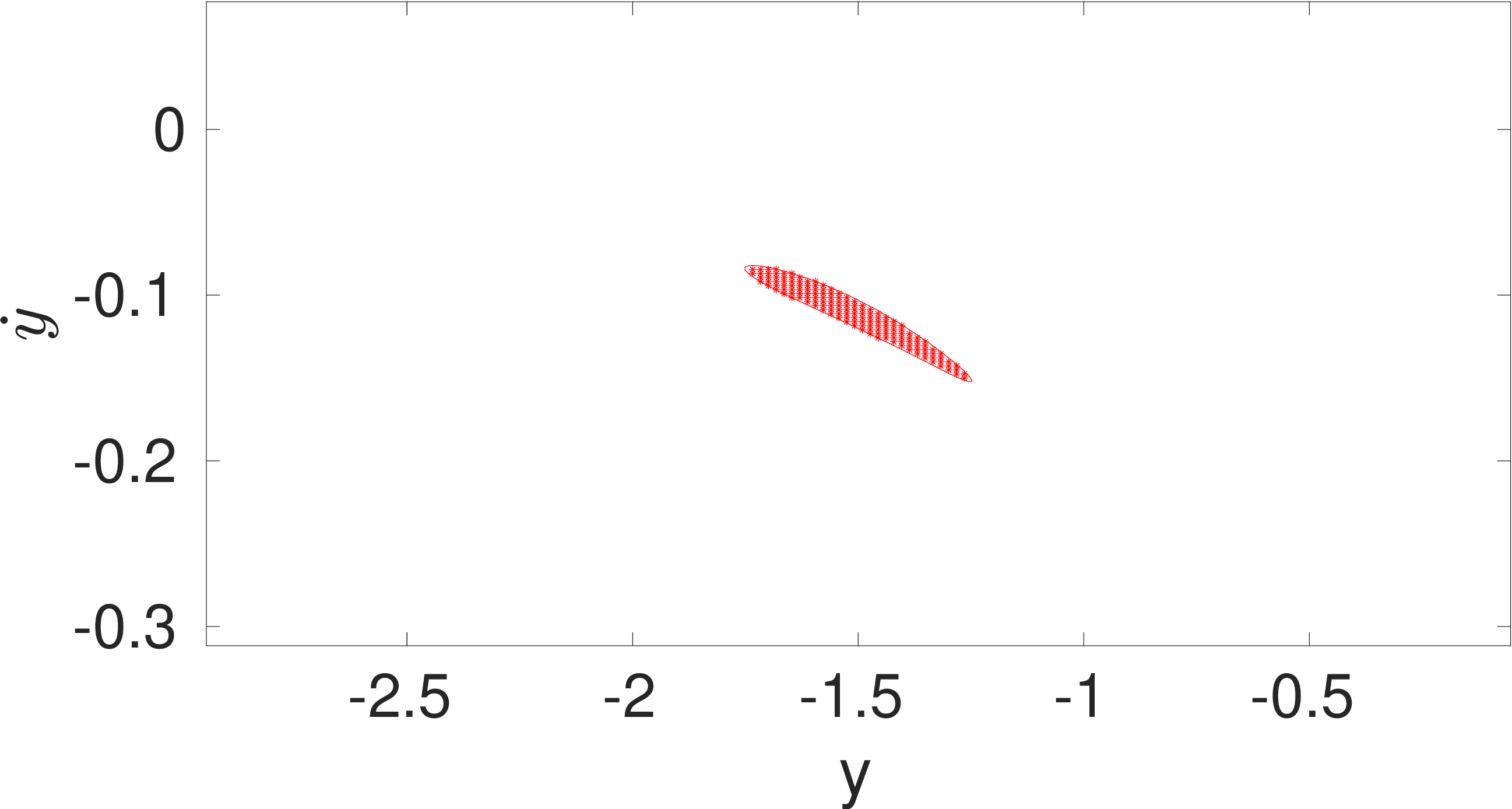}
\includegraphics[width=0.245\textwidth]{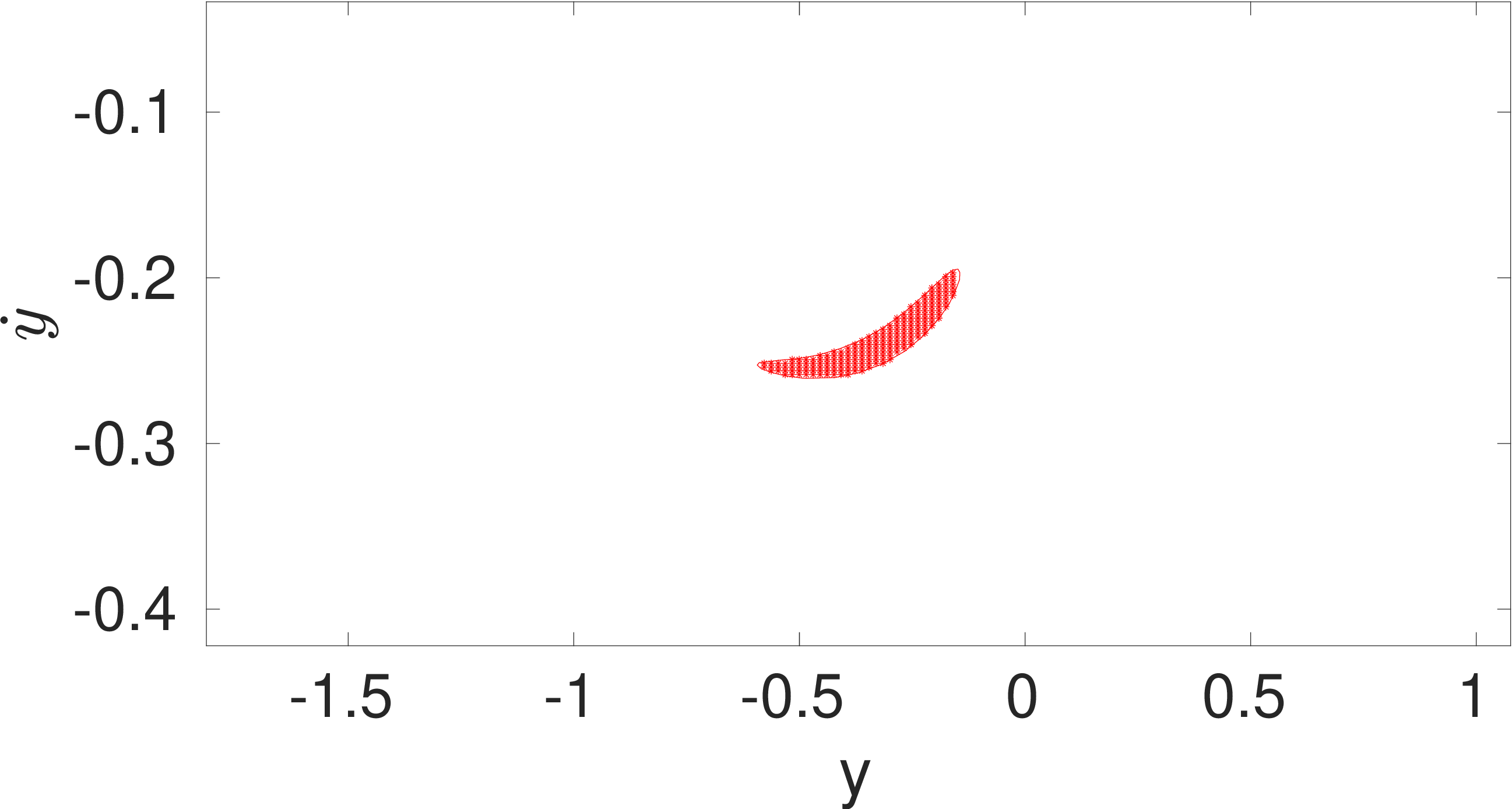}
\includegraphics[width=0.245\textwidth]{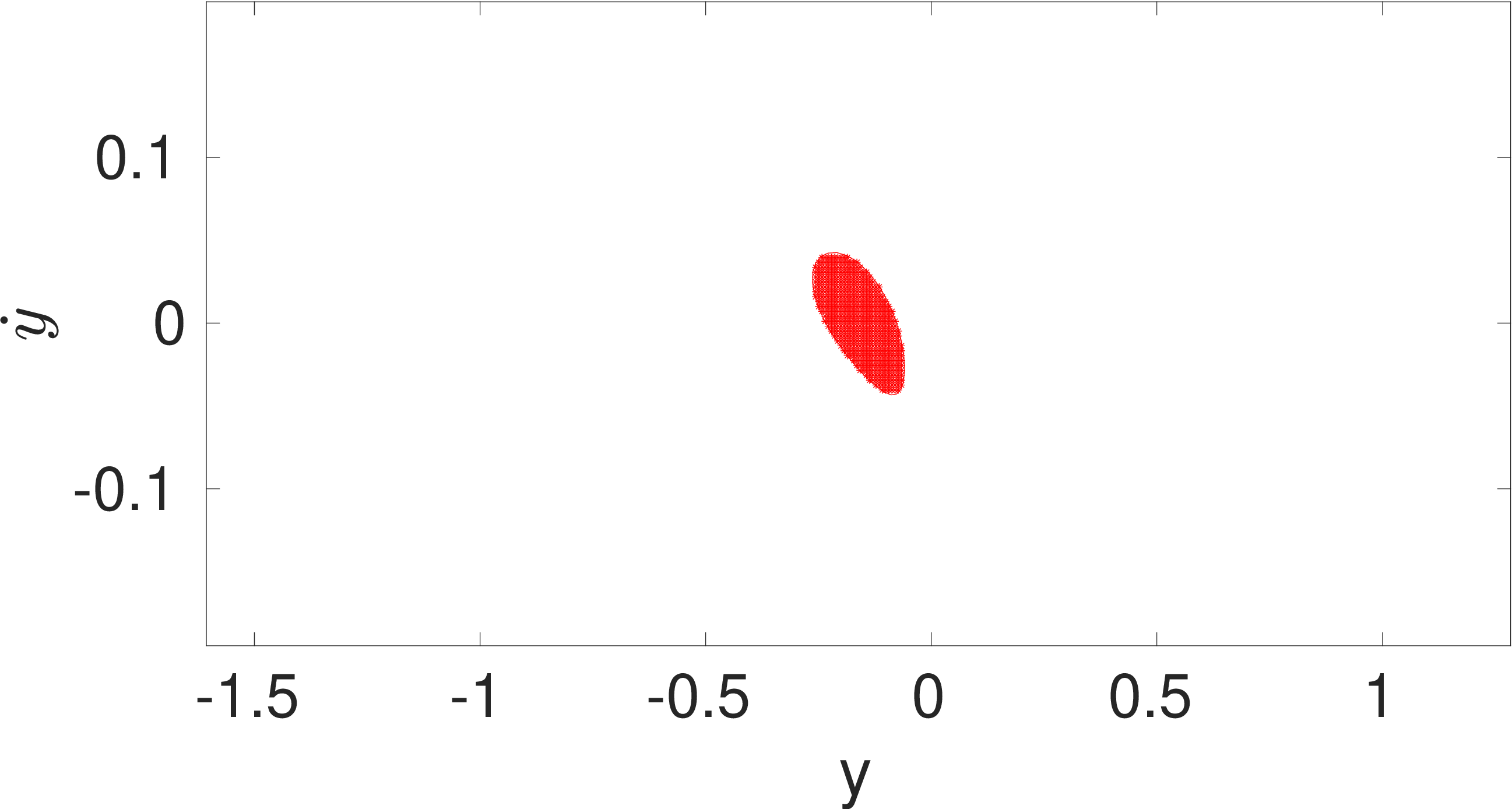}
\caption{Model A $(y,\dot{y})$ projection of the intersection of plane $S \equiv \{x=0\}$ with the stable manifold associated to the Lyapunov family around L$_2$ (top row) and L$_1$ (bottom row). From left to right: $\delta = 2.5, 3.5, 4.5, 5.5$ kpc. The curves show the narrowing and stretching of the tube as the asymmetric mass component moves away from L$_1$, marking the constriction of transport channels and differences in escape orbit populations for the two arms. Note that the axis limits are different but their scale and range length are constant in each row.}
\label{fig:ModelA_corteLi}
\end{figure*}

Figure~\ref{fig:ModelA_corteLi} displays the $(y,\dot{y})$ projections of the manifold intersections with the $x=0$ hyperplane for both L$_2$ (top) and L$_1$ (bottom). The key feature is the differential constriction of the curves: as $\delta$ increases and the asymmetric mass component moves away from L$_2$, the manifold tube associated with L$_2$ narrows significantly compared to that of L$_1$. This constriction indicates a reduced area of initial conditions available for transit orbits emanating from L$_2$, which directly translates into lower density in the arm associated with this unstable point. These results are consistent with previous work \citep{Asymmetry}, extending the analysis to the full asymmetric mass component displacement range.

\begin{figure}
\centering
\includegraphics[width=0.23\textwidth]{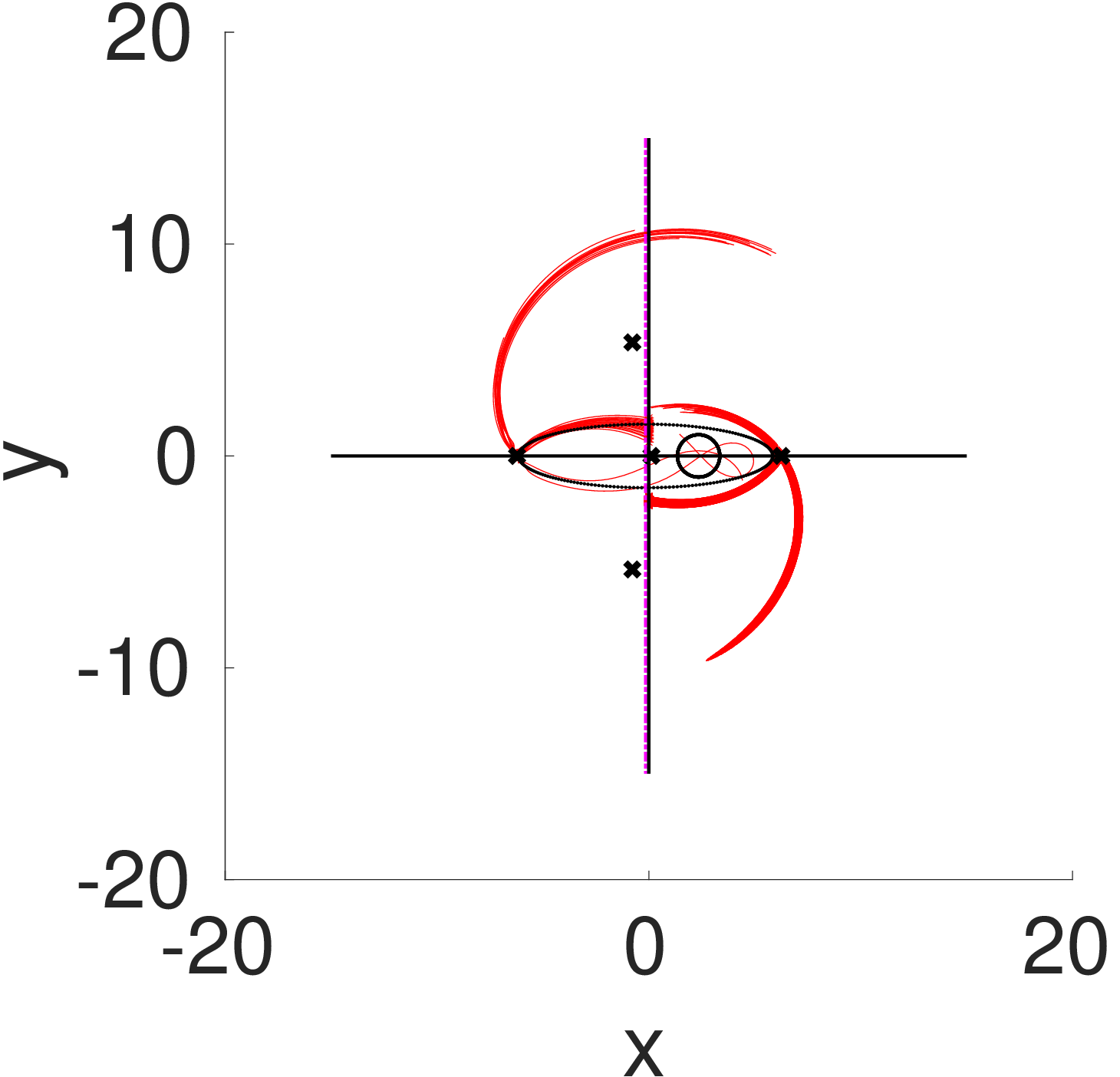}
\includegraphics[width=0.23\textwidth]{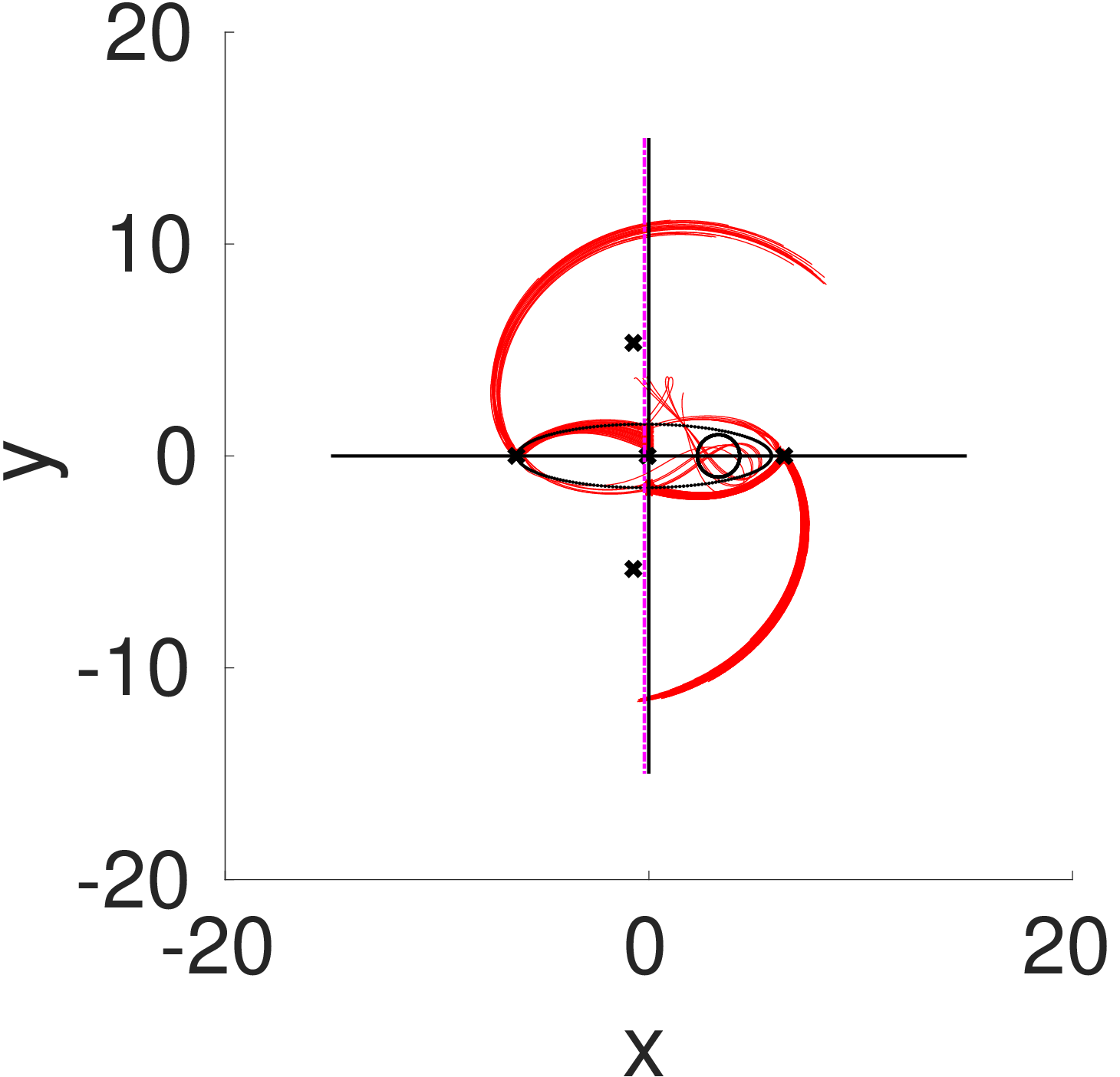}\\
\includegraphics[width=0.23\textwidth]{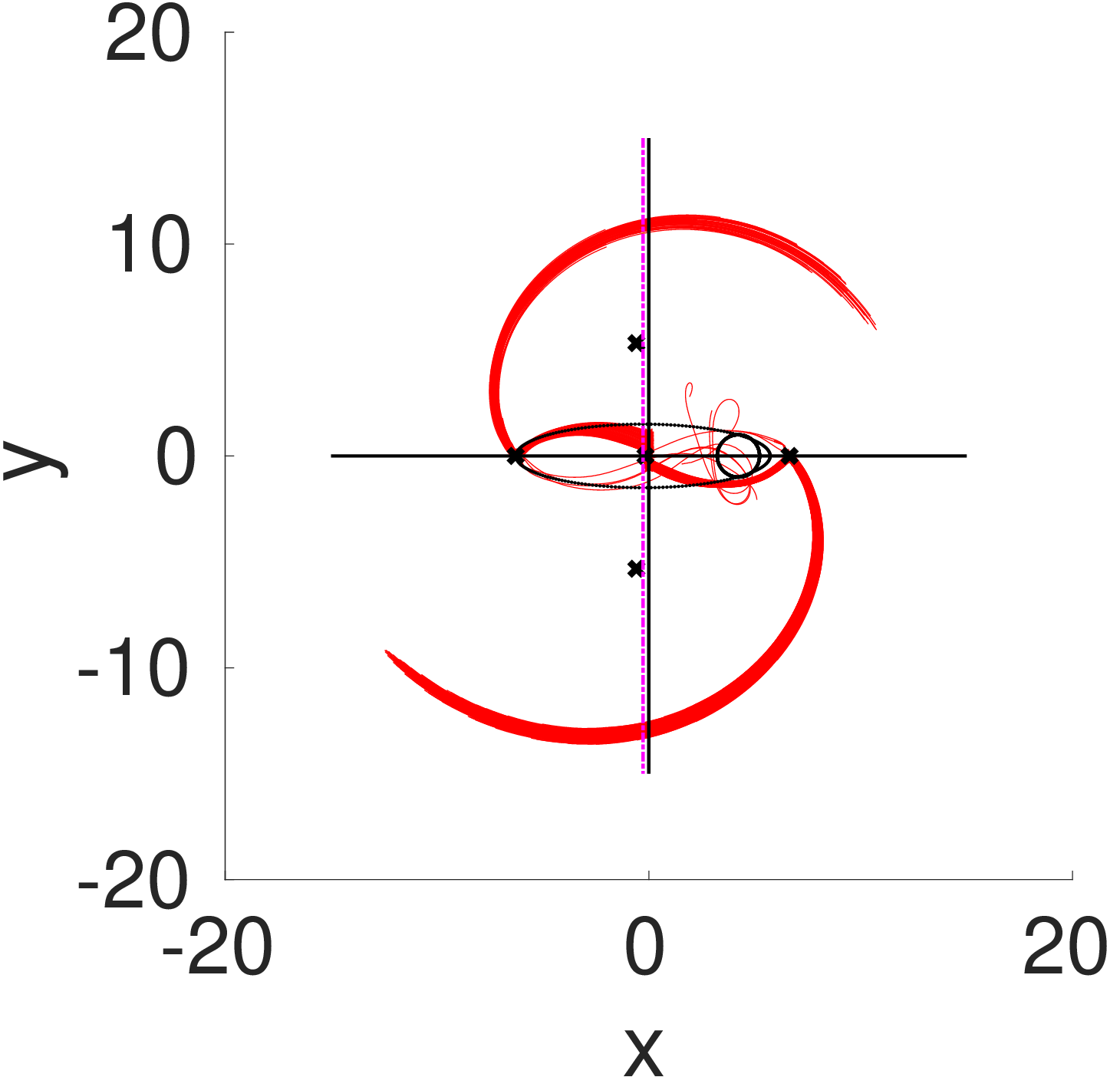}
\includegraphics[width=0.23\textwidth]{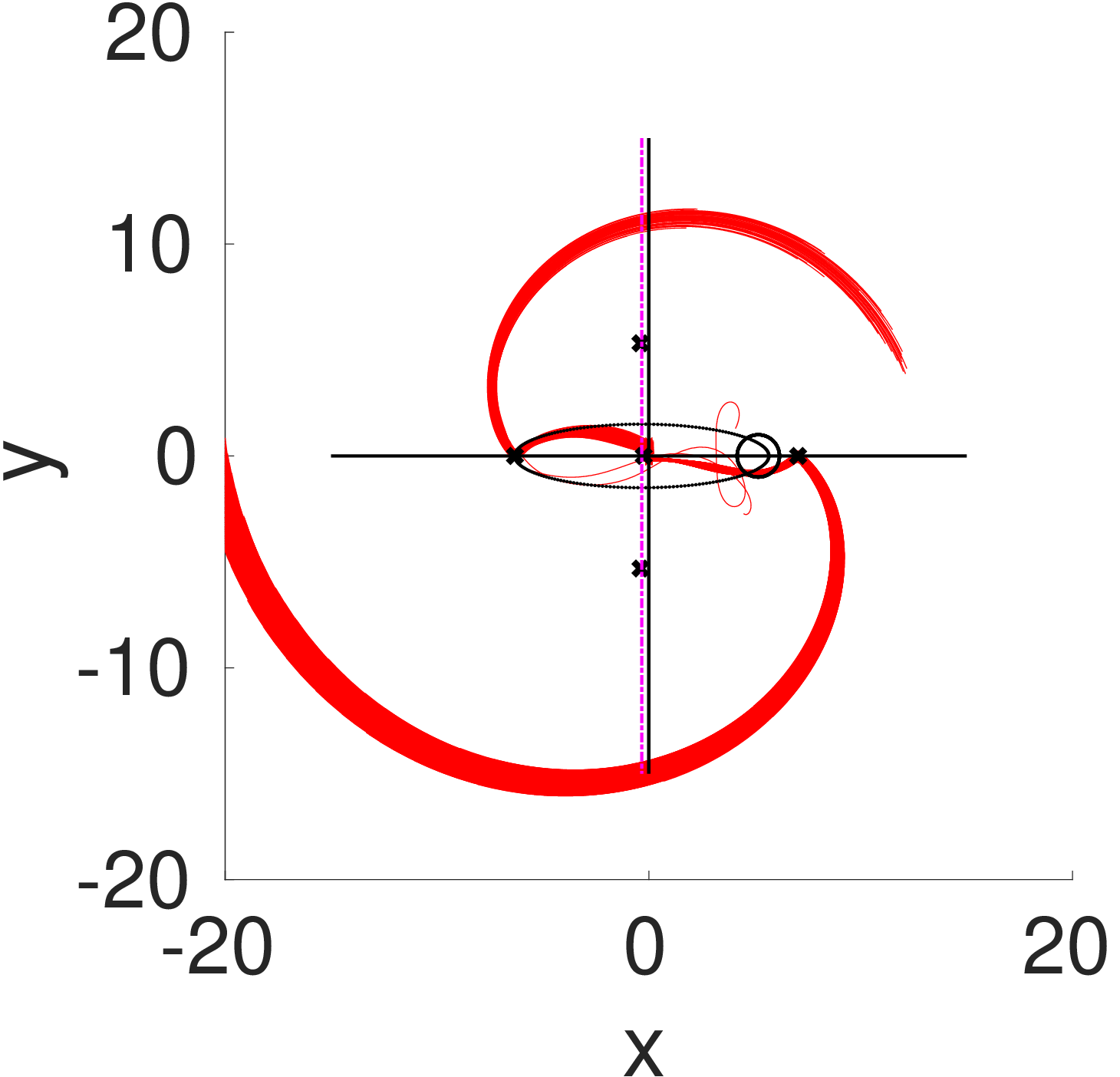}
\caption{Model A transit orbits resulting from integration of initial conditions from Fig.~\ref{fig:ModelA_corteLi}. Equilibrium points marked in red. Bar and the asymmetric mass component outlined by dotted black curves. Reference system marked with a solid black line and centre of the bar with a dotted magenta line. From top left to bottom right: $\delta = 2.5, 3.5, 4.5, 5.5$ kpc. The velocity of escaping trajectories depends on the position of the asymmetric mass component, demonstrating that manifold transport is sensitive to the details of the asymmetric potential.}
\label{fig:ModelA_orbits}
\end{figure}

The transit orbits (Fig.~\ref{fig:ModelA_orbits}) result from integrating the initial conditions from Fig.~\ref{fig:ModelA_corteLi}. Although the integration time is constant across all $\delta$ values, the escaping trajectories progress at different velocities and reach different radial distances, reflecting the asymmetry-dependent transport properties of the manifold system. This demonstrates that the asymmetric mass component displacement modulates not only the geometry of the manifolds but also the efficiency of matter transport through them.

For brevity, we only show these transit orbits for Model A, because the qualitative behaviour is analogous for all models. In the remaining models we display the invariant manifolds associated with the Lyapunov orbits, which confine these transit orbits.

\subsection{Model B: Small bar offset. Two-armed structure}

When a small bar offset ($\Lambda = 1.21$ kpc) is introduced while preserving the centre of mass interior to the bar, the equilibrium point configuration shifts but retains its five-point configuration without bifurcation (as shown in Section~\ref{sec:bifurcations}). This robustness has important consequences for the manifold structure and arm morphology.

\begin{figure}
\includegraphics[width=0.24\textwidth]{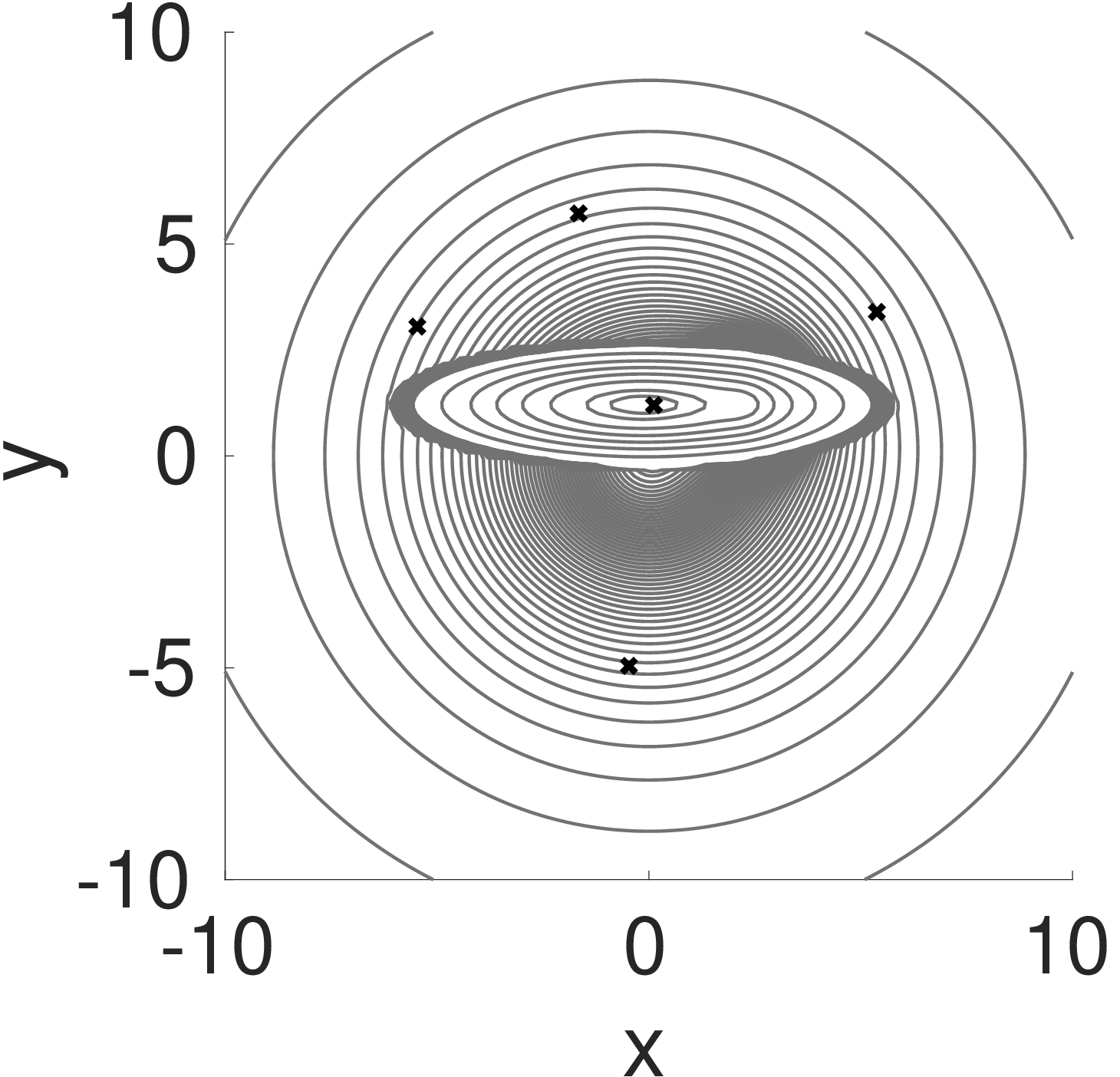}
\includegraphics[width=0.24\textwidth]{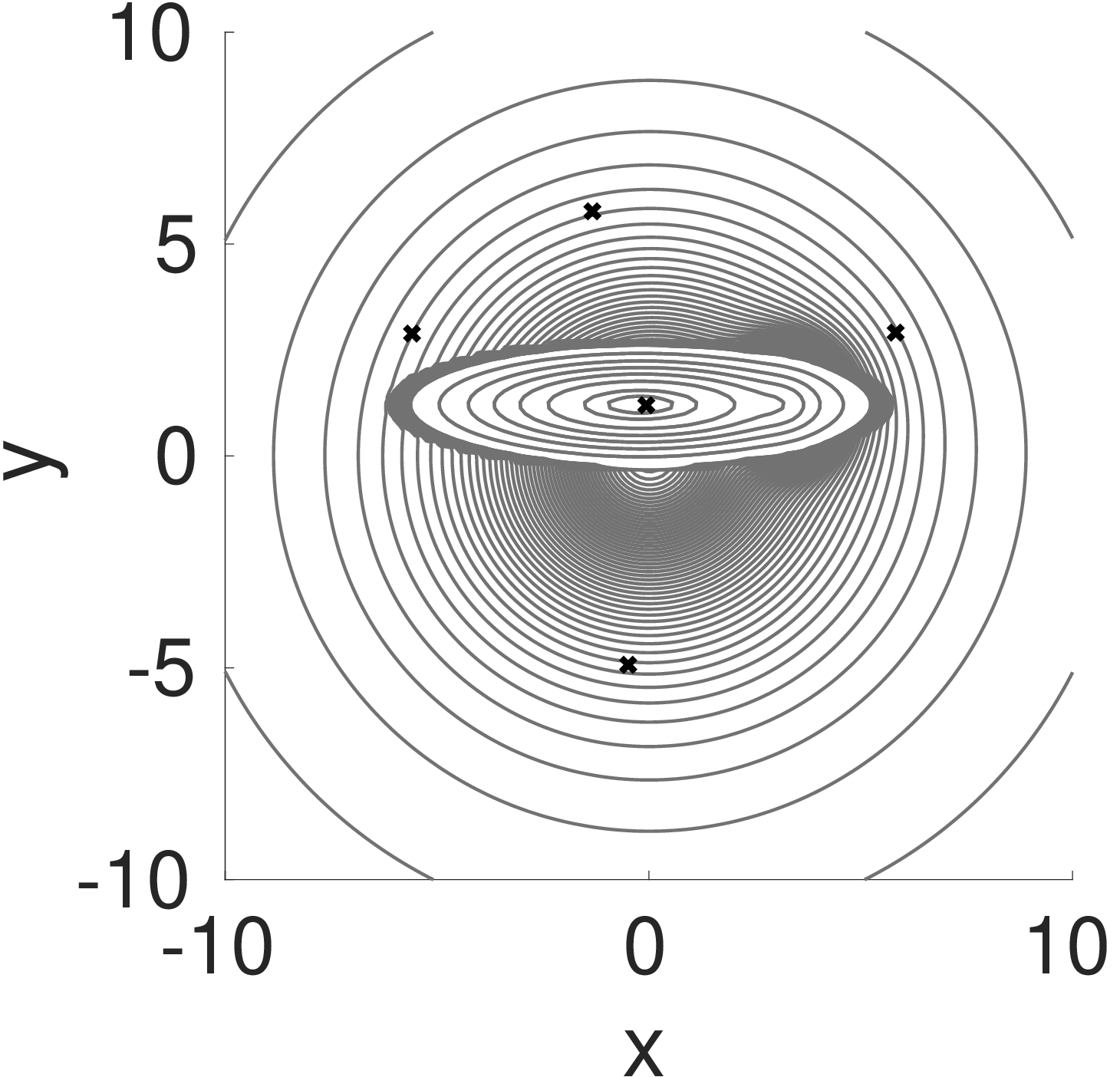} \\
\includegraphics[width=0.24\textwidth]{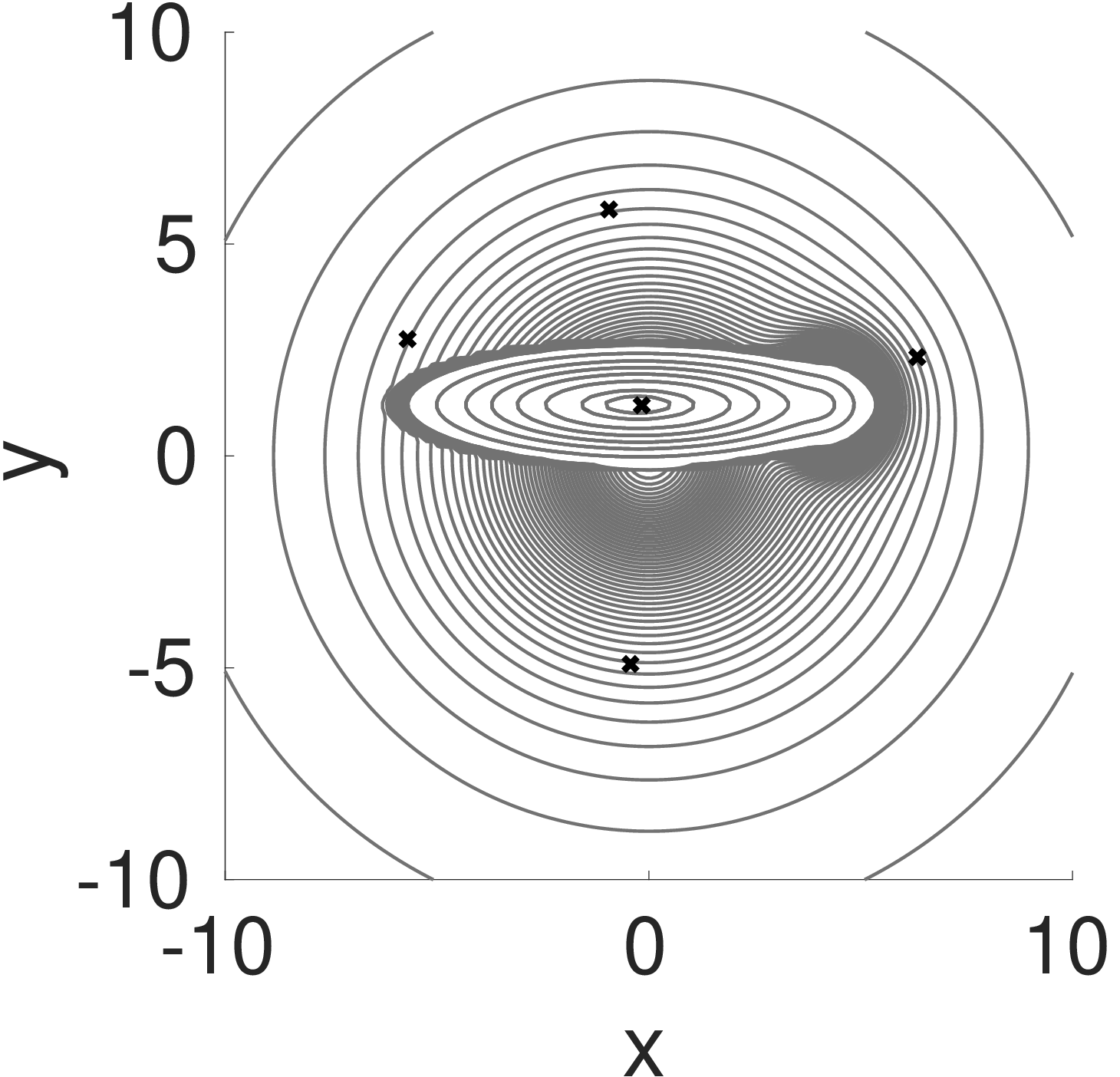}
\includegraphics[width=0.24\textwidth]{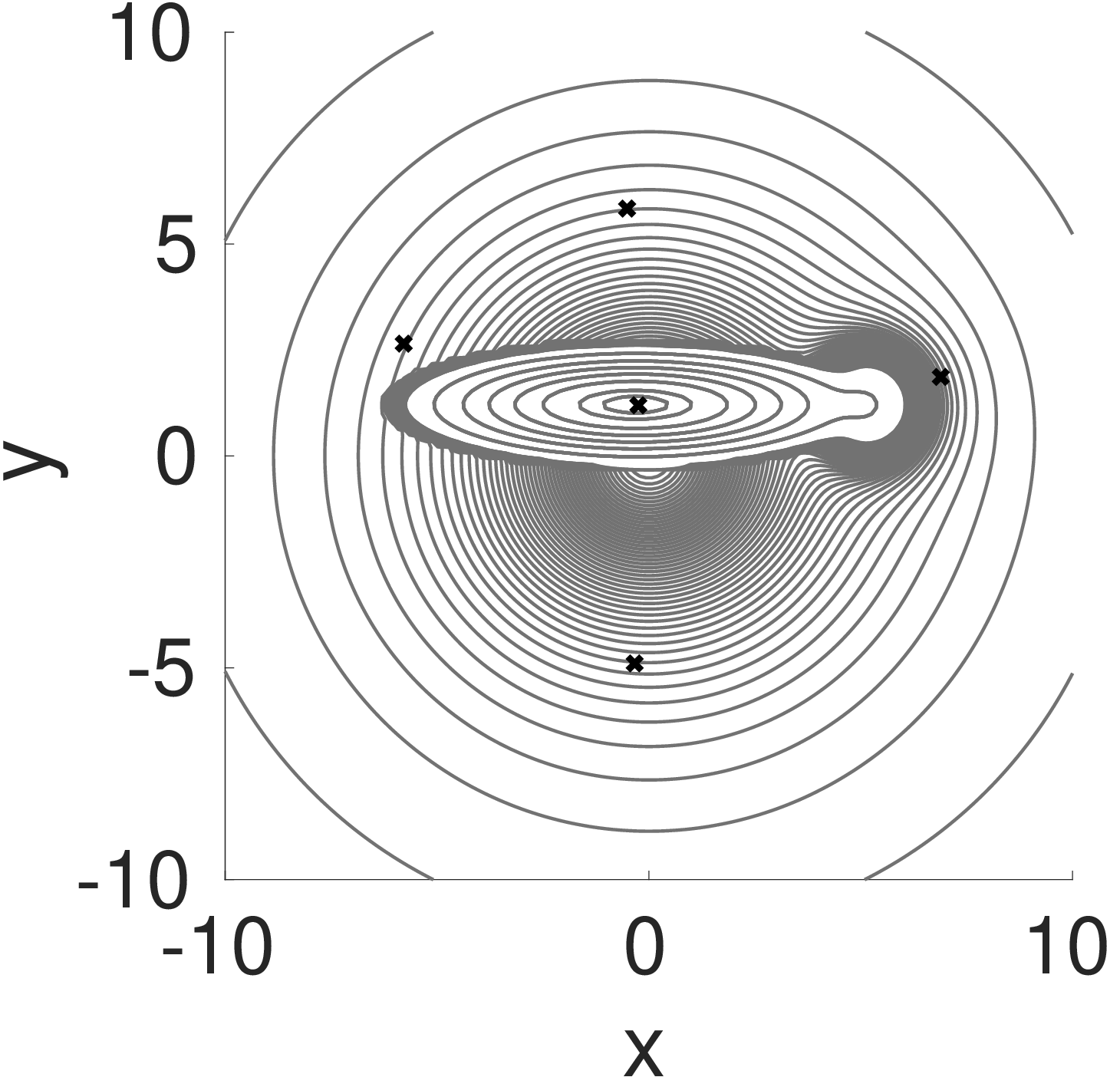}
\caption{Model B isodensity contours ($\Lambda = 1.21$ kpc, small offset). Equilibrium points marked by black crosses. From top left to bottom right: $\delta = 2.5, 3.5, 4.5, 5.5$ kpc. The offset shifts the entire configuration perpendicular to the bar, but the structure remains qualitatively similar to Model A.}
\label{fig:ModelB_isodens}
\end{figure}

The isodensity contours for Model B (Fig.~\ref{fig:ModelB_isodens}) show the configuration shifted by the bar offset. The density asymmetry pattern resembles Model A but is displaced relative to the centre of mass of the system.

\begin{figure}
\includegraphics[width=0.24\textwidth]{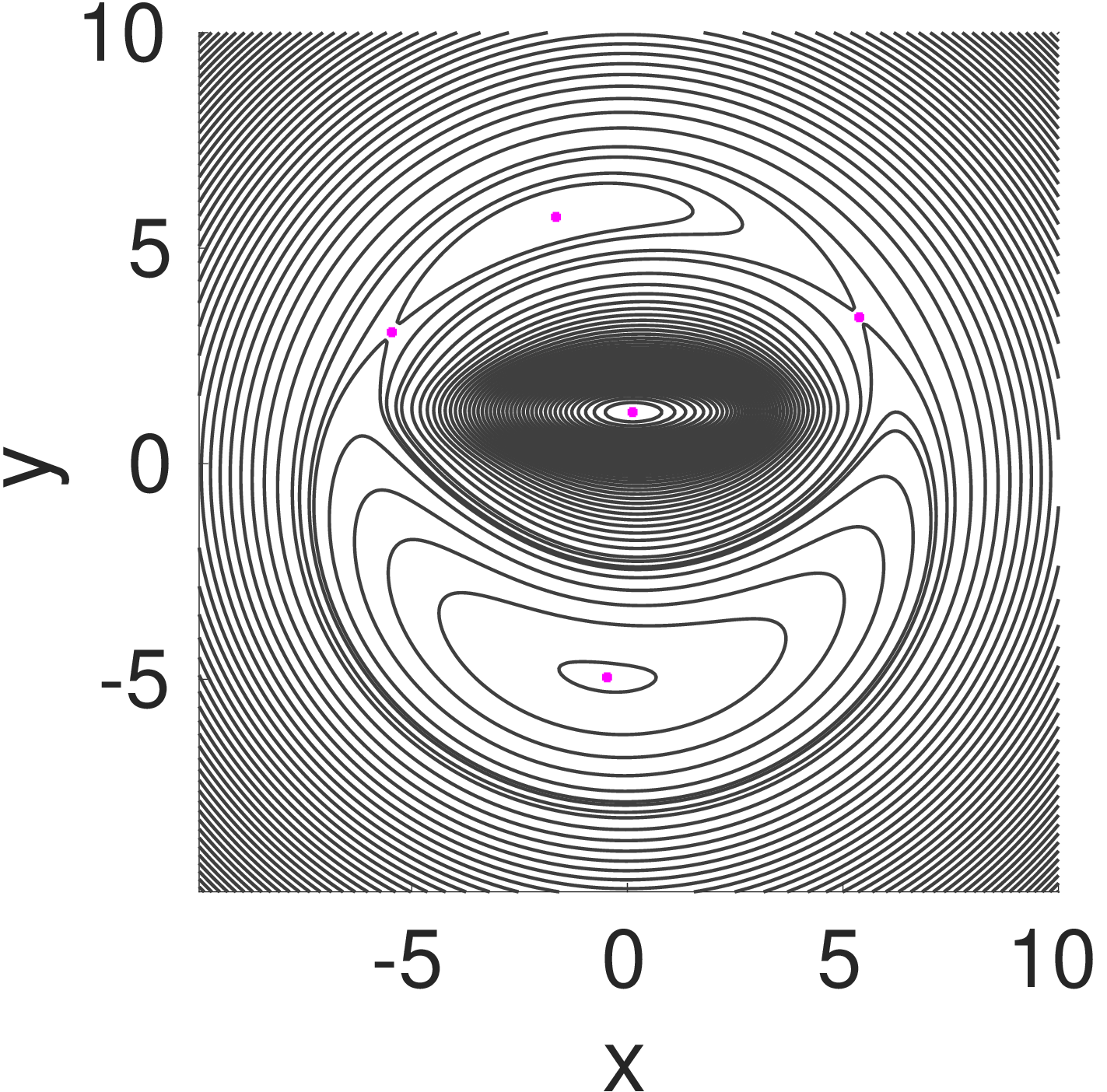}
\includegraphics[width=0.24\textwidth]{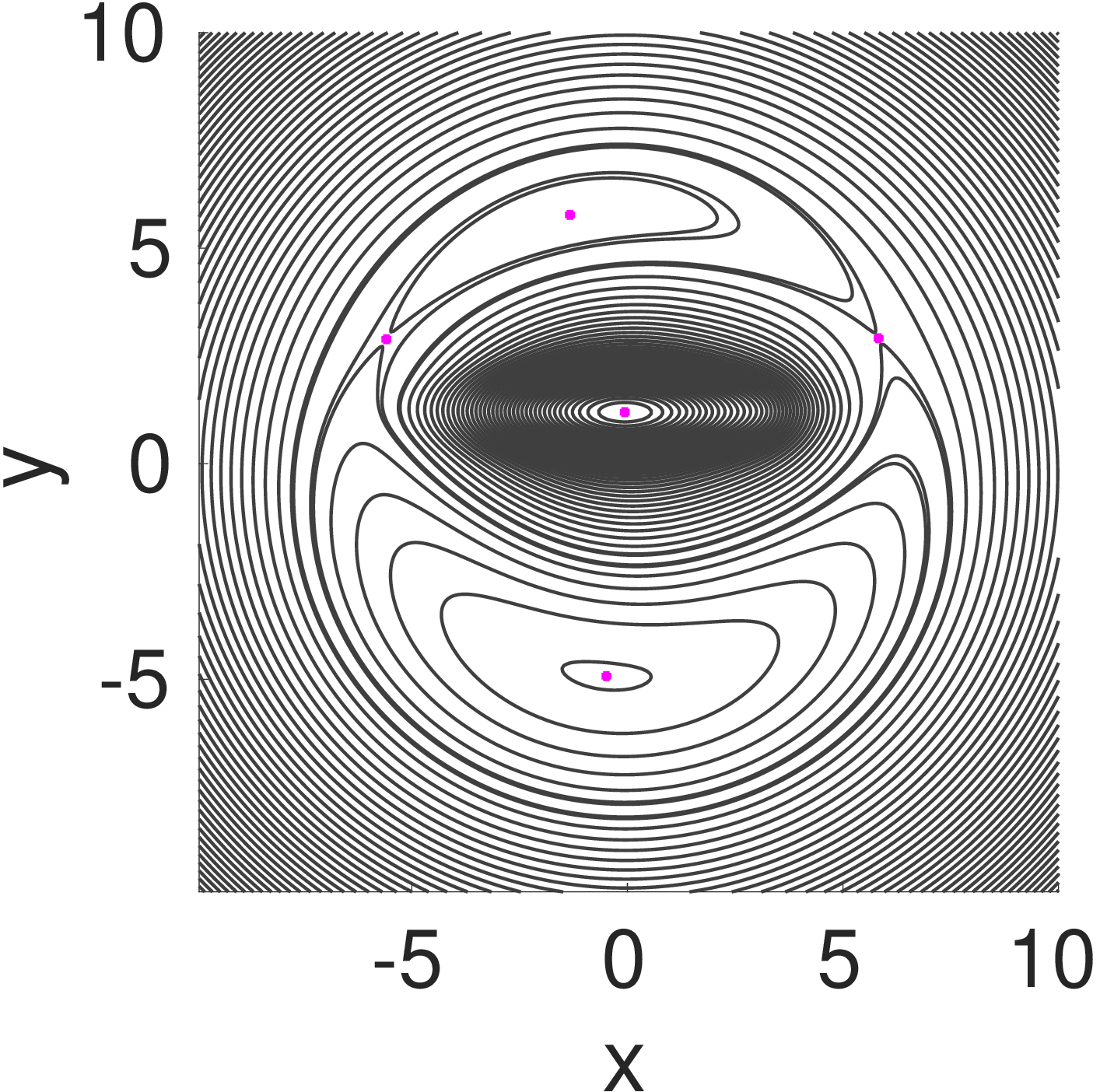} \\
\includegraphics[width=0.24\textwidth]{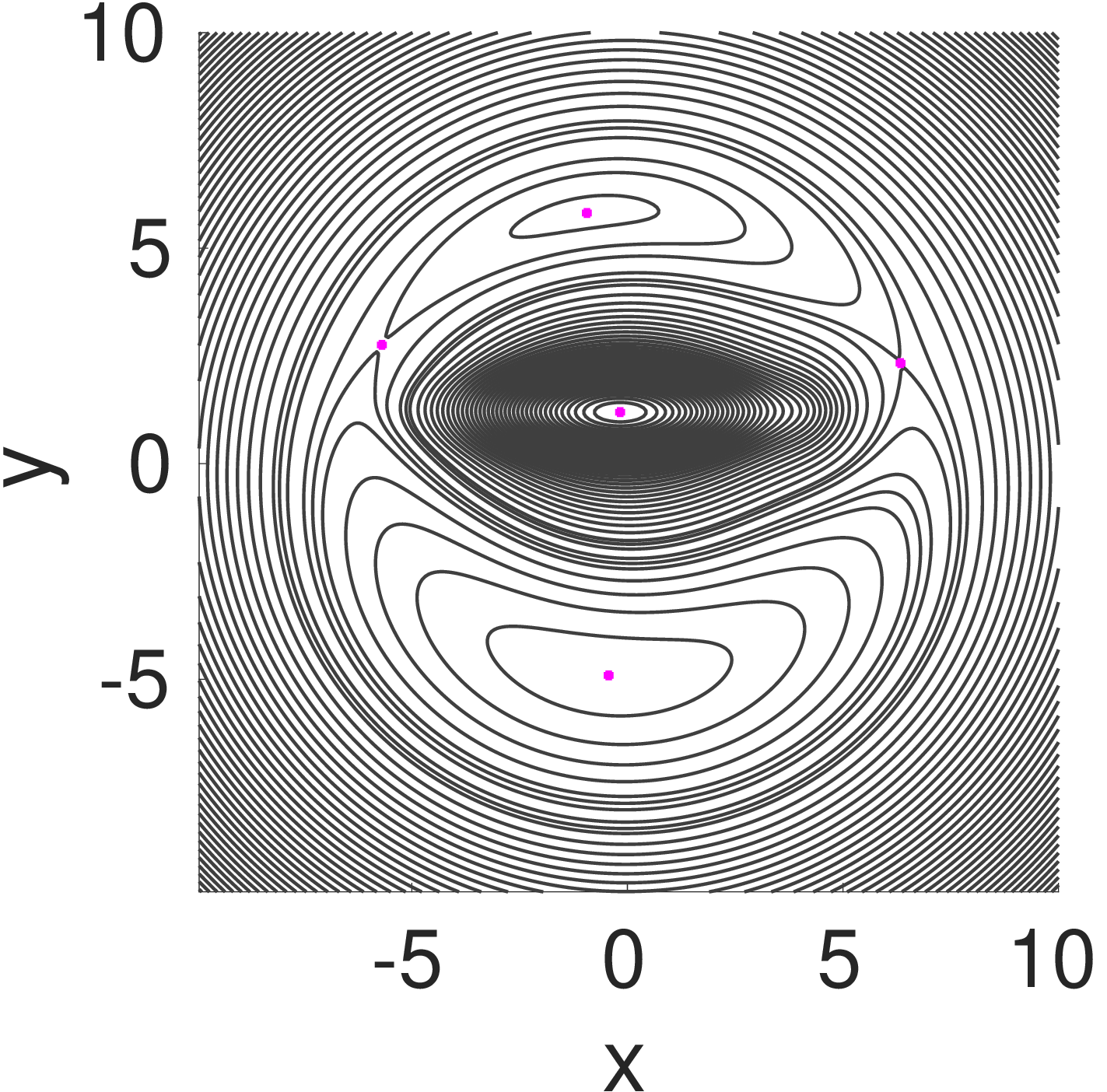}
\includegraphics[width=0.24\textwidth]{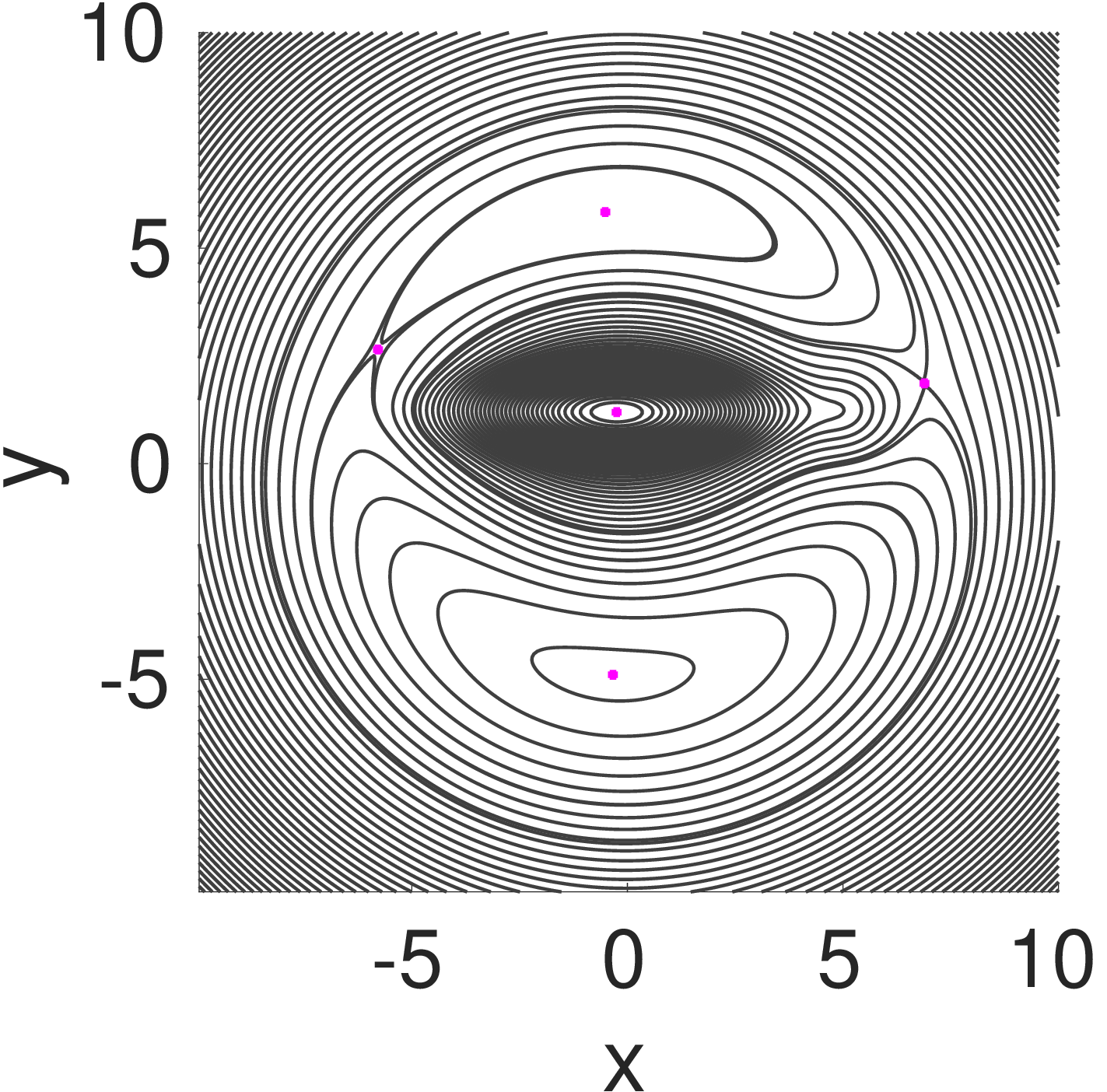}
\caption{Model B isopotential contours ($\Lambda = 1.21$ kpc). Equilibrium points marked by magenta dots. From top left to bottom right: $\delta = 2.5, 3.5, 4.5, 5.5$ kpc.}
\label{fig:ModelB_isopot}
\end{figure}

The isopotential curves (Fig.~\ref{fig:ModelB_isopot}) mark the effective potential structure. Although the level curves are distorted by the offset, the fundamental five-point equilibrium structure persists, maintaining the necessary unstable points for two-armed pattern.

\begin{figure}
\includegraphics[width=0.24\textwidth]{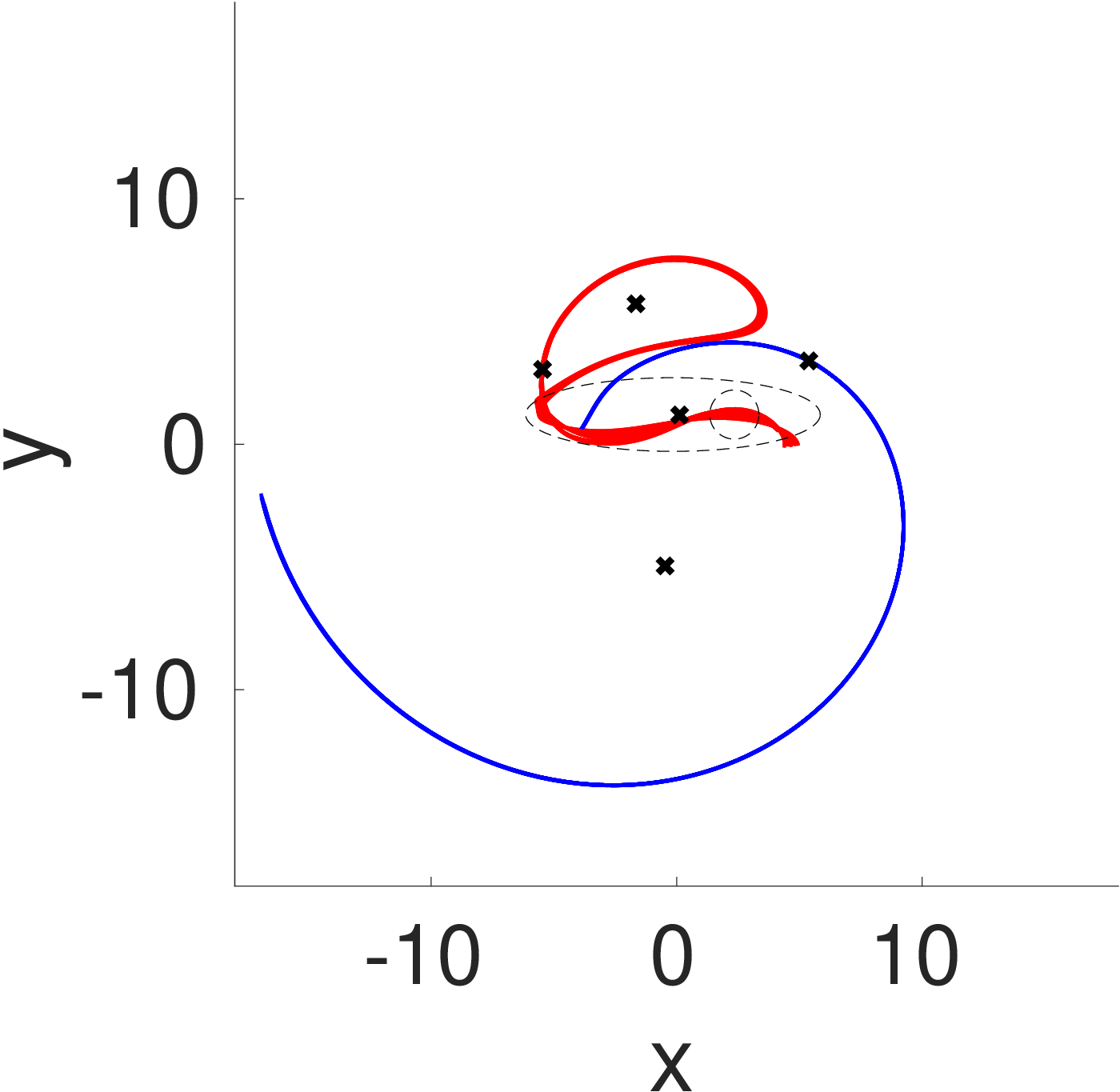}
\includegraphics[width=0.24\textwidth]{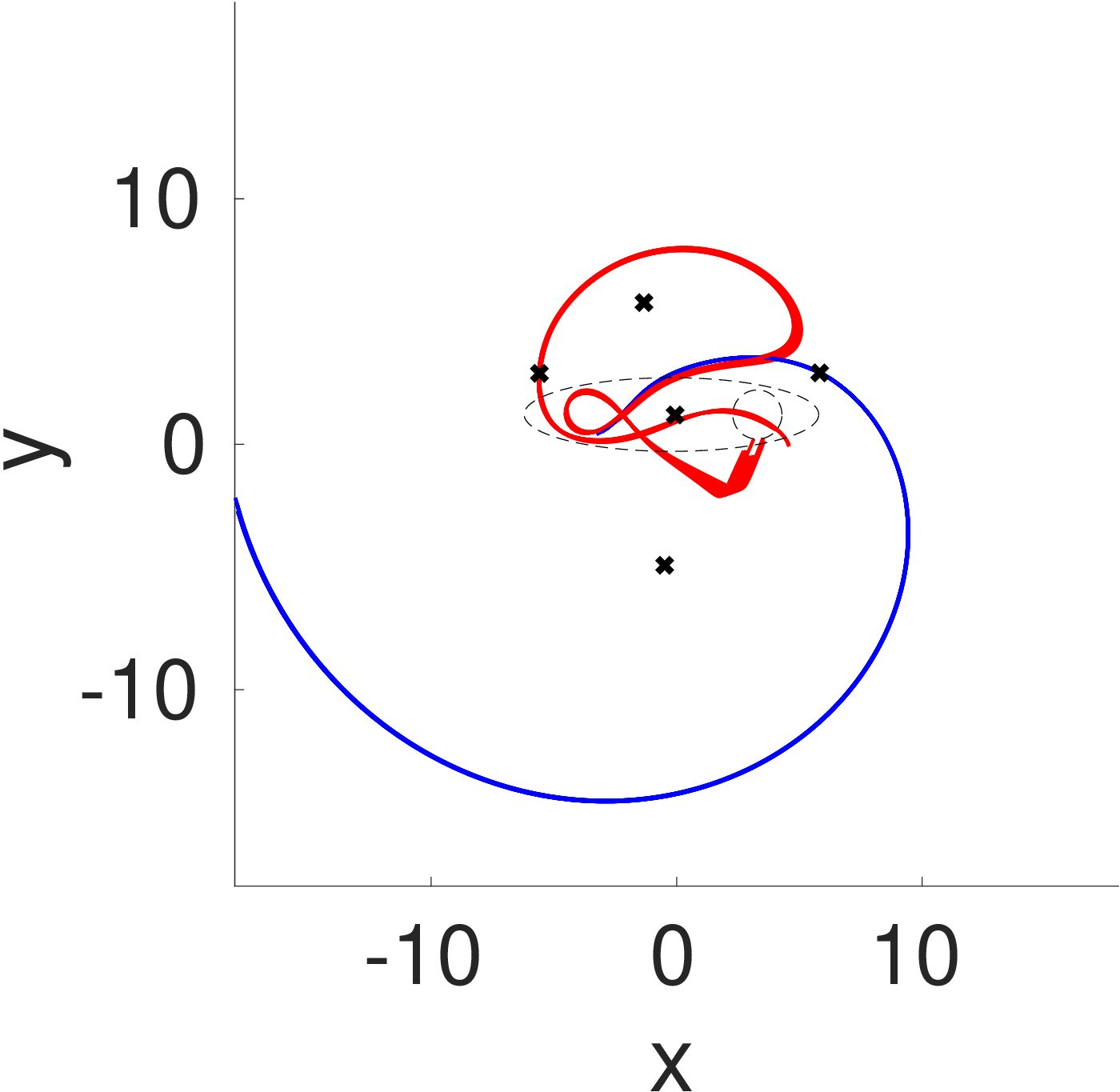}\\
\includegraphics[width=0.24\textwidth]{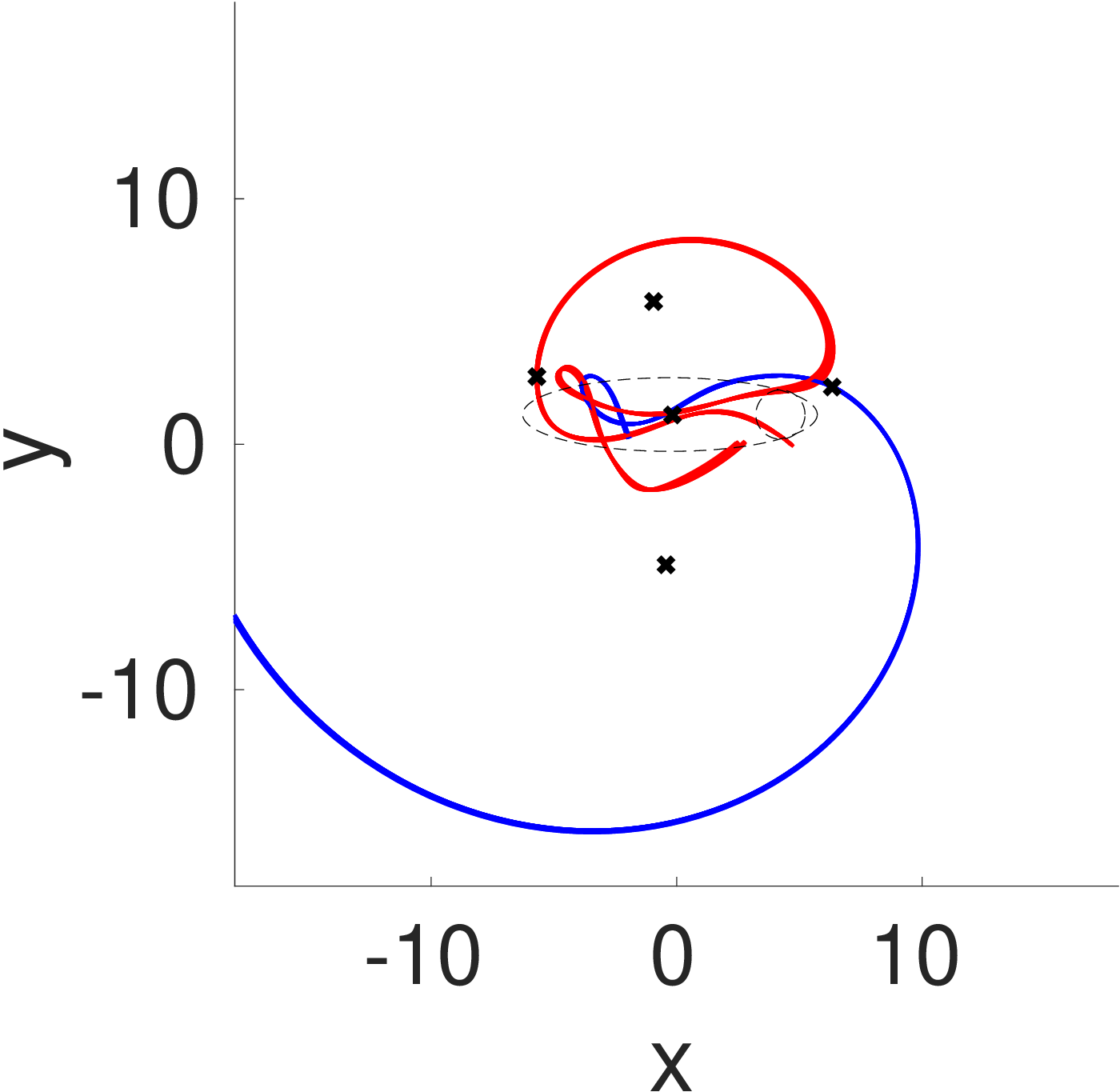}
\includegraphics[width=0.24\textwidth]{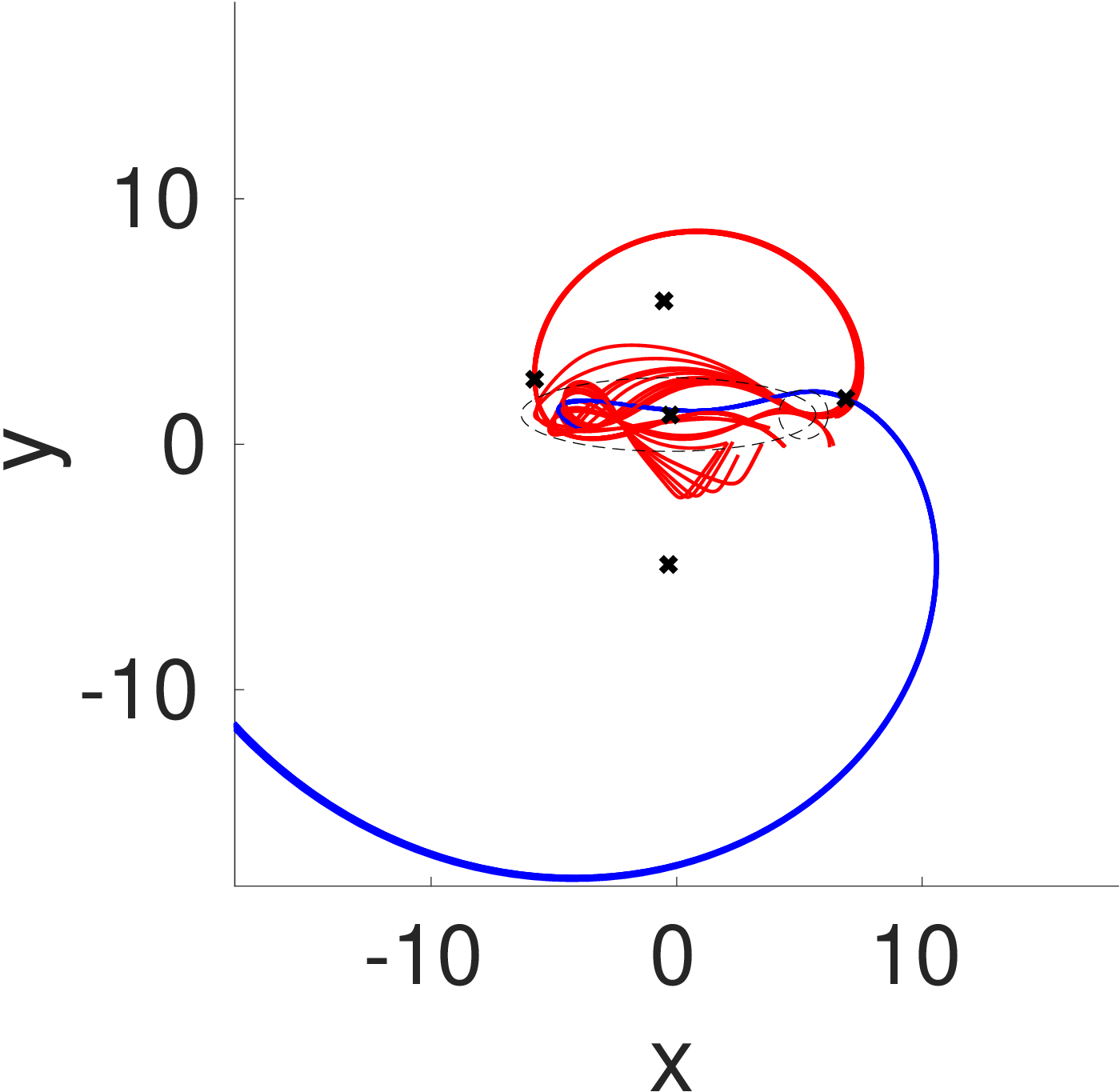}
\caption{Model B unstable invariant manifolds ($\Lambda = 1.21$ kpc). Bar and the asymmetric mass component outlined by dotted black curves. From top left to bottom right: $\delta = 2.5, 3.5, 4.5, 5.5$ kpc. The two-armed pattern is maintained, but the manifold shapes differ significantly from Model A: the manifold related to L$_2$ acquires a distinctly different pattern compared to that of L$_1$.}
\label{fig:ModelB_manifolds}
\end{figure}

The key observation in Fig.~\ref{fig:ModelB_manifolds} is that the invariant manifolds exhibit strongly asymmetric arm morphologies: the manifold emanating from L$_2$ shows a qualitatively different shape compared to that from L$_1$. This asymmetric arm structure is governed by the isopotential curves induced by the offset. 

\subsection{Model C: Large bar offset. One-armed structure}

For offsets sufficiently large to place the centre of mass exterior to the bar ($\Lambda = 2.42$ kpc), the bifurcation at $\Lambda \approx 1.5$ kpc eliminates the collinear unstable points (as detailed in Section~\ref{sec:bifurcations}), reducing the system to only three equilibrium points (two linearly stable, one unstable). This fundamental structure change has direct consequences for arm structure.

\begin{figure}
\includegraphics[width=0.24\textwidth]{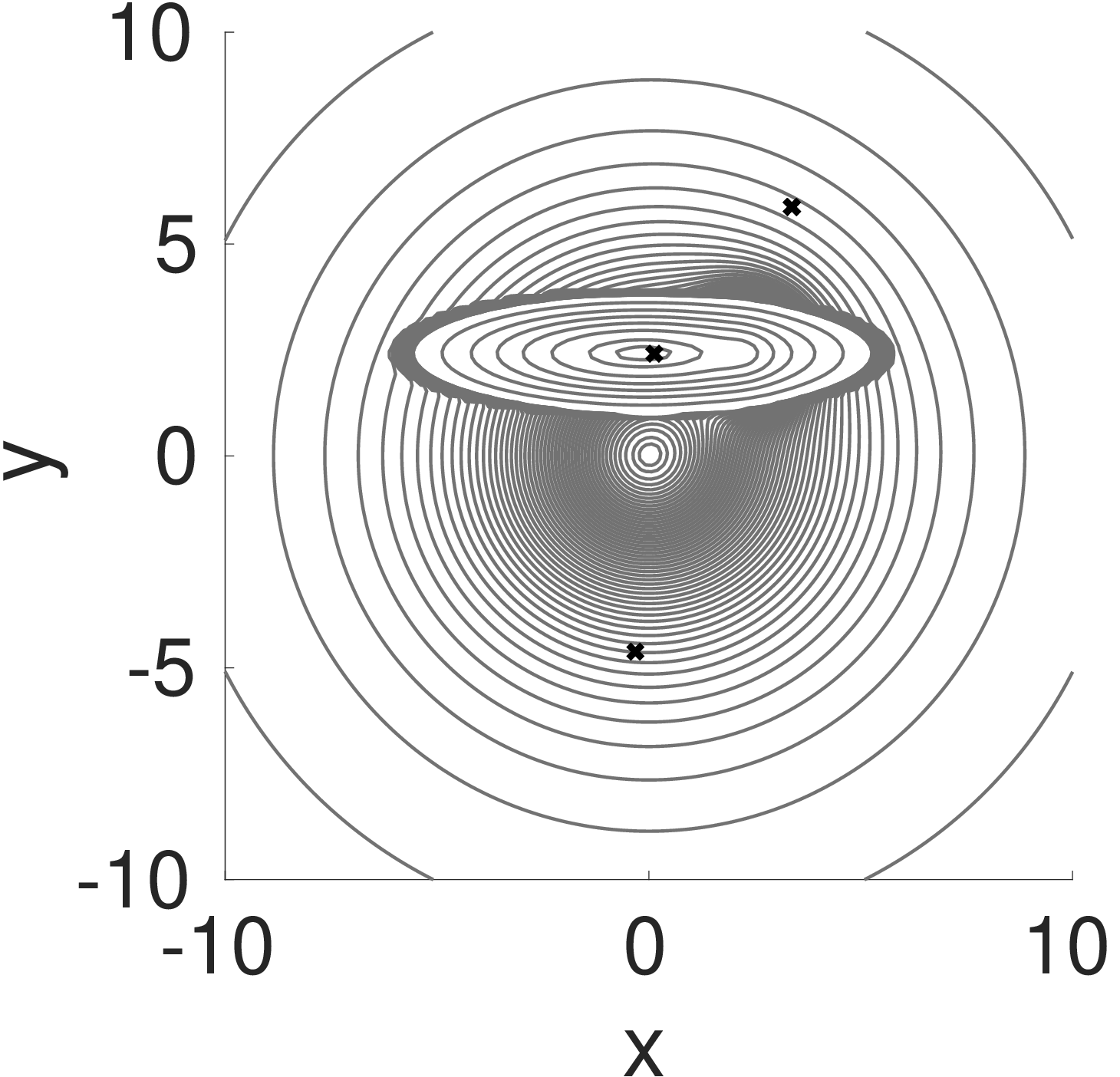}
\includegraphics[width=0.24\textwidth]{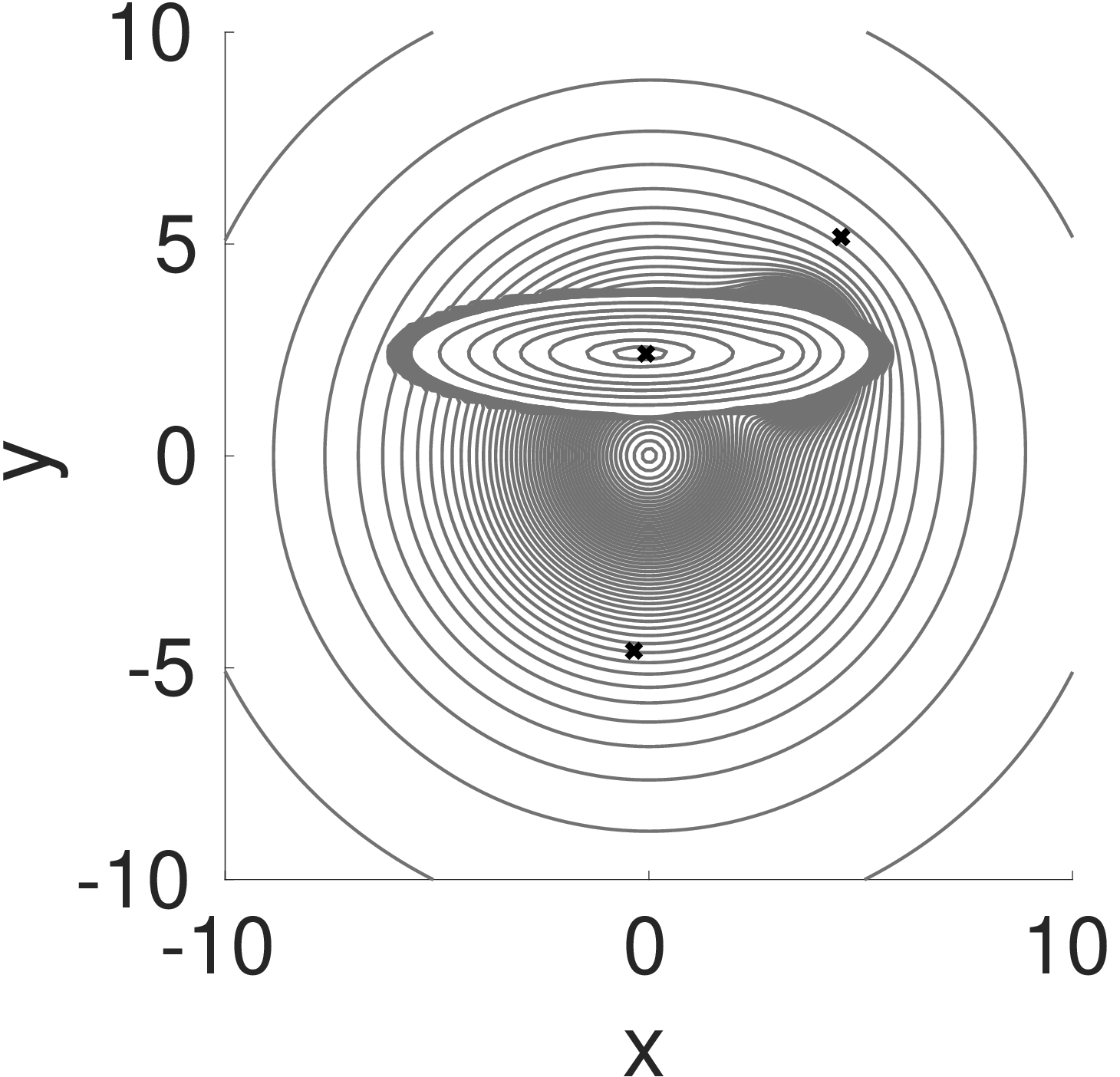}\\
\includegraphics[width=0.24\textwidth]{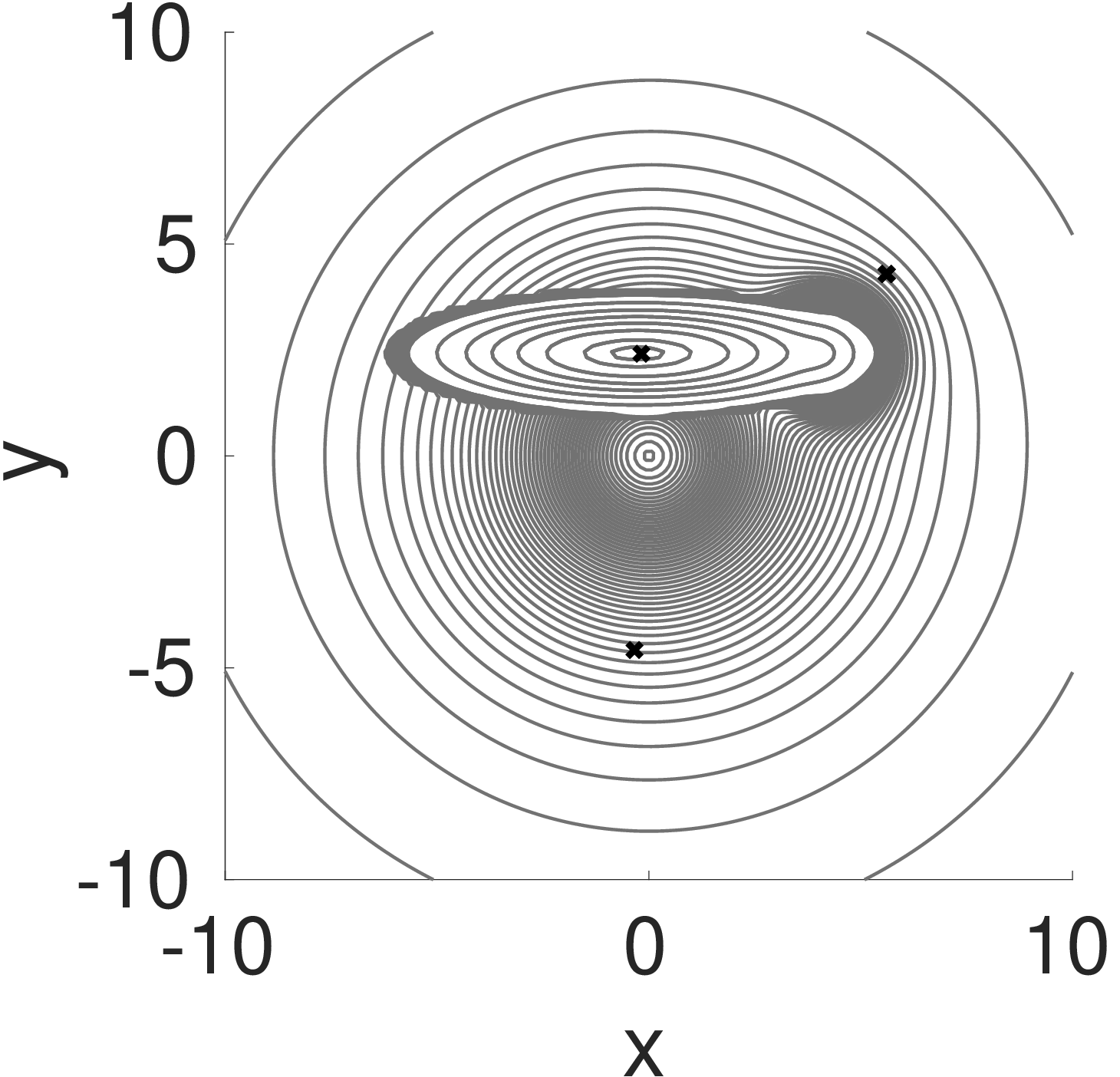}
\includegraphics[width=0.24\textwidth]{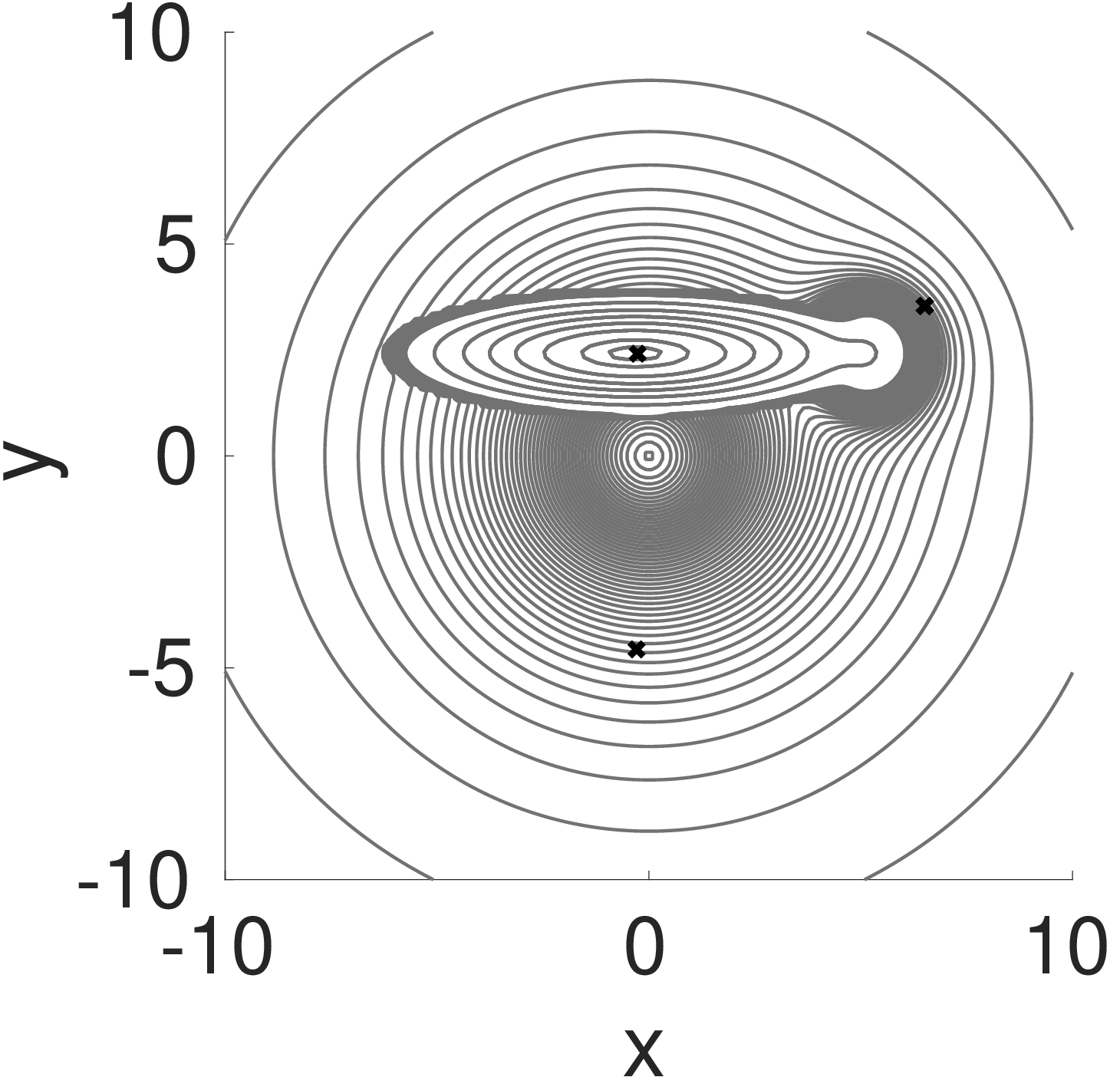}
\caption{Model C isodensity contours ($\Lambda = 2.42$ kpc, large offset with CM exterior to bar). Equilibrium points marked by black crosses. From top left to bottom right: $\delta = 2.5, 3.5, 4.5, 5.5$ kpc. The density distribution exhibits strong asymmetry.}
\label{fig:ModelC_isodens}
\end{figure}

The isodensity contours for Model C (Fig.~\ref{fig:ModelC_isodens}) reveal a configuration fundamentally different from Models A and B: the centre of mass of the system lies outside the bar, which affects the central density distribution.

\begin{figure}
\includegraphics[width=0.24\textwidth]{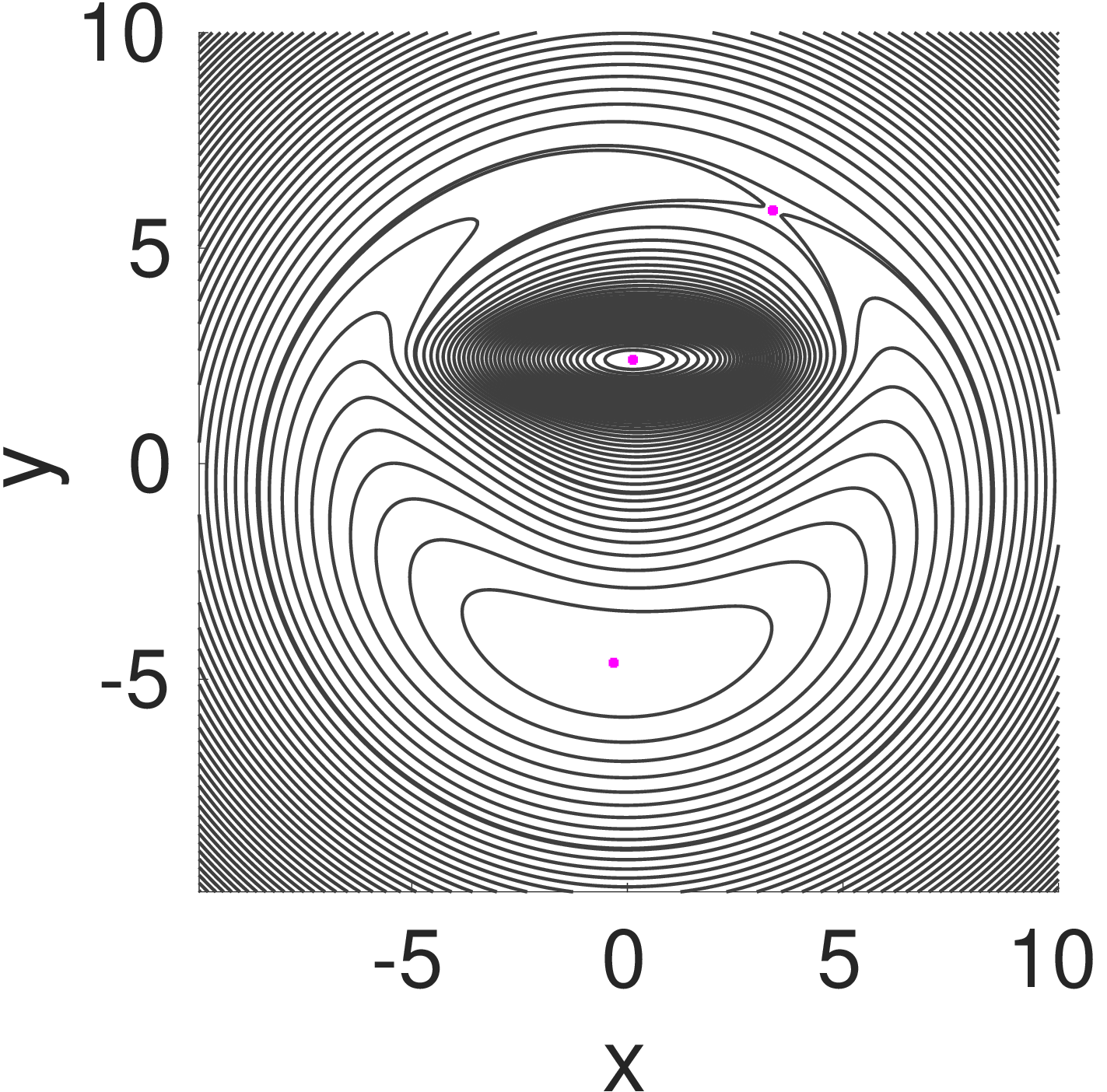}
\includegraphics[width=0.24\textwidth]{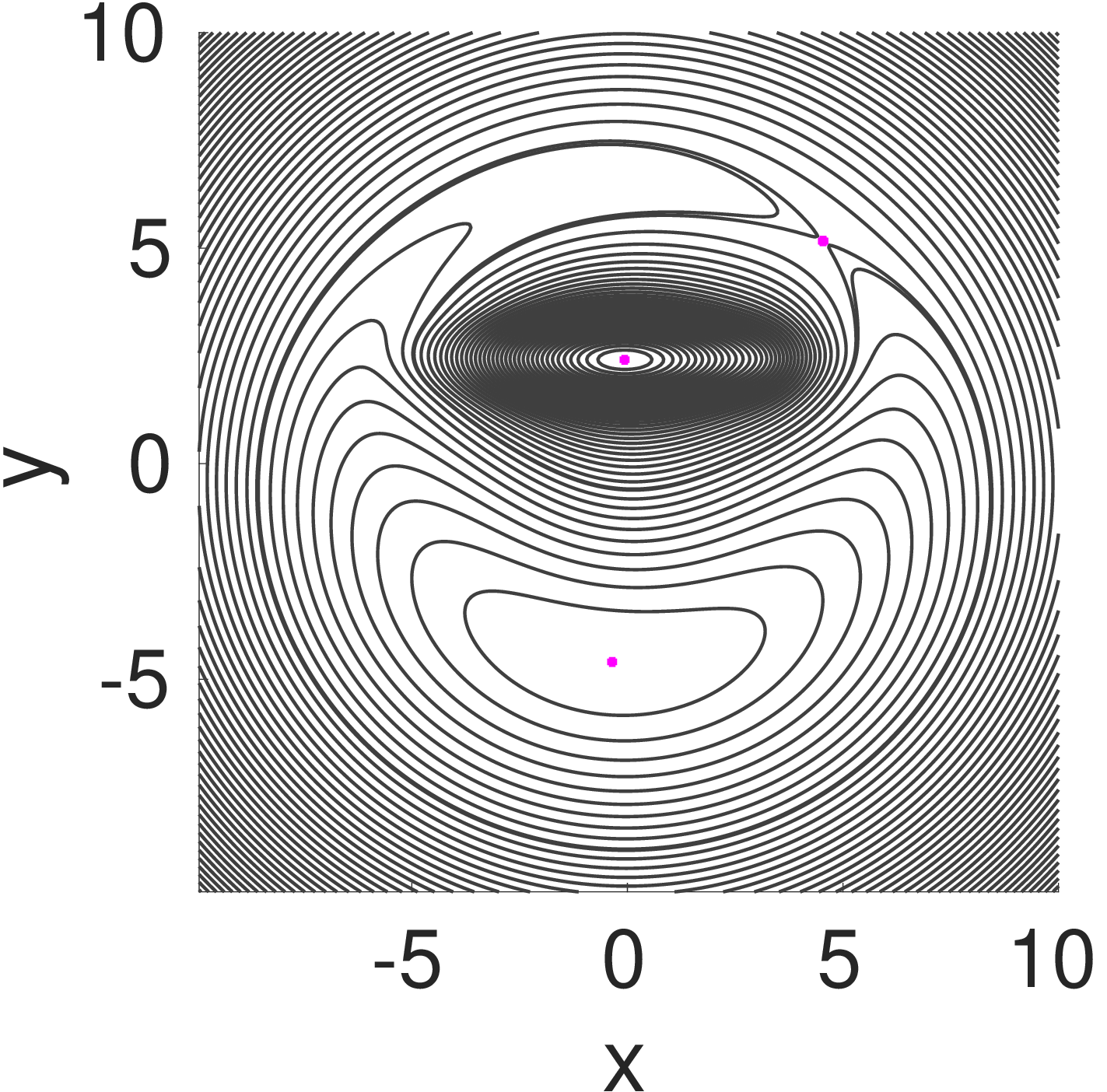}\\
\includegraphics[width=0.24\textwidth]{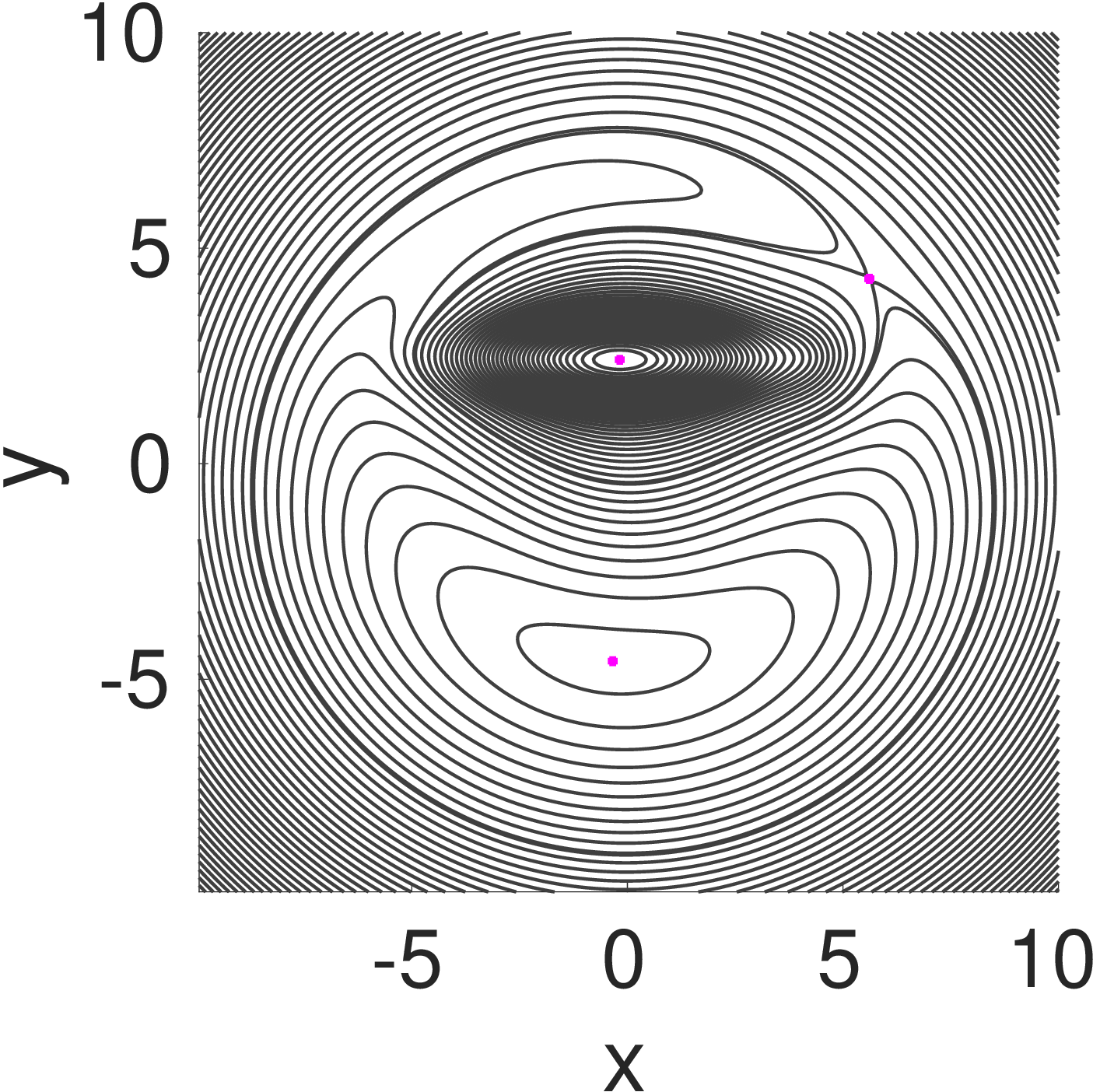}
\includegraphics[width=0.24\textwidth]{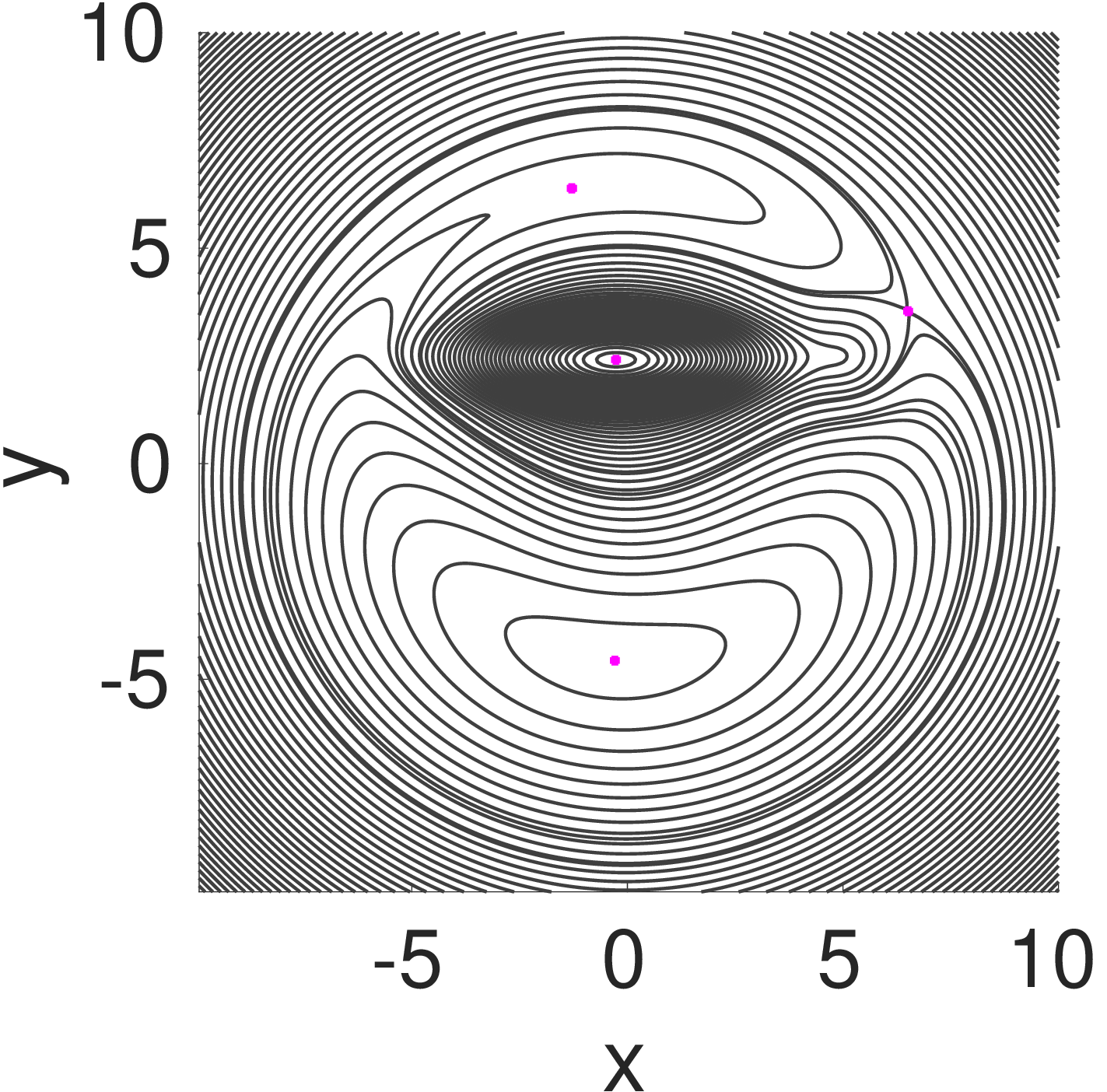}
\caption{Model C isopotential contours ($\Lambda = 2.42$ kpc). Equilibrium points marked by magenta dots. From top left to bottom right: $\delta = 2.5, 3.5, 4.5, 5.5$ kpc. The potential structure now supports only a single unstable equilibrium point.}
\label{fig:ModelC_isopot}
\end{figure}

The isopotential curves (Fig.~\ref{fig:ModelC_isopot}) reflect the critical transition: only one unstable equilibrium point L$_4$ remains, which has become unstable due to the bifurcation at $\Lambda = 1.5$ kpc.

\begin{figure}
\includegraphics[width=0.24\textwidth]{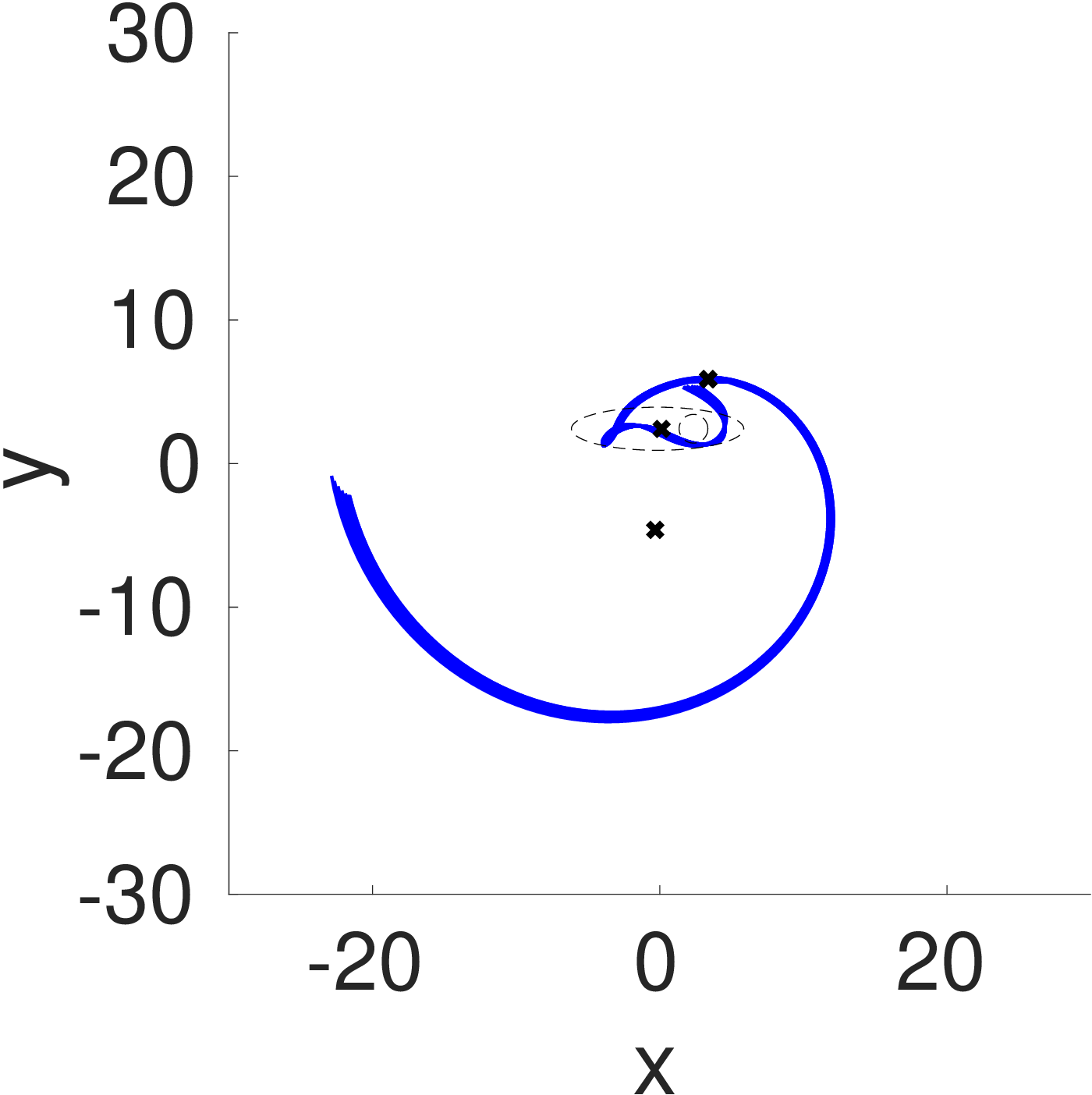}
\includegraphics[width=0.24\textwidth]{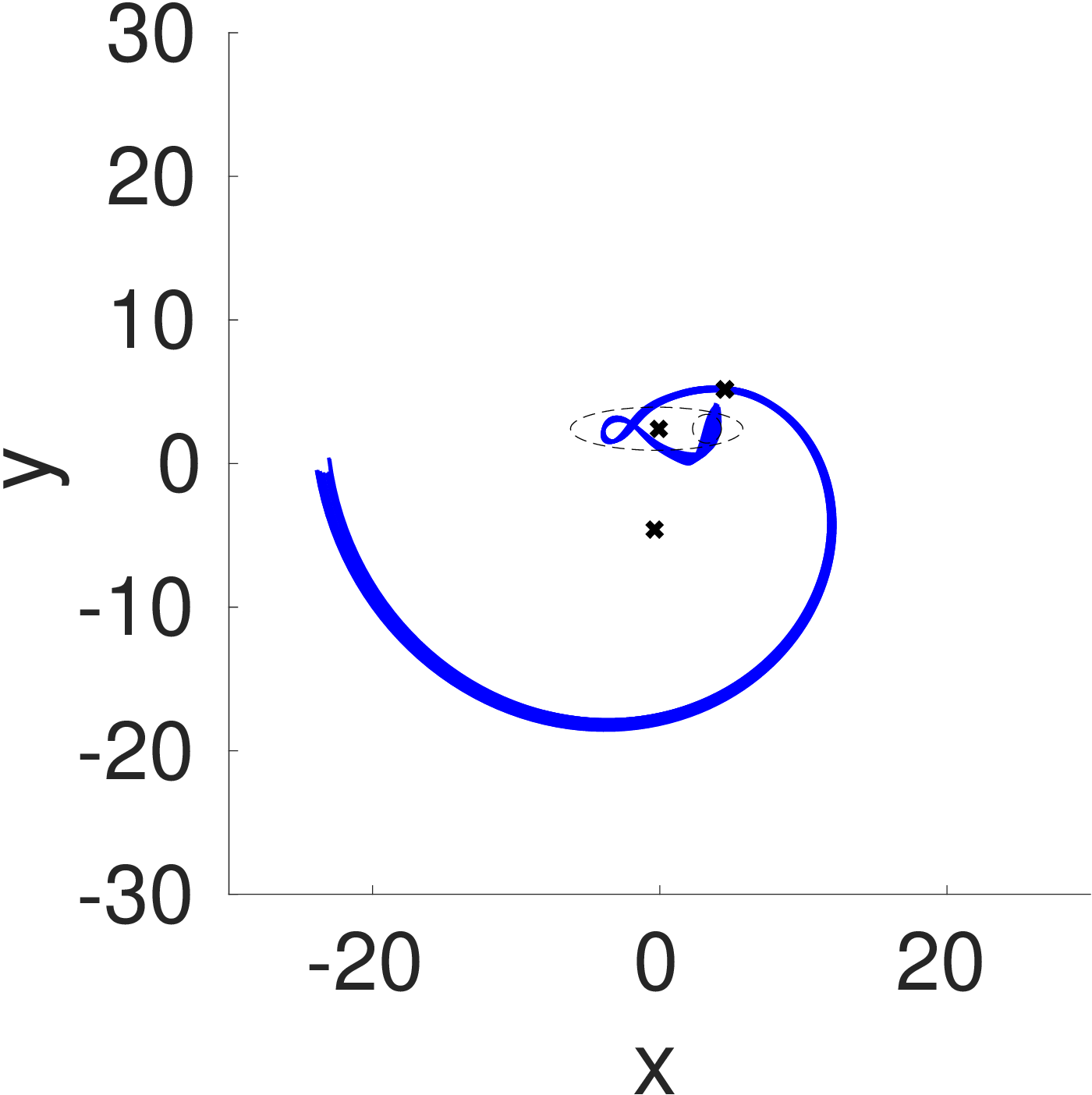}\\
\includegraphics[width=0.24\textwidth]{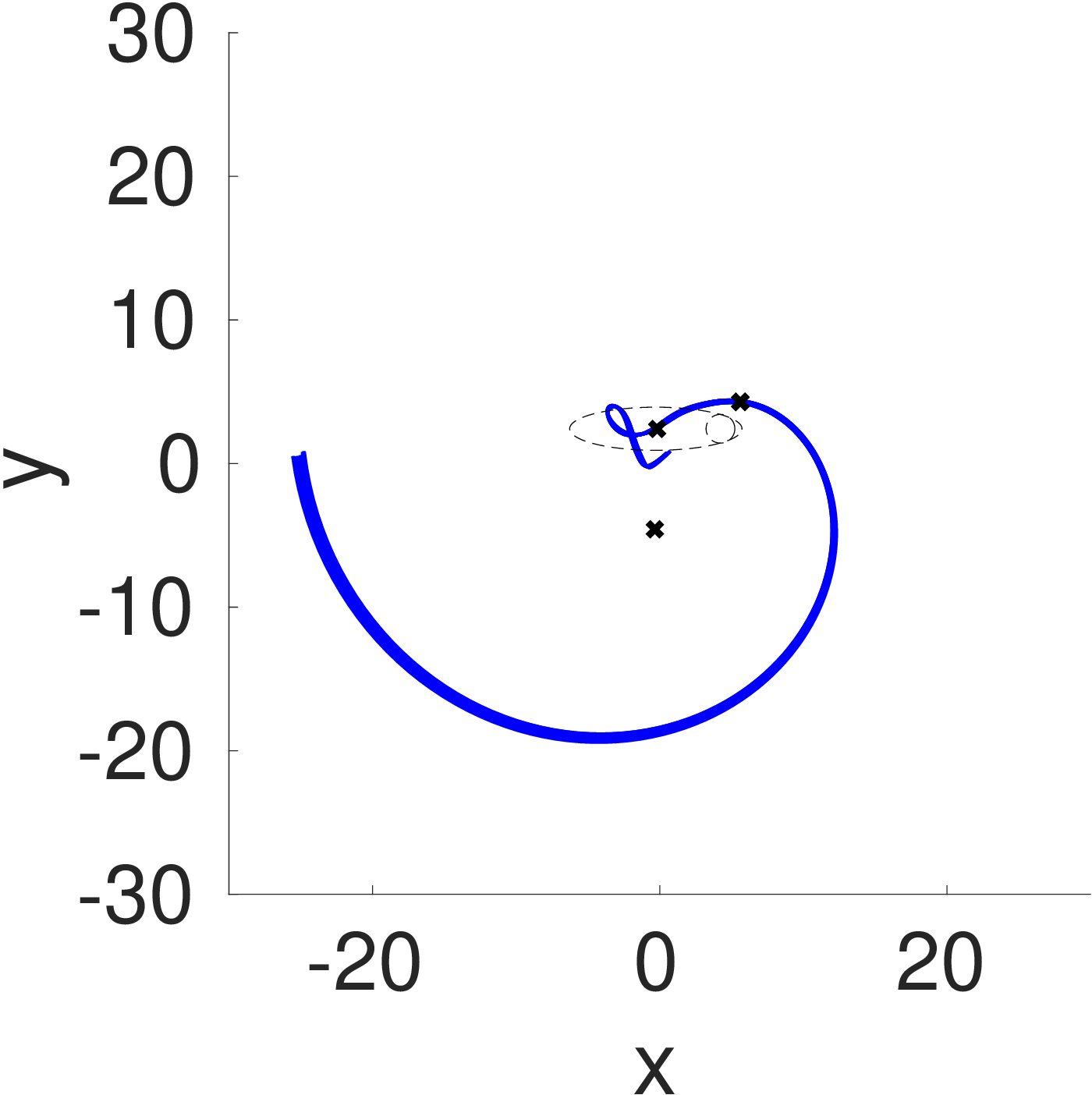}
\includegraphics[width=0.24\textwidth]{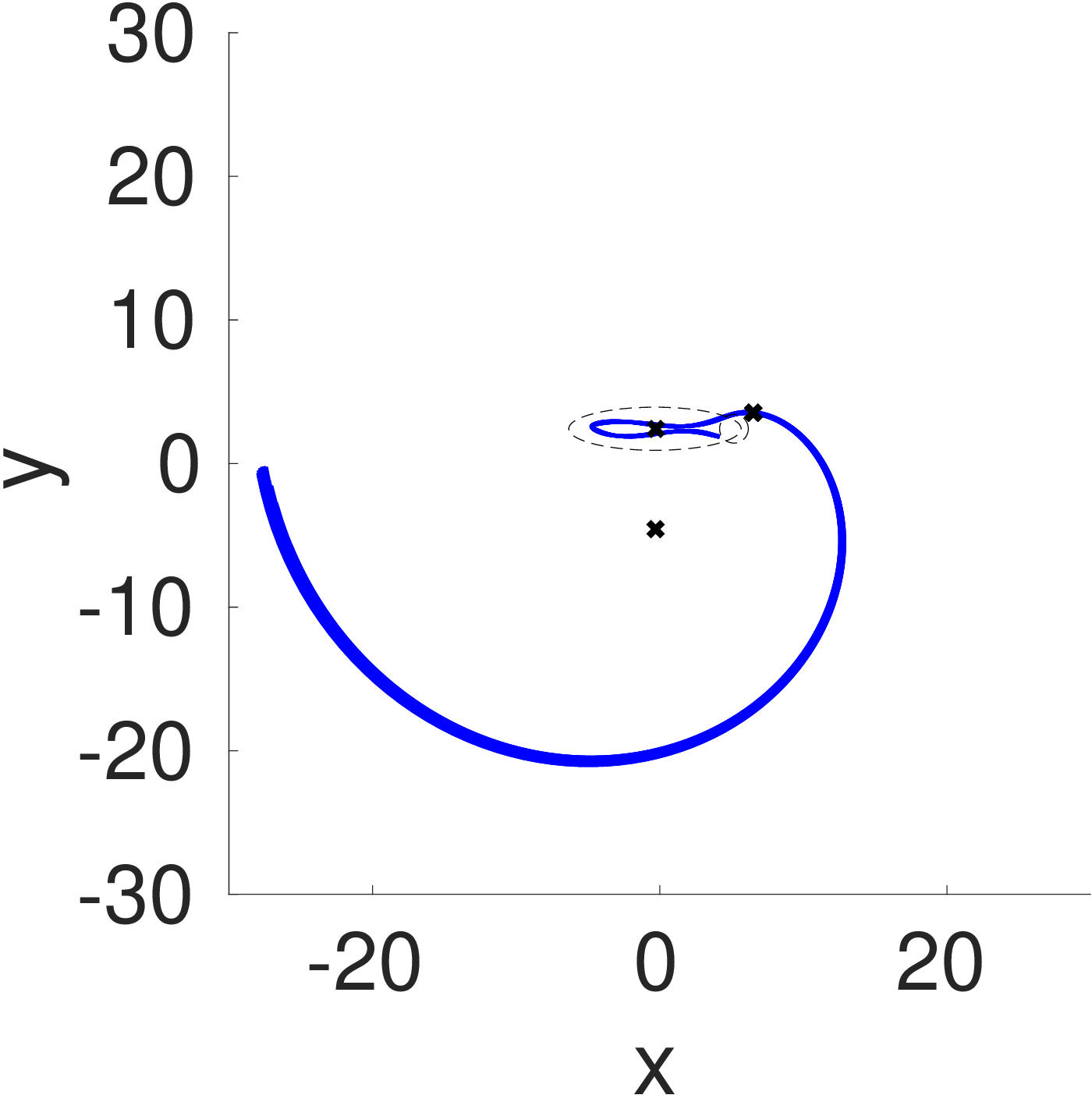}
\caption{Model C unstable invariant manifolds associated with the Lyapunov periodic orbits of the remaining unstable point L$_4$ ($\Lambda = 2.42$ kpc). Bar and the asymmetric mass component outlined by dotted black curves. From top left to bottom right: $\delta = 2.5, 3.5, 4.5, 5.5$ kpc. Only one arm is dynamically supported by the invariant manifold.}
\label{fig:ModelC_manifolds}
\end{figure}

The most significant result appears in Fig.~\ref{fig:ModelC_manifolds}: only a single invariant manifold emanates associated with Lyapunov orbits around the remaining unstable point L$_4$. This configuration fundamentally constrains the system to support only a one-armed morphology. The absence of the collinear unstable points L$_1$ and L$_2$, which existed in Models A and B, means that no second manifold can organize material transport. 

%%%%%%%%%%%%%%%%%%%%%%%%%%%%%%%%%%%%%%%%%%%%%%%%%%%%%%%%%%%%%%%%%%%%%%%%%%%%%%%%%%%%%%%
\section{Discussion}
\label{sec:discussion}

Observational studies indicate that these configurations are relevant in real systems. \citet{Kruk2017} report that barred galaxies with measurable offsets between the bar and the photometric centre of the disc tend, on average, to display stronger photometric lopsidedness than systems with bars aligned with the disc centre. They identify approximately 270 nearby barred galaxies with offsets in the range 0.2–2.5 kpc, which shows that displaced bars are relatively common. Within the framework considered here, it is plausible that a subset of these galaxies operates in or near the regime with three equilibrium points described by Model C, where only one unstable point survives and a single arm is dynamically supported.

\citet{Zaritsky2013} find that the degree of lopsidedness correlates with the character of the arms: galaxies with regular and well-defined arm patterns tend to exhibit weaker lopsided distortions, whereas systems with strong lopsidedness often show less symmetric arm structure. They also argue that lopsidedness is a generic property of galaxies and suggest that its amplitude is related to the underlying stellar and dark matter distributions. The models presented here provide a dynamical interpretation of these results. Moderate internal mass imbalances in a bar whose major axis passes through the system’s centre of mass (Model A)  already generate significant asymmetries in the density and transport efficiency of the manifolds associated with L$_1$ and L$_2$, leading to two arms with different densities.

When the bar is offset but the centre of mass remains inside the bar (Model B), the two unstable points are preserved but the manifolds associated with L$_1$ and L$_2$ become strongly asymmetric, which resembles galaxies with two discernible but unequal arms. For larger offsets (Model C), the system enters a regime in which a single unstable equilibrium point survives and there is only one invariant manifold supporting an arm. In this picture, galaxies with mild lopsidedness correspond to configurations where the structure with five equilibrium points and two manifolds is still present, while strongly lopsided systems are associated with configurations that have three equilibrium points and one arm.

The Large Magellanic Cloud (LMC) is a nearby system where these dynamical mechanisms may be operating. Analyses based on Gaia data and complementary photometric catalogues indicate a strongly asymmetric morphology, with a dominant one-armed structure and a bar that is both off-centre and intrinsically lopsided \citep[e.g.][]{JimenezArranz2023, Scholch2025}. In the context of our models, such a configuration is naturally associated with the large-offset regime represented by Models C, in which the centre of mass lies outside the bar ellipsoid. In this case the bifurcation at $\Lambda \approx 1.5$ kpc eliminates the collinear unstable points L$_1$ and L$_2$ and leaves only one unstable point, whose invariant manifold support a single arm (see Fig.~\ref{fig:ModelC_manifolds}).

Smaller offsets, such as those in Model B with $\Lambda < 1.5$ kpc, preserve the configuration with five equilibrium points and the two unstable points L$_1$ and L$_2$. The corresponding invariant manifolds support a two-armed pattern with distinctly asymmetric arms (see Fig.~\ref{fig:ModelB_manifolds}), one of which is weaker and lies very close to the bar, a configuration that could also be consistent with the observed morphology of the LMC \citep{ElYoussoufi2019, Grady2021, Ripepi2022}. This regime may likewise describe galaxies with moderate offsets and less extreme lopsidedness. The relative magnitude of the offset between bar and disc could therefore act as an indicator of the underlying configuration of equilibrium points and the associated manifold structure. A quantitative comparison between our models and the LMC requires applying bar detection and pattern speed estimation methods such as those of \citet{Dehnen2023, Pfenniger2023, BarDetection}. This lies beyond the scope of the present study and will be addressed in future work.

The bifurcation structure discussed in Section~\ref{sec:bifurcations}, together with the manifold organisation presented in Section~\ref{sec:manifolds}, defines a dynamical framework that links bar offsets, the configuration of equilibrium points, and the appearance of asymmetric arm morphologies. The results suggest that both internal mass imbalances and global misalignments between bar and disc are able to dynamically sustain asymmetric arm morphologies. From a dynamical systems perspective, an offset bar and a one- or two-armed spiral structure are not separate, competing effects that need to be disentangled, but are intrinsically linked features. The invariant manifolds that delineate the galactic arms are structures that are created by and dynamically tied to the bar, i.e., they do not form a posteriori as a delayed response of the material to the bar. Under this dynamical framework, regardless of the mechanism that displaces the bar, the associated invariant manifolds are displaced accordingly. The material guided by these shifted manifolds would then form the observed one- or two-armed structure. Overall, this paradigm provides a basis for interpreting the variety of asymmetric structures observed in barred galaxies.

\section{Conclusions}
\label{sec:conclusions}

In this work, we have analysed how internal mass imbalances within the bar and offsets between the bar and the disc modify the configuration of the equilibrium points and the associated invariant manifolds in barred galaxies. By combining numerical continuation of equilibrium points with a study of the invariant manifolds associated to Lyapunov orbits around the unstable points, we have identified dynamical regimes that support either two-armed structures with asymmetric properties or predominantly one-armed structures.

Our analysis demonstrates that the structure of equilibrium points determines which invariant manifolds exist in the system, and consequently, which arm morphologies can be supported. Through numerical continuation, we revealed a hierarchical structure of bifurcations under parameter variation. For purely internal mass imbalances (Model A), the system creates a subsystem with three equilibrium points via a pair of saddle-node bifurcations (at $\delta \approx 5.7$ and $\delta \approx 8.6$~kpc) when the displaced mass moves beyond the physical extent of the bar, but ultimately returns to a configuration with five equilibrium points. More importantly, the system is robust under small bar-disc offsets (Model B): the five-point configuration persists without bifurcation, preserving the two unstable collinear points required to sustain a two-armed, albeit asymmetric, spiral structure.

However, a qualitative transition occurs for larger offsets. When the offset exceeds the bar's semi-minor axis ($\Lambda \approx 1.5$~kpc, Model C), the collinear unstable points L$_1$ and L$_2$ are removed, reducing the system to only three equilibrium points. This change suppresses the transport channels associated with two-armed configurations and leaves only a single unstable equilibrium point. The location of this transition depends only moderately on the pattern speed $\Omega$, although high values of $\Omega$ can produce additional transient equilibrium points.

%%%%%%%%%%%%%%%%%%%%%%%%%%%%%%%%%%%%%%%%%%%%%%%%%%%%%%%%%%%%%%%%%%%%%%%%%%%%%%%%%%%
%\input{6-acknow}
\begin{acknowledgements}
P.S.M. thanks the Spanish Ministry of Economy grant and PID2021-123968NB-I00. M.R.G. acknowledges that this work was (partially) supported by the Spanish MICIN/AEI/10.13039/501100011033 and by "ERDF A way of making Europe" by the European Union through grants PID2021-122842OB-C21 and PID2024-157964OB-C21, the Institute of Cosmos Sciences University of Barcelona (ICCUB, Unidad de Excelencia María de Maeztu) through grant CEX2024-001451-M and the project 2021-SGR-00679 GRC de l’Agència de Gestió d’Ajuts Universitaris i de Recerca (Generalitat de Catalunya). J.J.M. thanks MINECO-FEDER for the grant PID2024-158570NB-I00. We thank the anonymous referee for their constructive comments, which helped to improve the manuscript.
\end{acknowledgements}

\bibliographystyle{aa} % style aa.bst
\bibliography{biblio} % your references Yourfile.bib

\begin{appendix}

\section{Appendix A: Influence of the disc centre location}

\begin{figure}[ht]
\centering
\includegraphics[width=0.24\textwidth]{images/curvas_varios_isopotencial_halomass_0_4_bulgeM_0_04_halo_xd_0_yd_-2_bulgexd_2_5.eps}
\includegraphics[width=0.24\textwidth]{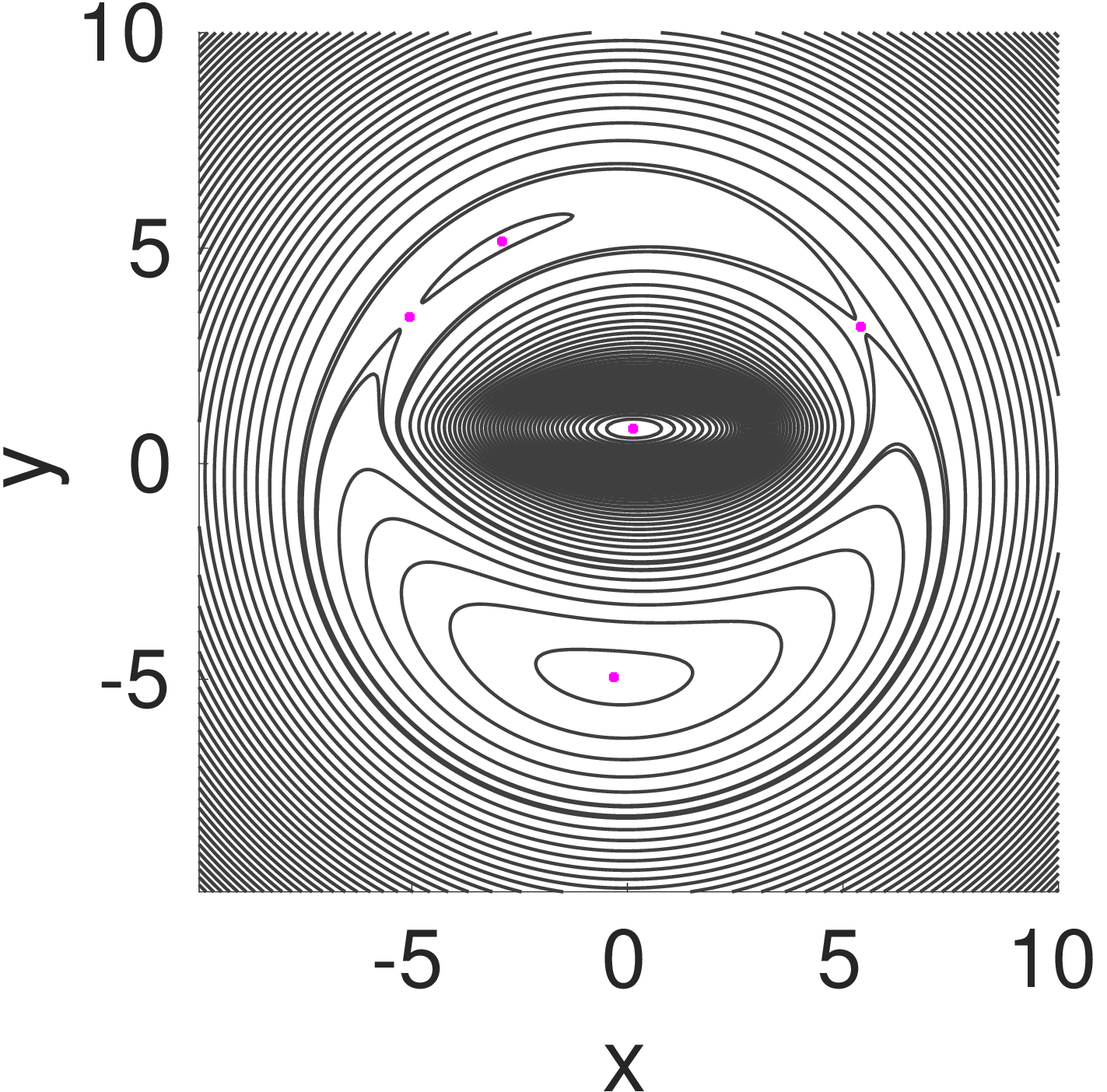}

\includegraphics[width=0.24\textwidth]{images/curvas_varios_isopotencial_halomass_0_4_bulgeM_0_04_halo_xd_0_yd_-4_bulgexd_2_5.eps}
\includegraphics[width=0.24\textwidth]{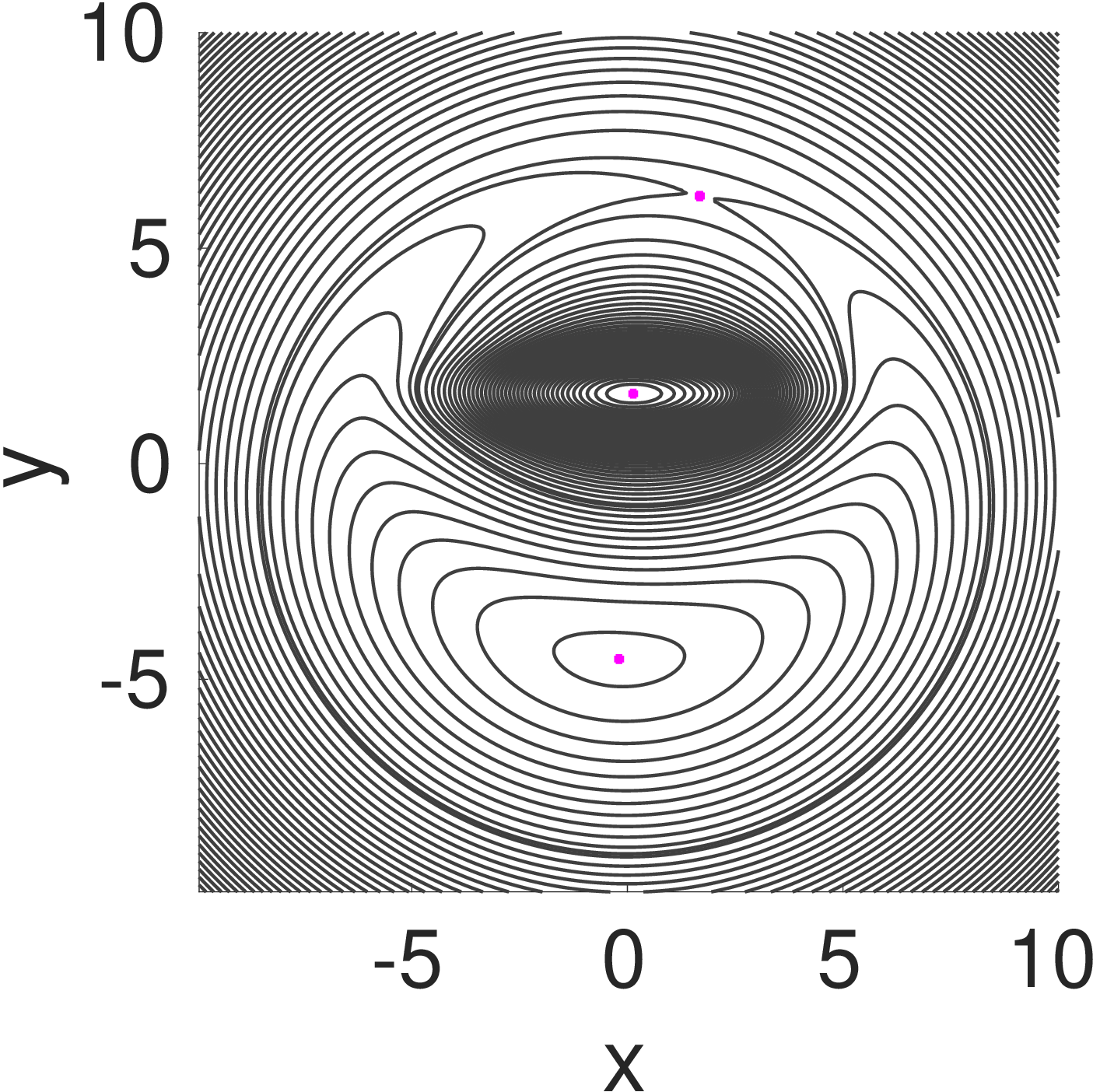}
\caption{Isopotential curves for Models B (top) and C (bottom) with $\delta=2.5$ kpc. Left: Original models used in the main text with the disc centre located at the origin of the reference system. Right: Modified models with the disc centre shifted to the bar centre.}
\label{fig:isop}
\end{figure}

To ensure that our results are robust with respect to the choice of placing the geometric centre of the disc at the coordinate origin, we tested an alternative setup where the disc centre is shifted to coincide with the geometric centre of the bar. In this configuration, both components are offset with respect to the system's centre of mass. To demonstrate that the results undergo only minor quantitative changes, leaving the underlying qualitative dynamics completely unaffected, Fig.~\ref{fig:isop} compares the isopotential contours of Models B and C (with $\delta=2.5$ kpc). The left column displays the original models used in the main text, while the right column displays the same models but with the disc centre shifted to the bar centre. The potential landscape, which dictates the Lagrangian point configuration and the structure of the invariant manifolds, remains qualitatively the same, consequently, the resulting arm morphology is unaffected.

\end{appendix}

%====================================

% Don't change these lines
% \bsp	% typesetting comment
% \label{lastpage}
\end{document}